\documentclass[twocolumn]{aastex61}
\submitjournal{ApJ}

\shorttitle{Galaxies with prolate rotation in Illustris}
\shortauthors{I. Ebrov\'{a} and E. L. {\L}okas}

\begin{document}

\title{Galaxies with prolate rotation in Illustris}

\correspondingauthor{Ivana Ebrov\'{a}}
\email{ebrova.ivana@gmail.com}

\author{Ivana Ebrov\'{a}}
\affil{Nicolaus Copernicus Astronomical Center, Polish Academy of Sciences, Bartycka 18, 00-716 Warsaw, Poland}

\author{Ewa L. {\L}okas}
\affil{Nicolaus Copernicus Astronomical Center, Polish Academy of Sciences, Bartycka 18, 00-716 Warsaw, Poland}

\begin{abstract}
Tens of early-type galaxies have been recently reported to possess prolate rotation of the stellar component,
i.e. significant amount of rotation around the major axis, including two cases in the Local Group. Although expected theoretically, this phenomenon is rarely observed and remains elusive.
We study its origin using the population of well-resolved galaxies in the Illustris cosmological simulation.
We identify 59 convincing examples of prolate rotators at the present time,
more frequently among more massive galaxies, with the number varying very little with redshift. We follow their evolution back
in time using the main progenitor branch galaxies of the Illustris merger trees. We find that the emergence of prolate
rotation is strongly correlated with the time of the last significant merger the galaxy experienced, although other
evolutionary paths leading to prolate rotation are also possible. The transition to prolate rotation most often happens
around the same time as the transition to prolate shape of the stellar component. The mergers leading to prolate
rotation have slightly more radial orbits, higher mass ratios, and occur at more recent times than mergers in the
reference sample of twin galaxies we construct for comparison. However, they cover a wide range of initial conditions
in terms of the mass ratio, merger time, radiality of the progenitor orbits, and the relative orientations of progenitor spins
with respect to the orbital angular momenta. About half of our sample of prolate rotators were created
during gas-rich mergers and the newly formed stars usually support prolate rotation.
\end{abstract}


\section{Introduction} \label{sec:intro}

Peculiar kinematics of galaxies, such as a kinematically decoupled component or misalignment between kinematic and
photometric axes, can carry valuable information on the evolution of galaxies. In this paper, we focus on
the phenomenon of prolate rotation of the stellar component. 
This type of rotation is also often referred to as `minor-axis rotation', since observationally it is
detected as a gradient of the mean line-of-sight velocity along the minor axis of the projected image of the galaxy.
Such a gradient indicates rotation around the 3D major axis and an ellipsoid rotationally symmetric around the major
axis is by definition prolate. This is the origin of the designation `prolate rotation', although the situation is
probably more complicated as galaxies are generally expected to be triaxial. Theoretical studies
of the dynamics of triaxial ellipsoids have demonstrated that rotation around the major axis is expected in terms of
the family of orbits called long axis tubes in addition to the short axis tube orbits that rotate around the minor
axis \citep{dz85}.

In the present era of large spatially resolved spectroscopic surveys the number of galaxies with
prolate rotation detected in stellar kinematics 
is rapidly growing and therefore the stronger is the motivation to better understand this phenomenon.
In a recent work, \cite{tsa17} list 12 early-type galaxies (ETGs) known to be prolate rotators and
additional 4 galaxies with some evidence for prolate rotation, although not so clear. They also supply 8
newly-reported ETGs with prolate rotation, and one candidate for a prolate rotator, from the CALIFA Survey.
About 8 more prolate rotators were reported in the preliminary results of the M3G (MUSE Most Massive Galaxies)
survey, an ongoing project targeting the most massive galaxies in the Shapley Supercluster and brightest cluster
galaxies in rich clusters (Emsellem, Krajnovi{\'c}, conference presentations
\footnote{\url{http://www.astroscu.unam.mx/galaxies2016/presentaciones/Miercoles/Cozumel2016_Emsellem.pdf}}
\footnote{\url{https://www.eso.org/sci/meetings/2015/StellarHalos2015/talks_presentation/emsellem_M3G.pdf}}
\footnote{\url{http://www.sr.bham.ac.uk/footsteps15/talks/day4/Krajnovic_footsteps2015.pdf}}).
In total we now have 20 confirmed known prolate rotators and further 13 promising candidates among massive galaxies.
Prolate rotation, or minor-axis velocity gradients, are also observed in the bulges of early-type spirals \citep[e.g.,][]{ber99,fa03} or barred galaxies \citep[e.g.][]{ja88}.

\begin{figure*}
\resizebox{\hsize}{!}{\includegraphics{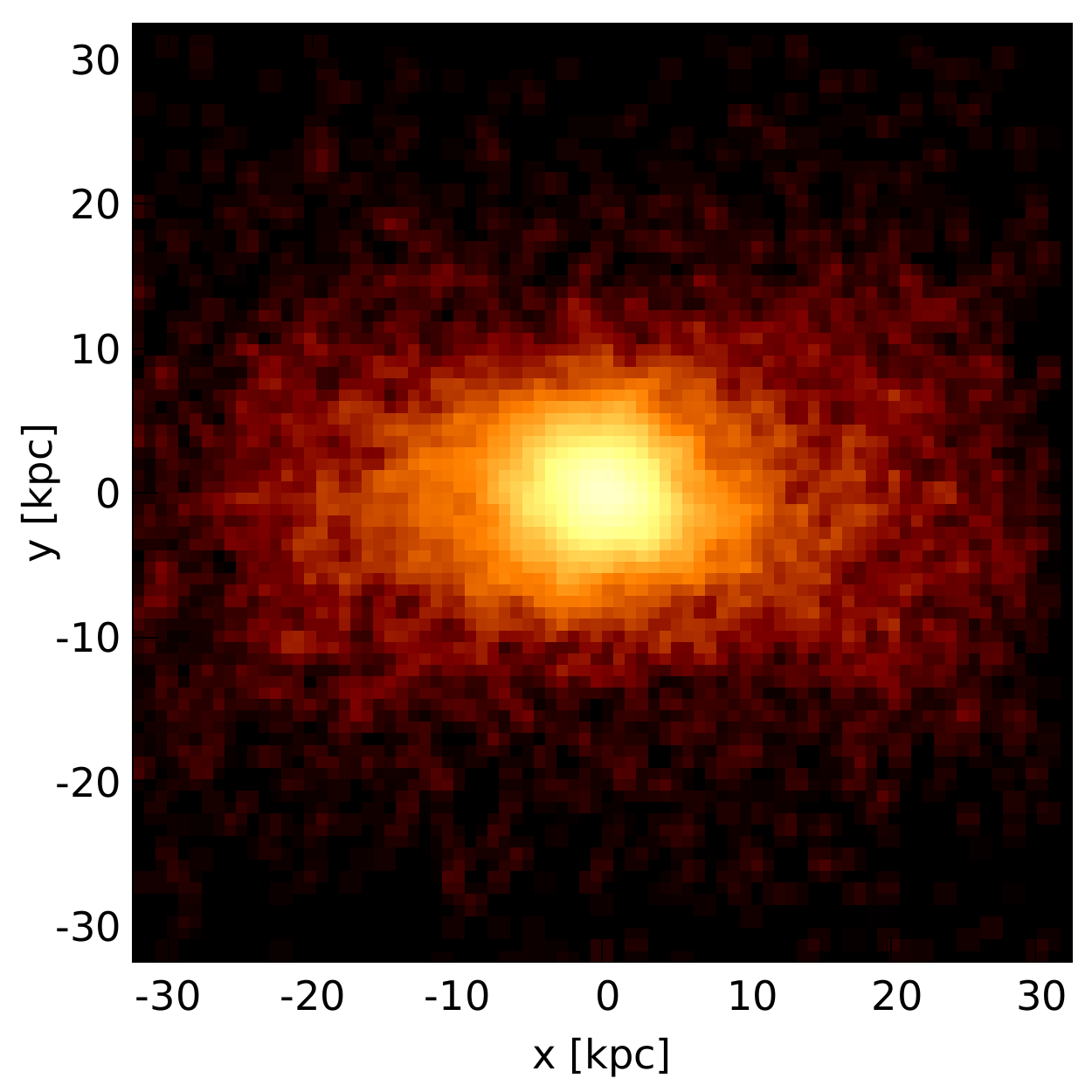}\includegraphics{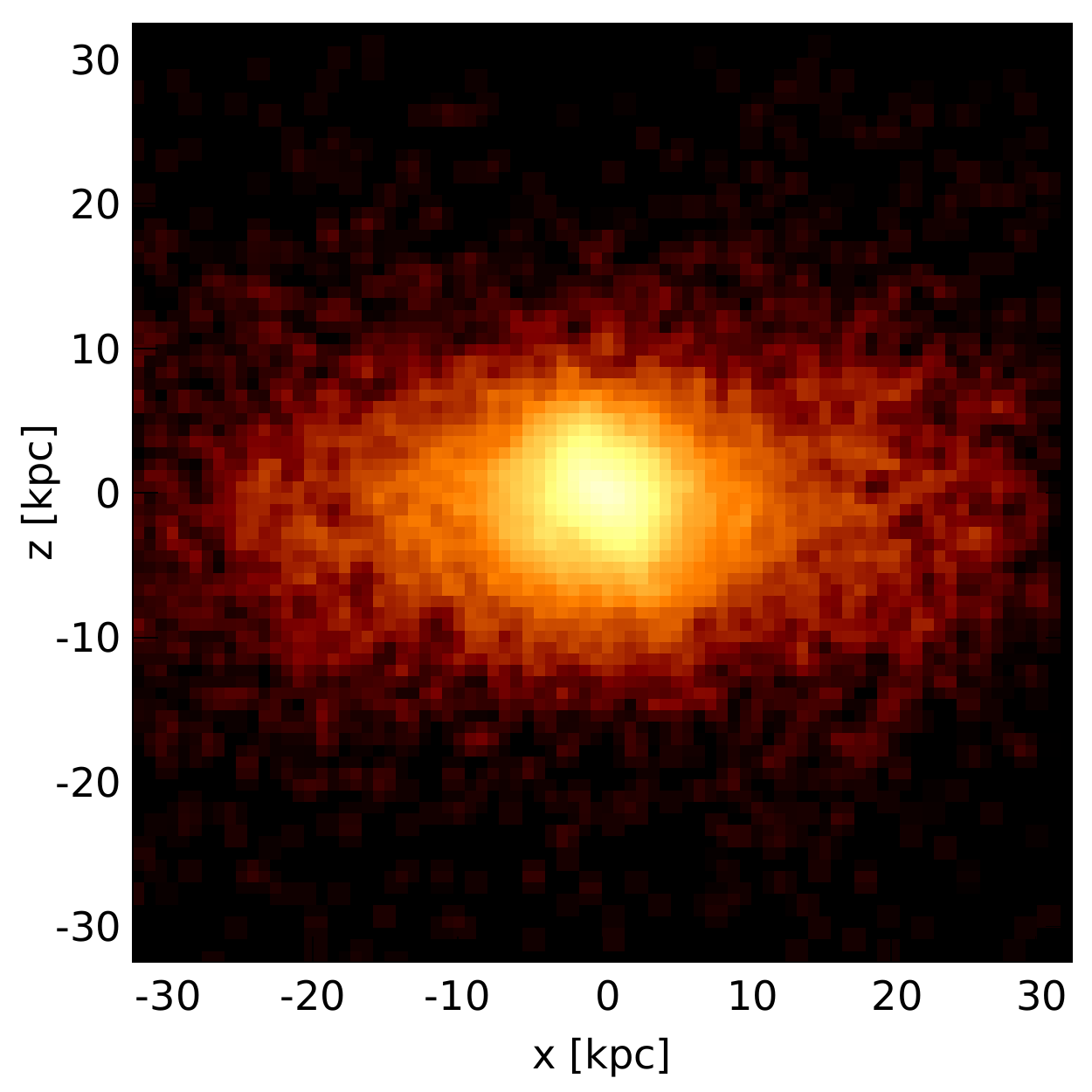}\includegraphics{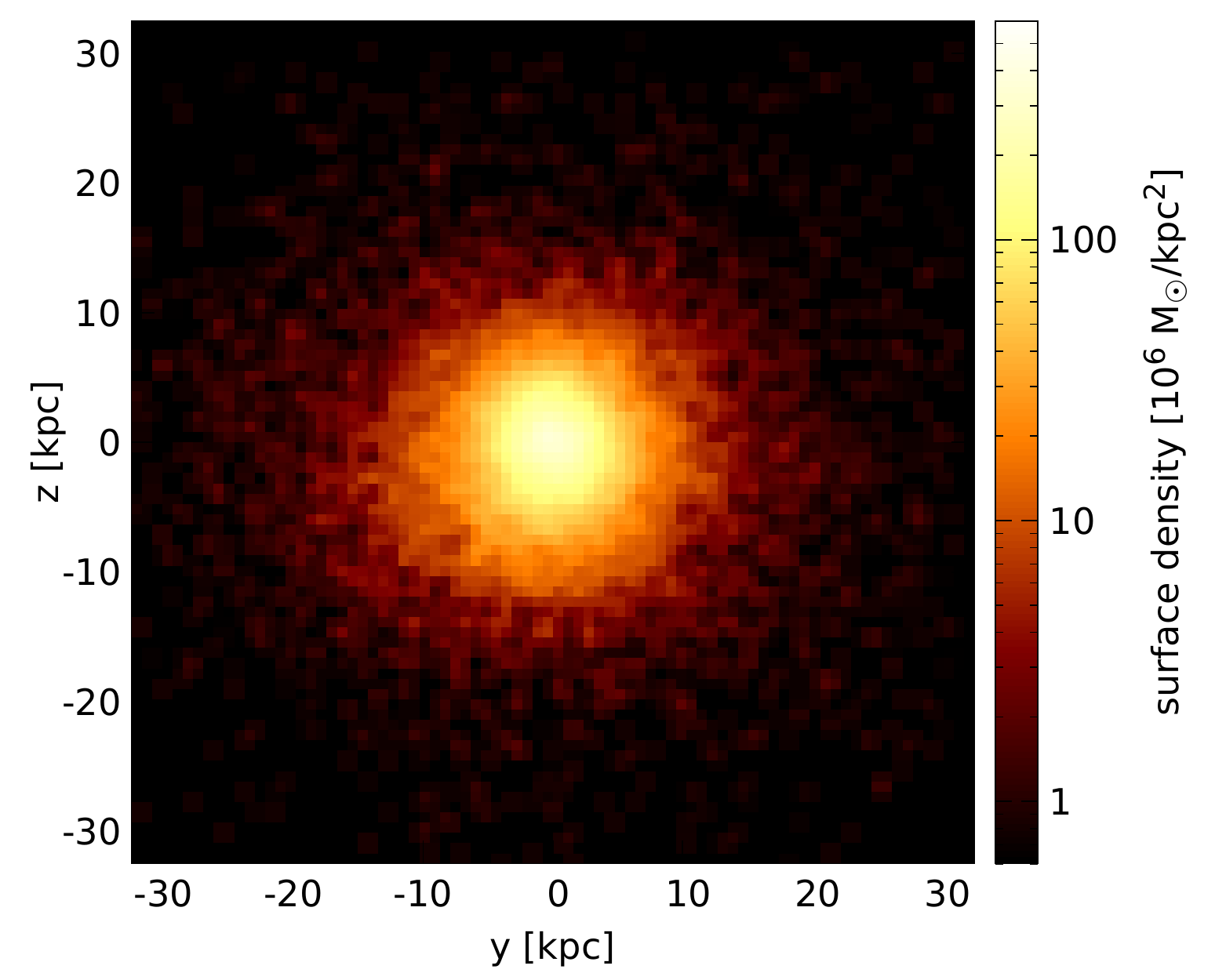}}
\resizebox{\hsize}{!}{\includegraphics{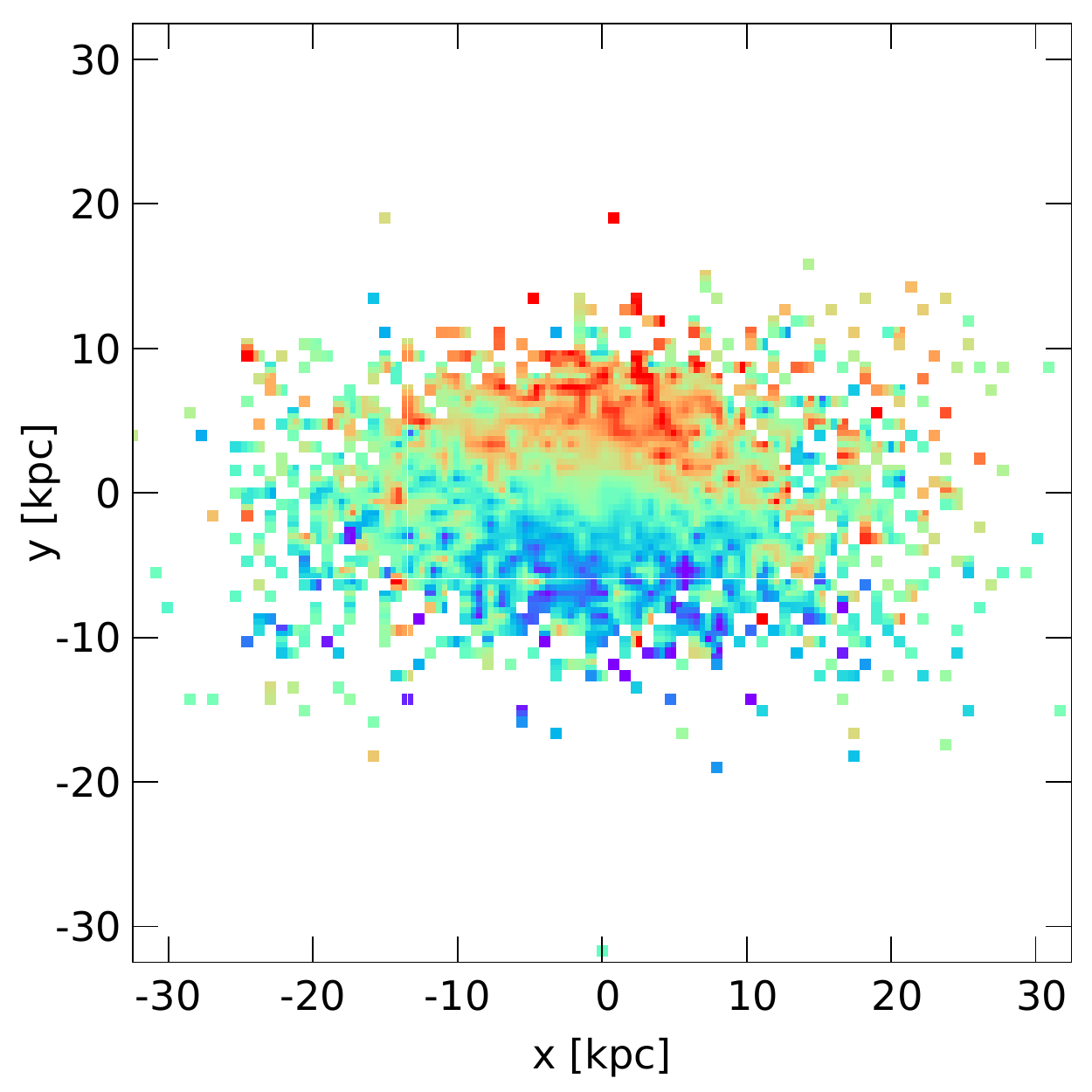}\includegraphics{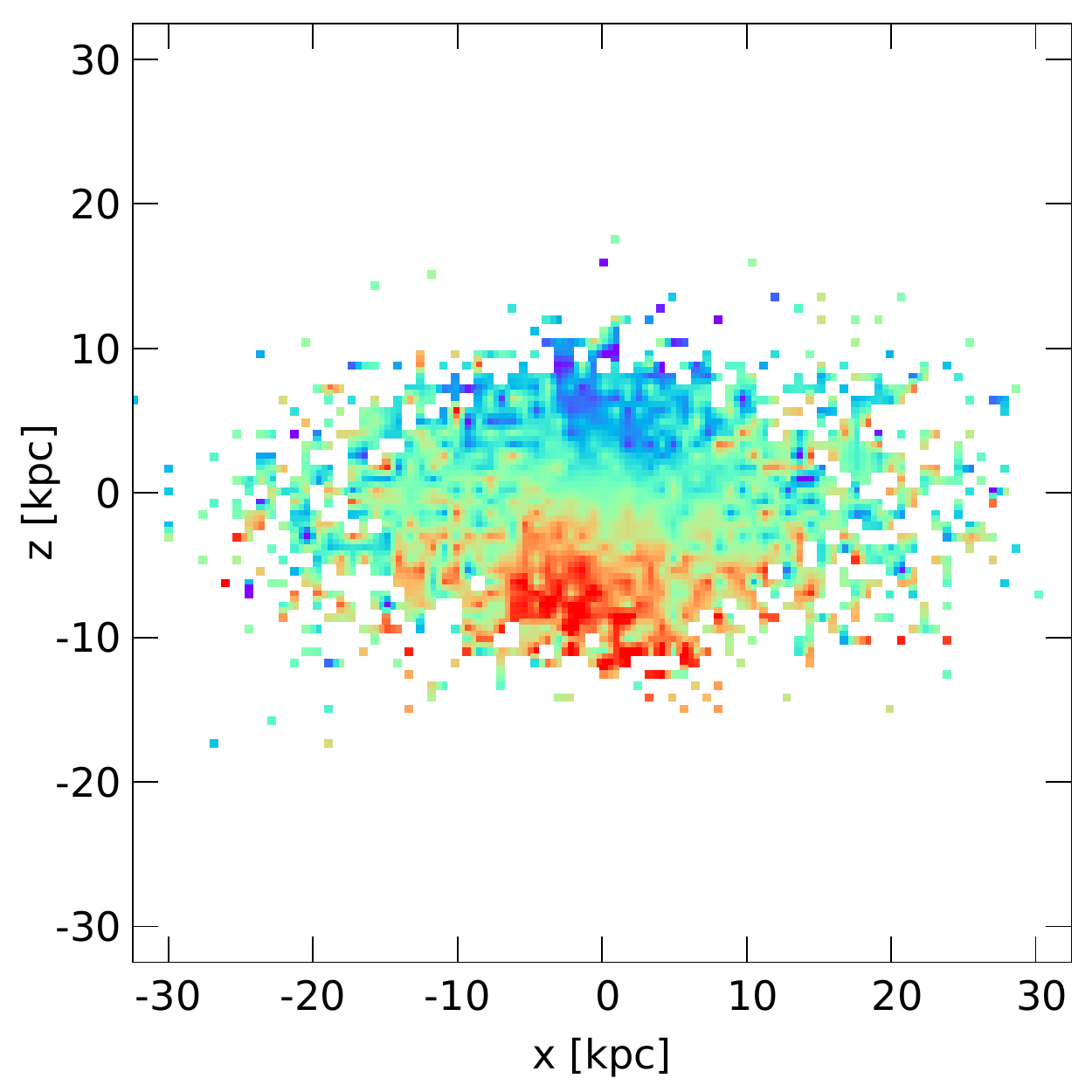}\includegraphics{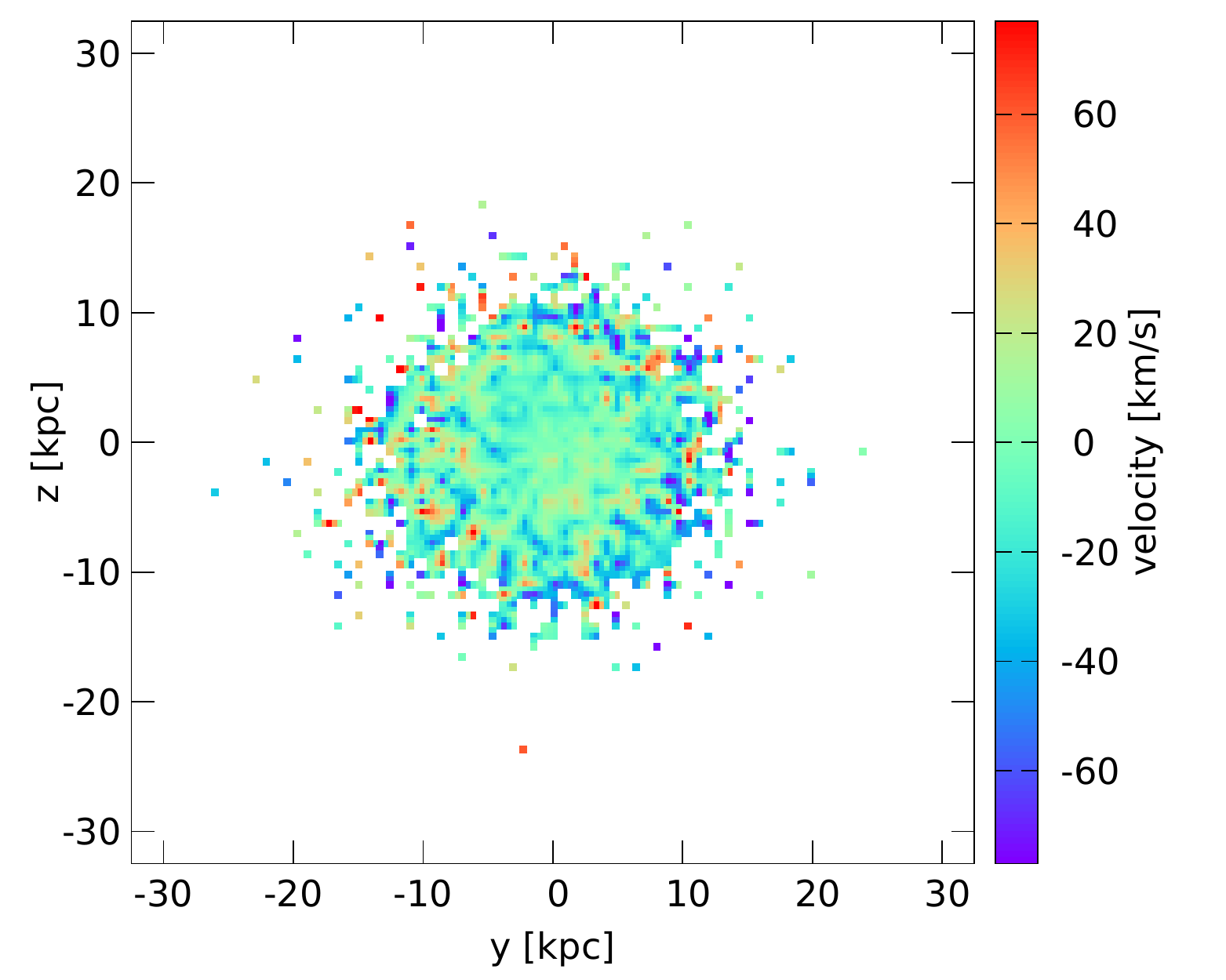}}
\caption{
Maps of the stellar surface density and the mean line-of-sight velocity for galaxy-1, our best example of prolate
rotator selected from Illustris. The columns show the views along the minor (left), intermediate (middle) and major
(right) axis of the stellar component.
The range on the axes corresponds to [-$r_{\rm max}$, $r_{\rm max}$], see Table~\ref{tab:g1-6}.
\label{fig:map-g1} }
\end{figure*}

\begin{figure}
\resizebox{\hsize}{!}{\includegraphics{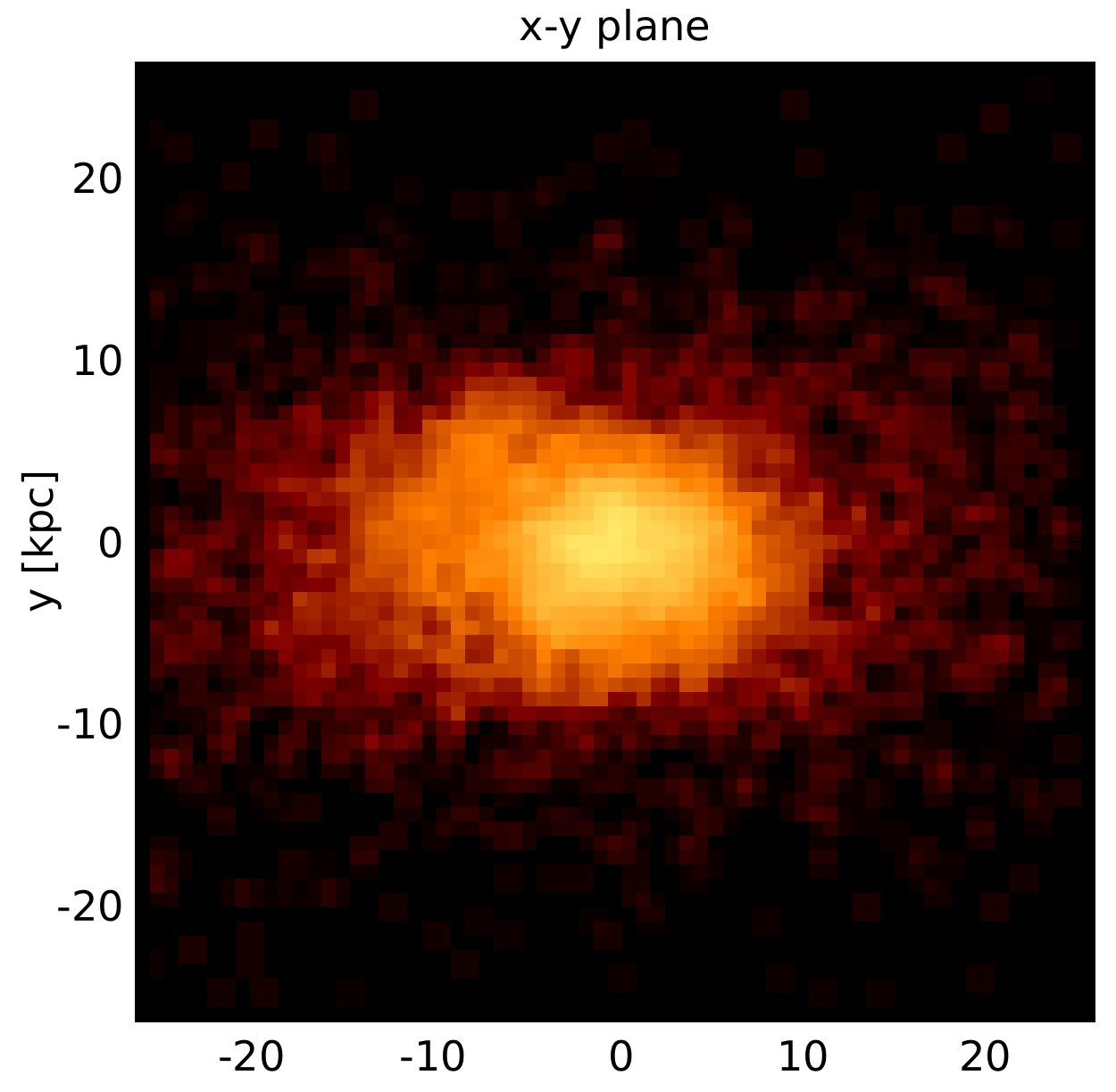}\includegraphics{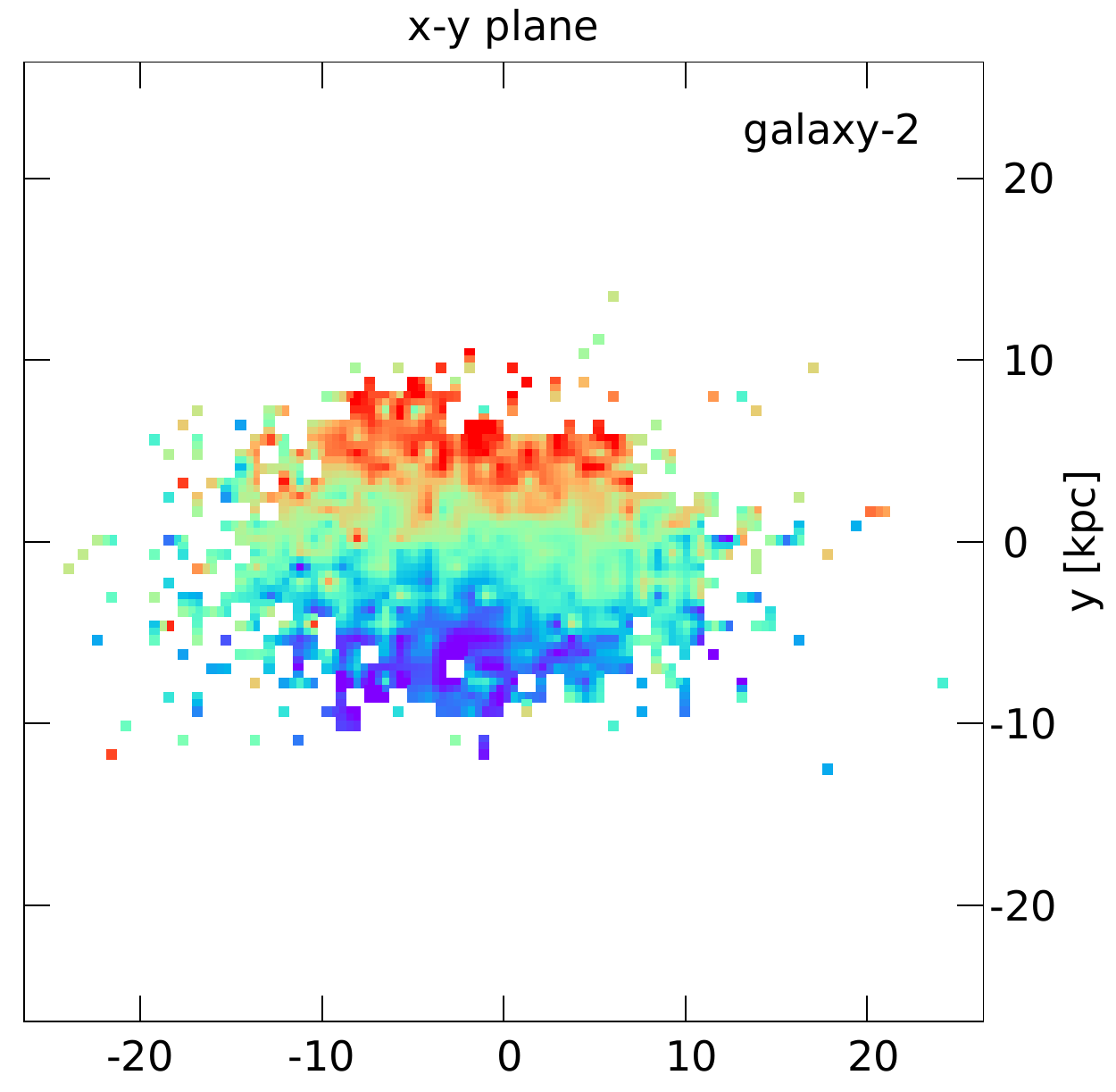}}
\resizebox{\hsize}{!}{\includegraphics{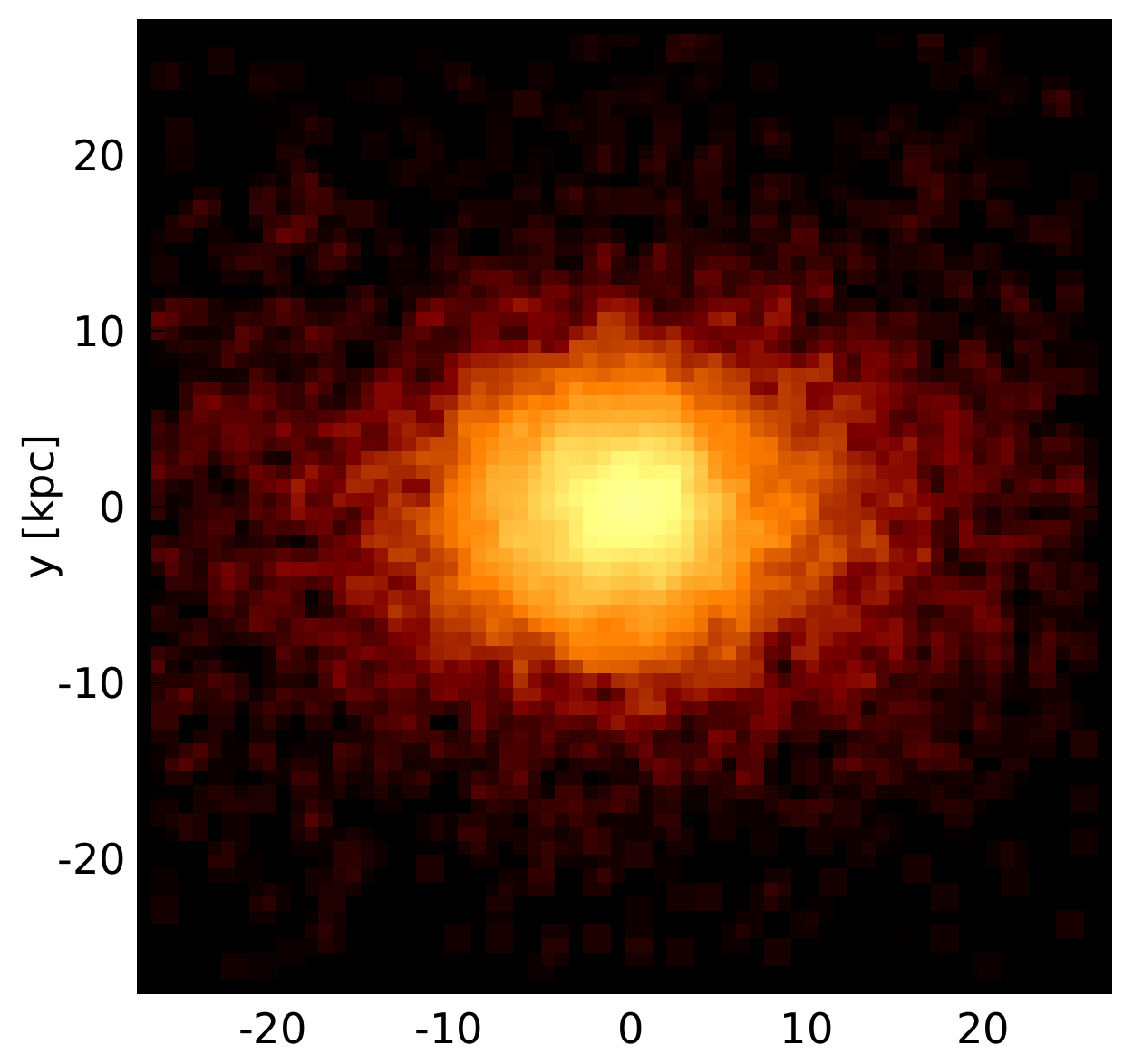}\includegraphics{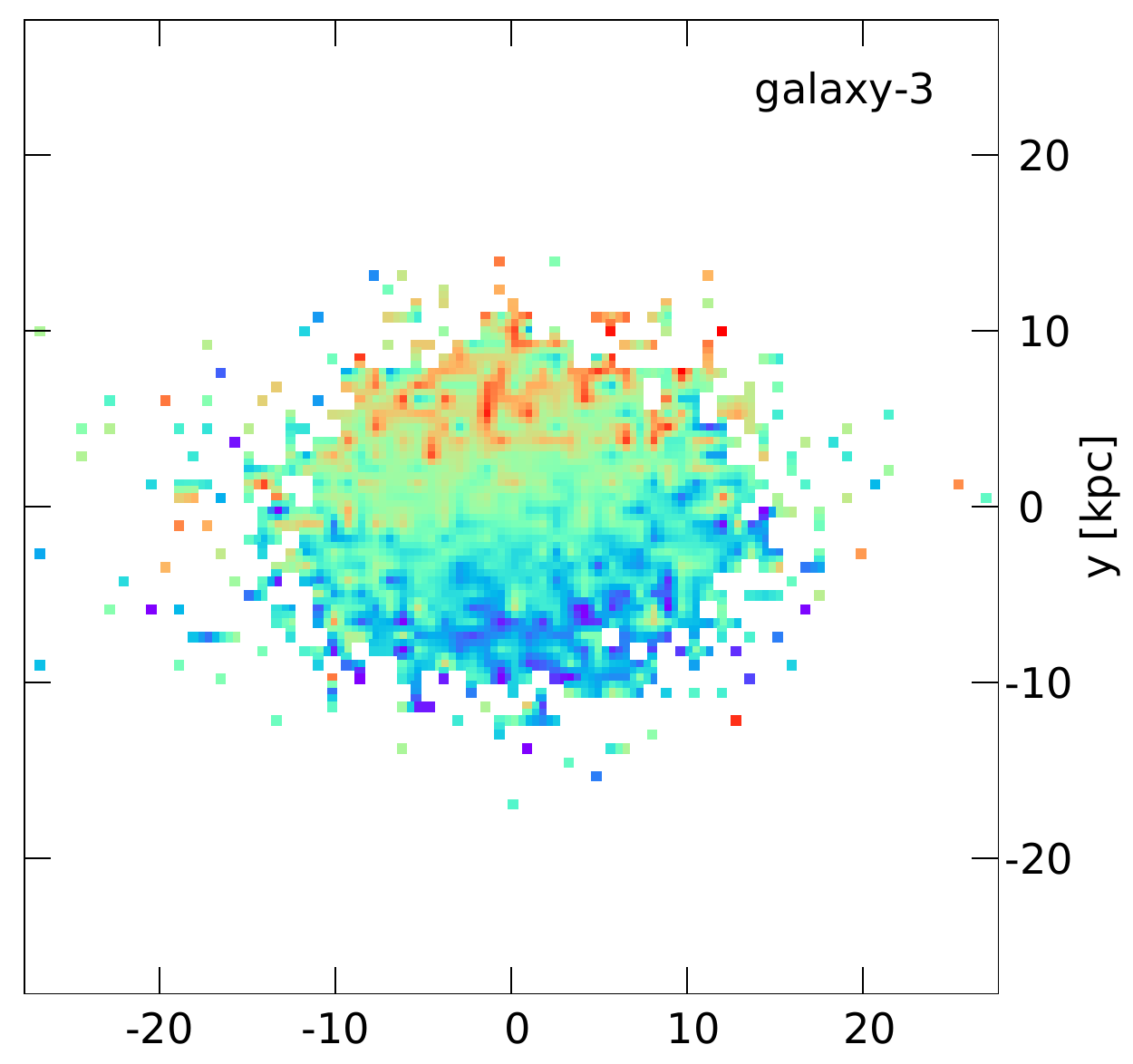}}
\resizebox{\hsize}{!}{\includegraphics{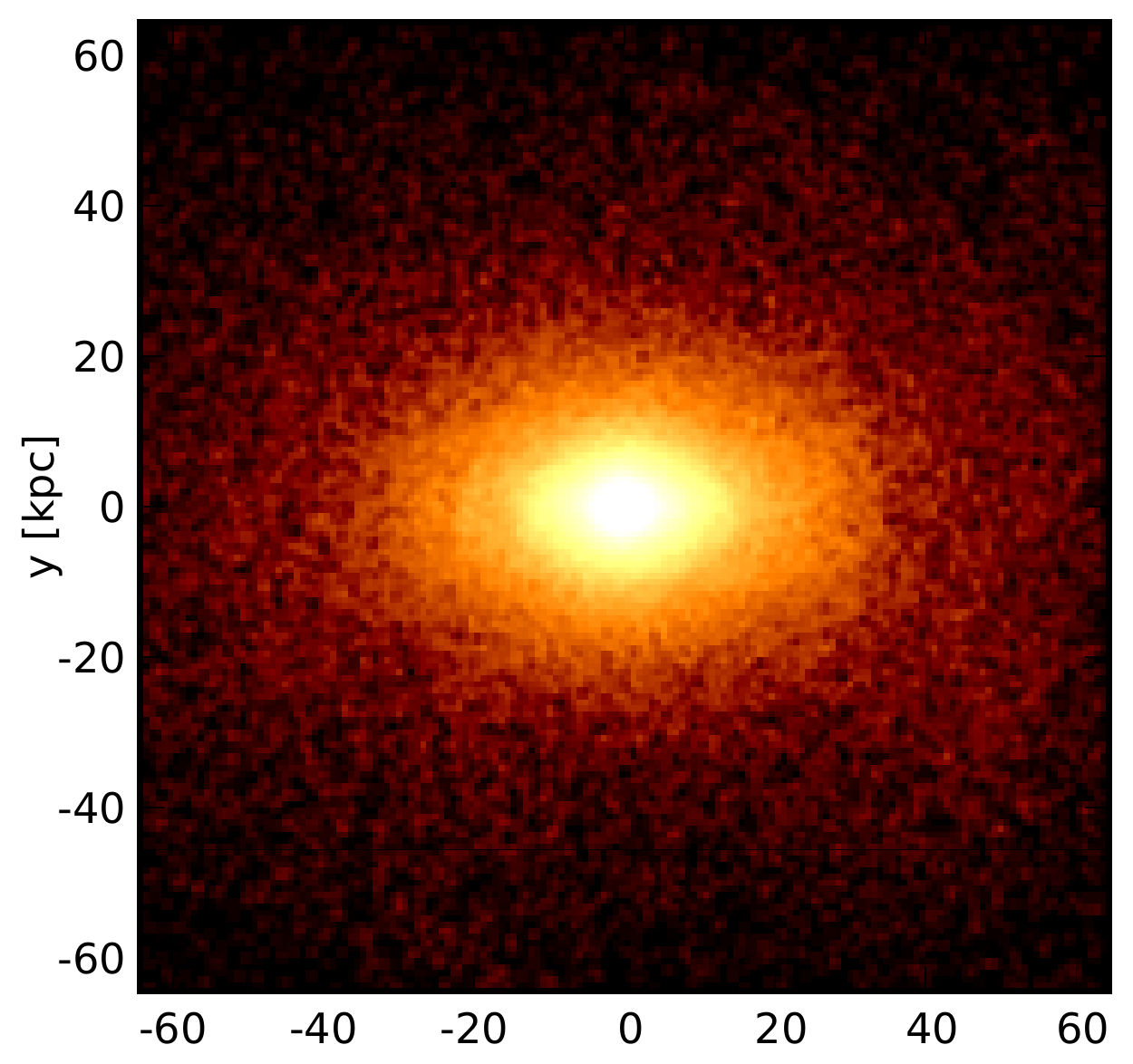}\includegraphics{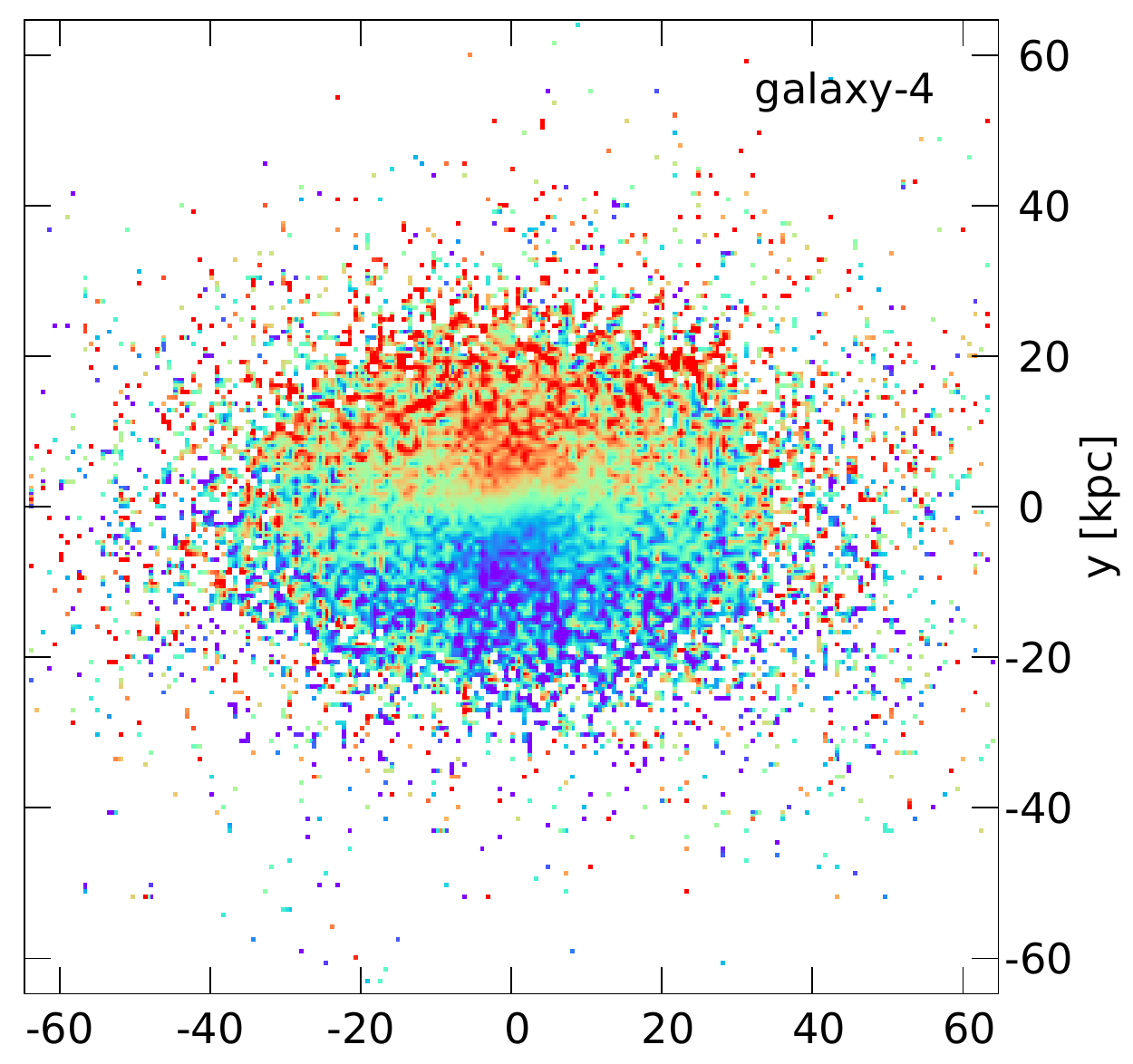}}
\resizebox{\hsize}{!}{\includegraphics{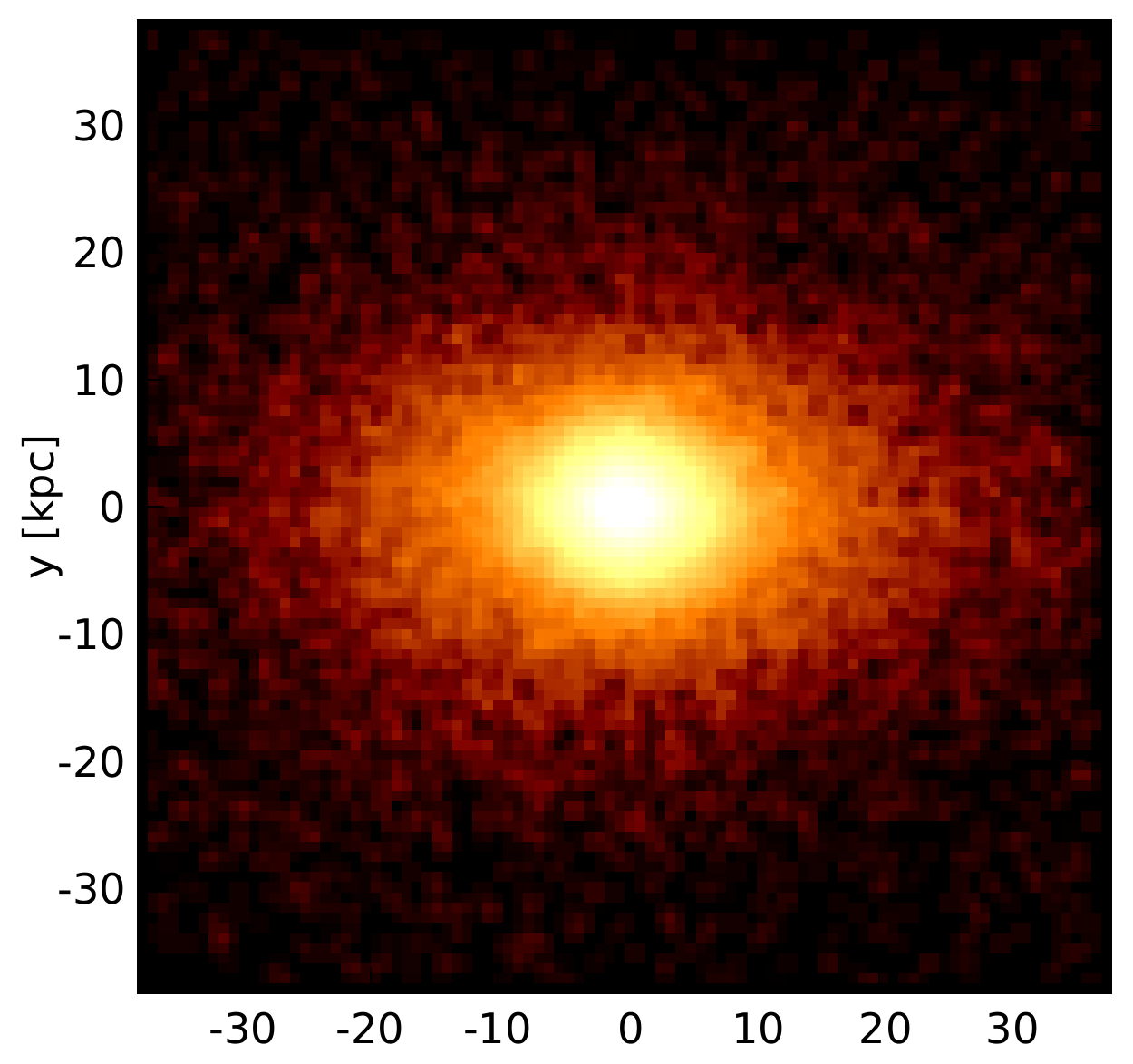}\includegraphics{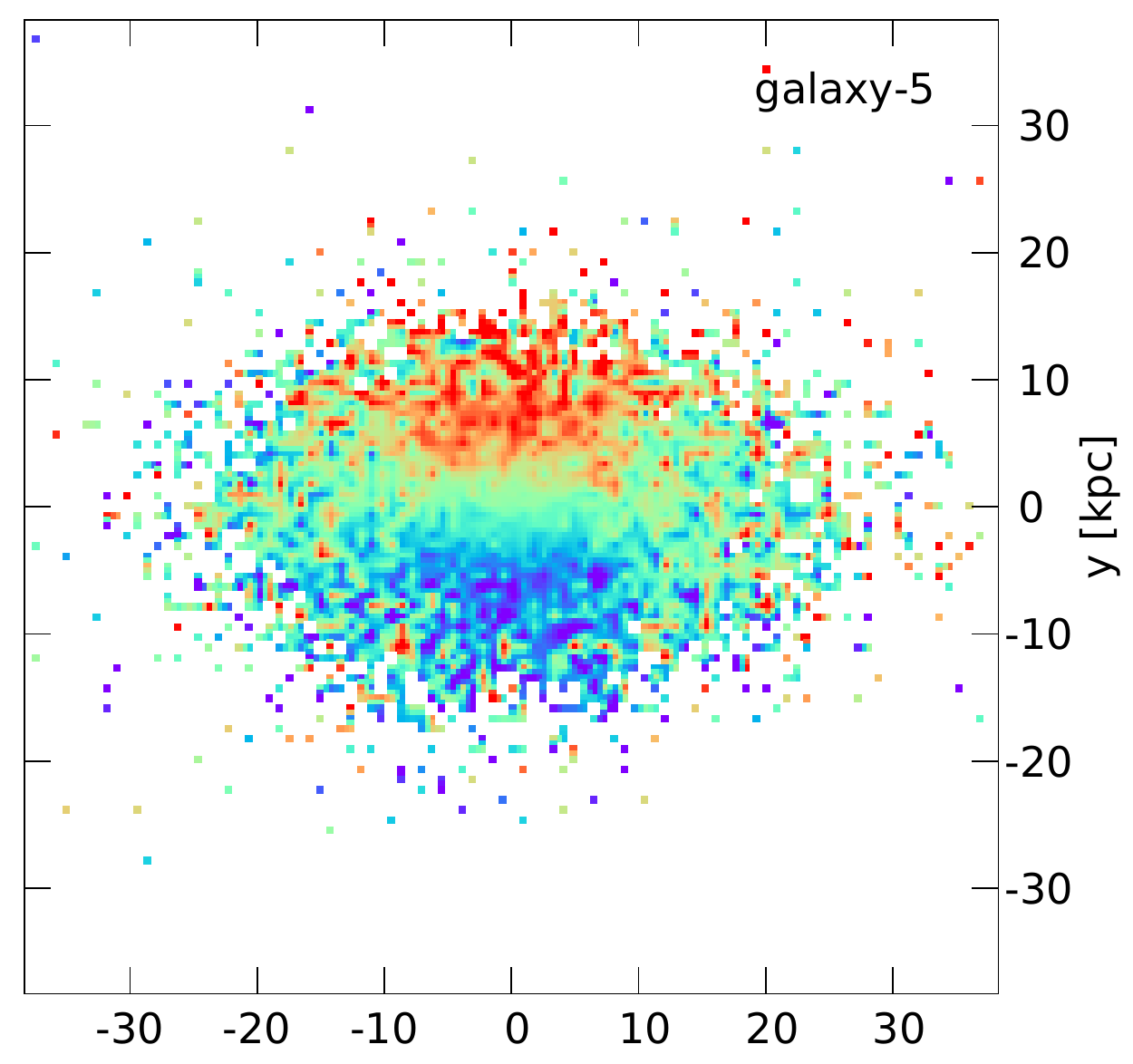}}
\resizebox{\hsize}{!}{\includegraphics{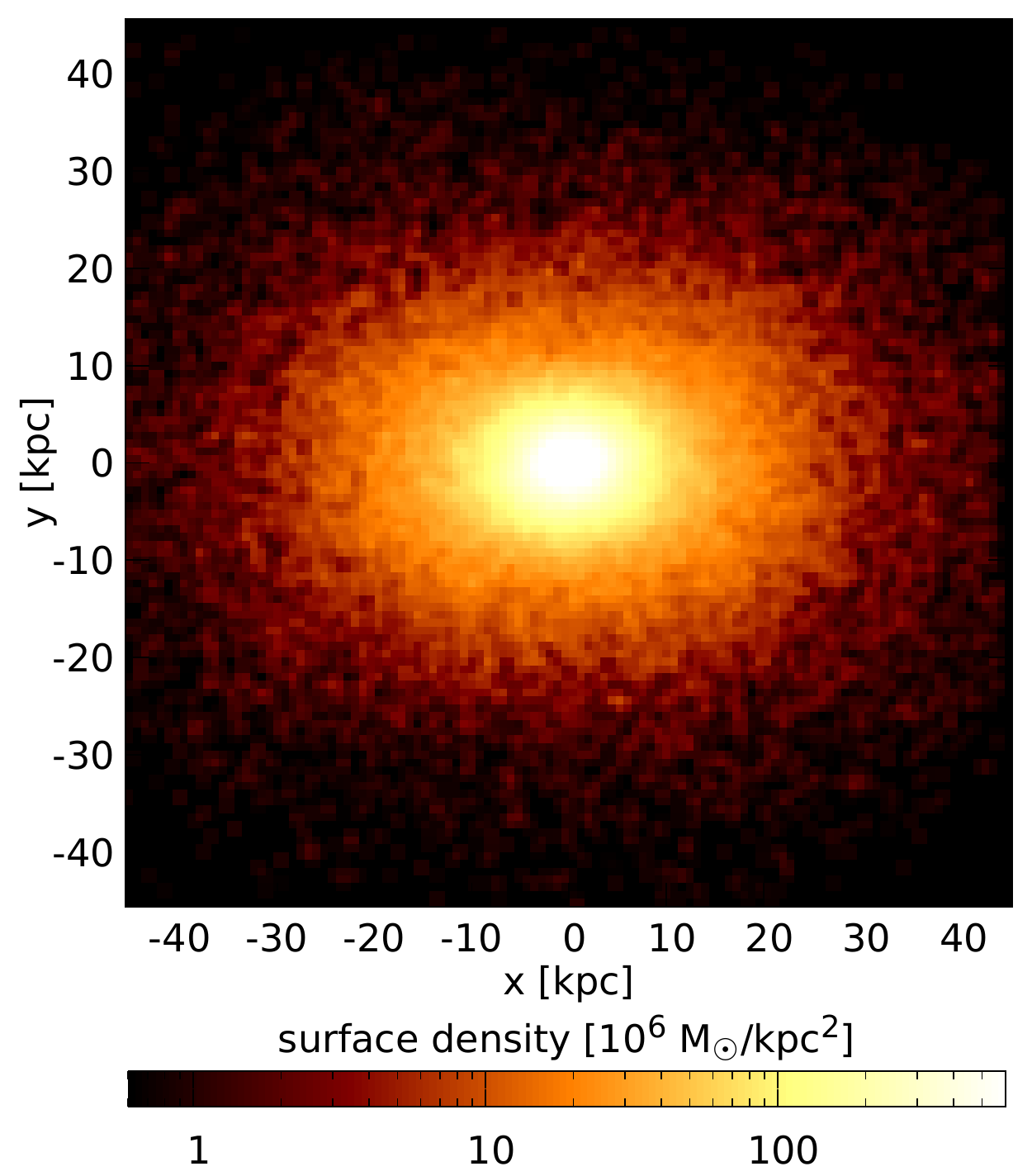}\includegraphics{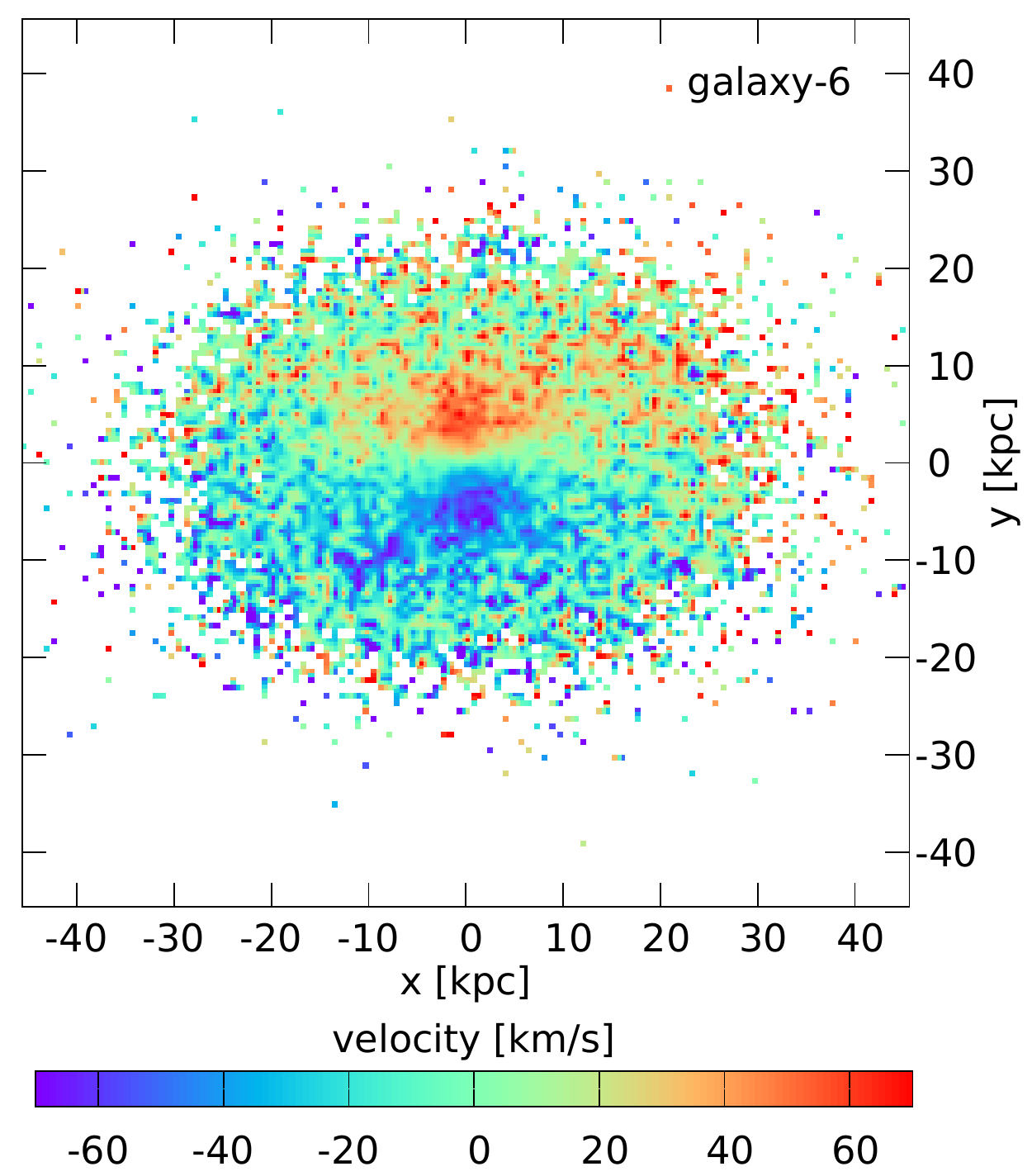}}
\caption{
Maps of the stellar surface density and the mean line-of-sight velocity for five additional galaxies with
prolate rotation selected from Illustris, viewed along the 3D minor axis.
The range on the axes corresponds to [-$r_{\rm max}$, $r_{\rm max}$] for the respective galaxy, see Table~\ref{tab:g1-6}.
\label{fig:map-g2-6}
}
\end{figure}

Additional cases of prolate rotators have been found among the Local Group dwarf galaxies in which the kinematics has
been determined using spectroscopic studies of resolved stars. The satellite of M31, Andromeda\,II is a classical dwarf
spheroidal galaxy for which \cite{ho12} reported a strong rotation signal along the minor axis, comparable in magnitude
to the velocity dispersion. In a more thorough analysis of the same data, \cite{dp17} identified two kinematically
different populations in this dwarf, one dominated by old stars with clear prolate rotation and another, composed
mostly of intermediate-age stars, that appears to rotate around the optical minor axis. Recently, \cite{kach17}
detected prolate rotation in Phoenix, a transition type dwarf (i.e. a galaxy that displays intermediate properties
between a dwarf irregular and a dwarf spheroidal).

The origin of dwarf spheroidals is still not well understood. Due to their low surface brightness, spatially and
kinematically resolved observations are restricted to the limited number of Local Group spheroidals and cosmological
simulations usually do not have enough resolution on such small scales. This motivated the studies of the origin of
dwarf spheroidals based on controlled simulations of their tidal evolution around a bigger galaxy \citep{ma01}.
While mergers between normal-size galaxies are
a generally accepted channel of their formation \citep[e.g.][]{naa14},
collisions between dwarf galaxies are believed to be rather rare, but
still expected from the cosmological simulations of the Local Group \citep{kli10,dea14}. The observed prolate rotation
of Andromeda\,II and Phoenix could be an exceptional indicator of a past major merger of dwarf galaxies.

In our previous work, \cite{lo14andii} and \cite{e15andii}, we showed that it is rather impossible to obtain prolate
rotation in pure tidal stirring scenario in which the satellite is transformed from disky to spheroidal by the tidal
forces of the host galaxy \citep[see also][]{lok15ts}.
However, we reproduced the observed kinematics of Andromeda\,II in controlled, self-consistent
simulations of mergers between equal-mass disky dwarf galaxies on a radial, or close-to-radial orbit. \cite{f17andii},
reproduced even more observables by including gas dynamics, star formation and ram pressure stripping. Recently
\cite{tsa17} examined the formation of prolate rotators in dry polar mergers of disky progenitors with different
bulge-to-disk mass ratios.

It is desirable to understand not only the possible mechanisms of the genesis of prolate rotation, but also to identify
the pathways that actually take place in the Universe. For this purpose, we take advantage of publicly available data
from the Illustris project \citep{vog14illpreintro,nel15illpub}, a large-scale ($106.5^3$\,Mpc$^3$ periodic box)
cosmological simulation following the evolution of dark matter and different baryonic components from redshift
$z=127$ up to the present time, using AREPO moving-mesh code \citep{sp10}. We use only the Illustris-1 run which has
$6\times10^9$ initial hydrodynamic cells with the mean mass of a baryonic particle $1.6\times10^6$\,M$_{\sun}$ at the
end of the simulation and  $6\times10^9$ dark matter particles with a particle mass $6.3\times10^6$\,M$_{\sun}$. The
particle data for 134 snapshots between redshifts 14 and 0 are publicly available, as well as subhalo catalogs and
merger trees \citep{rg15illmer}. The access to the formation and merger history of the simulated galaxies enables us to
explore the origin of prolate rotation in the cosmological context.

The paper is organized as follows. In Sect.~\ref{sec:selection} we describe our procedure to select prolate rotators
among Illustris galaxies and basic properties of the obtained sample. In Sect.~\ref{sec:evo} we examine the
evolution and merger history of our sample of Illustris prolate rotators. Properties of the mergers are analyzed in
Sect.~\ref{sec:merger} and compared with mergers of the twin reference sample in Sect.~\ref{sec:others}. In
Sect.~\ref{sec:Lx} we investigate the origin of the stellar particles that contribute to the prolate rotation.
The connection between prolate rotators and shell galaxies is studied in Sect.~\ref{sec:shells} and the detectability
of prolate rotation in different projection planes in Sect.~\ref{sec:los}. The results are discussed and summarized in
Sects.~\ref{sec:dis} and \ref{sec:con}, respectively.

\section{Results} \label{sec:2}

\subsection{Sample selection} \label{sec:selection}

In this work, we are interested in examining Illustris galaxies with most of the 3D rotation of stellar
particles around the 3D major axis. We construct our selection procedure according to this requirement, rather than in
a way directly comparable with observations. We shall call them galaxies with prolate rotation or prolate rotators
although their shape may not be close to prolate in all cases.

For the sample selection, we use the SubFind Subhalo catalog \citep{vog14illintro,nel15illpub} of Illustris. In order to
have sufficient resolution for the stellar kinematics, we analyze all 7697 subhalos with more than $10^4$
stellar particles in the final output (redshift $z=0$) of the Illustris-1 run. We refer to those subhalos simply as
galaxies and the set of 7697 galaxies as the global sample.

Since in our general approach we do not intend to examine the galaxies viewed from a particular line of sight, we need
to define a unified 3D radius, within which we are going to analyze the kinematics of the galaxies. A sufficient and
simple way, which gives reasonable values of radius for all galaxies, is to choose some limiting density. For each
galaxy we define $r_{\rm max}$ as the radius at which the average 3D stellar density in a shell between $0.8r_{\rm
max}$ and $1.25r_{\rm max}$ is equal to $1.4\times10^4$\,M$_{\sun}$\,kpc$^{-3}$. To eliminate the effects of different
projection planes, we took 31 most spherical galaxies and computed their surface brightness in V band including all
stellar particles associated with the galaxy. The typical surface brightness at the projected radius $r_{\rm max}$ is
29.5 (29.0) mag\,arcsec$^{-2}$ for the mass-to-light ratio of 5 (3). We do not consider any type of extinction. This
value of the surface brightness is beyond the capability of the present-day IFU spectroscopic surveys, but stars at
higher radii, that are along the line of sight, also contribute to the spectroscopic measurements and we do not want to
exclude them all by cutting at some lower 3D radius.

\begin{figure*}
\plottwo{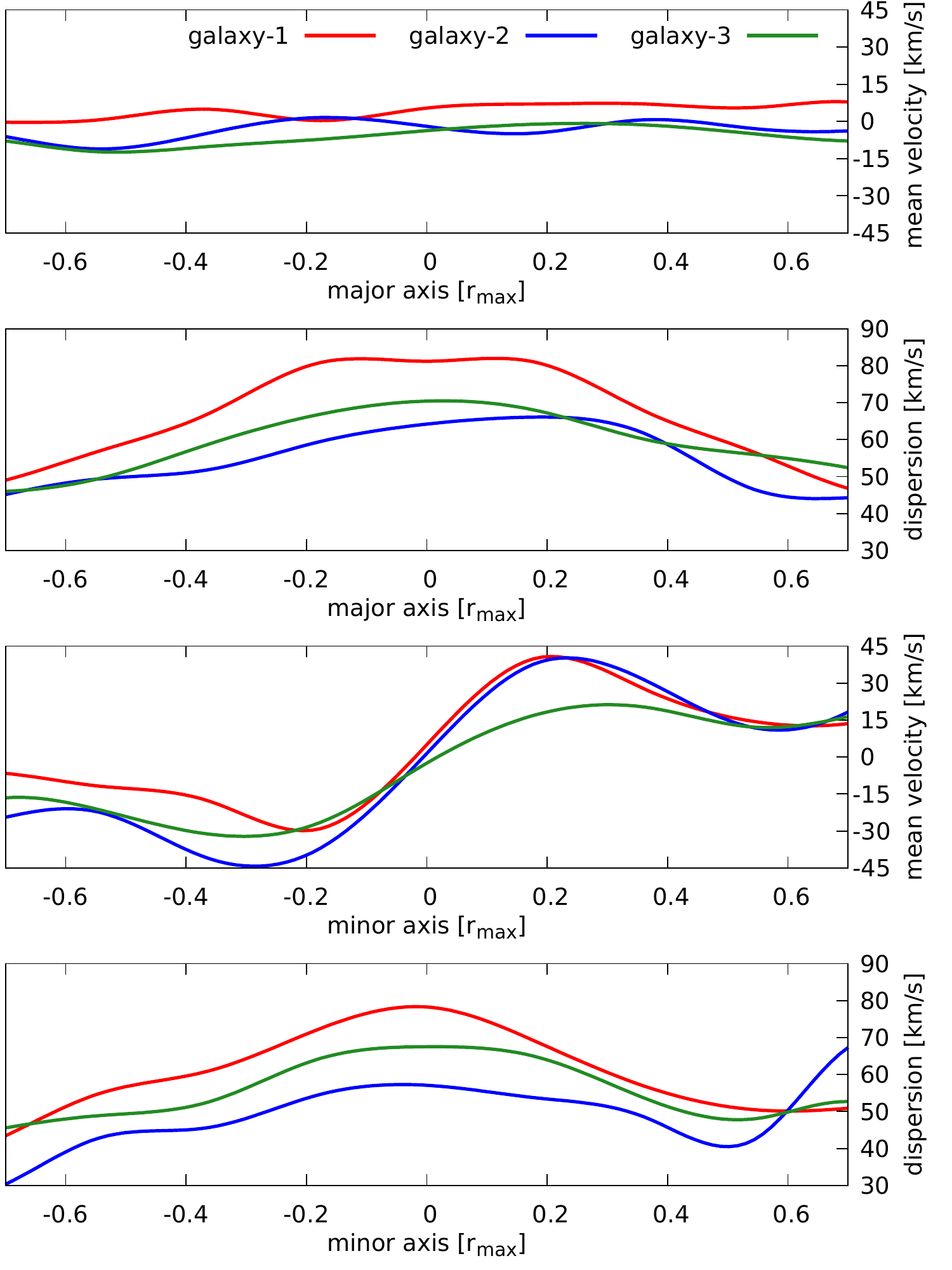}{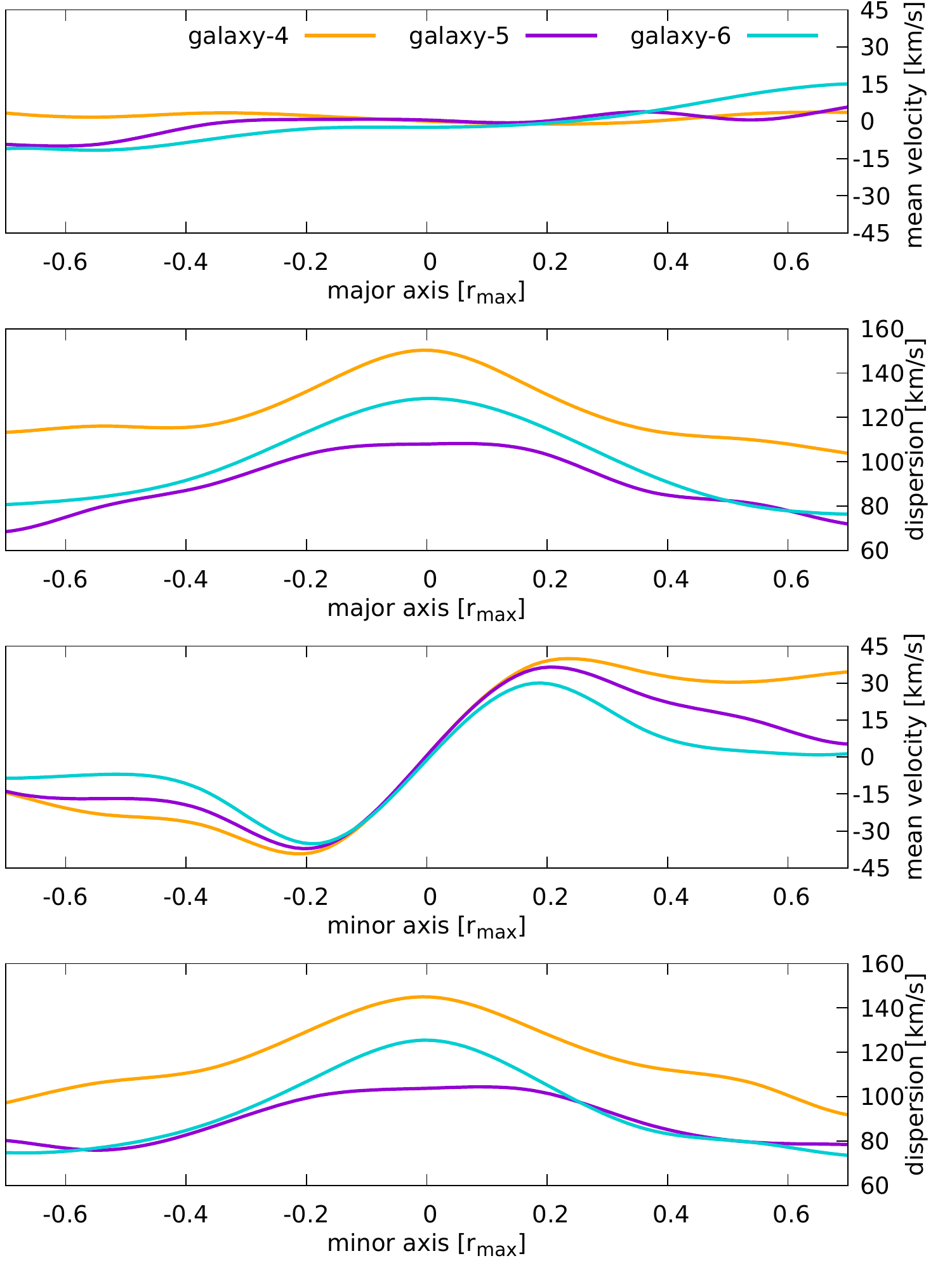}
\caption{
Mean line-of-sight velocity and velocity dispersion profiles measured along the major and minor projected axis in the
$xy$ plane for our six selected Illustris galaxies with prolate rotation.
For each panel, the horizontal axis is plotted in the units of the value of $r_{\rm max}$ for the respective galaxy,
see Table~\ref{tab:g1-6}.
\label{fig:axvel}}
\end{figure*}

For the global sample of 7697 galaxies, $r_{\rm max}$ ranges from 188 to 5\,kpc with the median at 29.2 kpc; the
stellar mass inside $r_{\rm max}$ falls between $208.8\times10^{10}$ and $0.6\times10^{10}$\,M$_{\sun}$ with the
median at $2.1\times10^{10}$\,M$_{\sun}$. Note that while at the high-mass end the size distribution of the
simulated galaxies matches the observed one quite well, the physical extent of less massive galaxies in Illustris can
be a factor of a few larger than observed \citep{syn15illmorph}.

\begin{deluxetable*}{cccccccccccccccccccccccccccccc}
\tablecaption{Properties of six selected Ilustris galaxies with prolate rotation \label{tab:g1-6}}
\tablecolumns{14}
\tablewidth{0pt}
\tablehead{
\colhead{(1)} & \colhead{(2)} & \colhead{(3)} & \colhead{(4)} & \colhead{(5)} & \colhead{(6)} &
\colhead{(7)} & \colhead{(8)} & \colhead{(9)} & \colhead{(10)} & \colhead{(11)} & \colhead{(12)} &
\colhead{(13)} & \colhead{(14)} & \colhead{(15)}\\
\colhead{galaxy} & \colhead{line color} & \colhead{\#} & \colhead{$r_{\rm max}$} &
\colhead{$M(r_{\rm max})$} & \colhead{$f_{\rm g0}$} & \colhead{$T$} & \colhead{$L_{\rm min}$} &
\colhead{$t_{\rm KT}$} & \colhead{$t_{\rm ST}$} & \colhead{$t_{\rm M}$} &
\colhead{$M_{\rm s}/M_{\rm p}$} & \colhead{$f_{\rm gM}$} & \colhead{$N$} & \colhead{$P$}\\
\colhead{} & \colhead{} & \colhead{} & \colhead{[kpc]} & \colhead{[$10^{10}$\,M$_{\sun}$]} &
\colhead{} & \colhead{} & \colhead{} & \colhead{[Gyr]} & \colhead{[Gyr]} & \colhead{[Gyr]} &
\colhead{} & \colhead{}
}
\startdata
galaxy-1 & red & 451304 & 32.5 & 2.2 & 0.38 & 0.90 & 0.89 & 4.17 & 4.34 & 4.17 & 0.69 & 0.64 & 1 & 0.93\\
galaxy-2 & blue & 510126 & 26.4 & 1.0 & 0.59 & 0.96 & 0.89 & 5.44 & 13.00 & 4.17 & 0.35 & 0.81 & 2 & 0.87\\
galaxy-3 & green & 466861 & 27.7 & 1.5 & 0.44 & 0.93 & 0.90 & 6.89 & 13.00 & 7.02 & 0.93 & 0.72 & 2 & 0.84\\
galaxy-4 & orange & 273741 & 64.7 & 17.9 & 0.01 & 0.88 & 0.91 & 3.09 & 4.17 & 2.25 & 0.41 & 0.04 & 2 & 0.70\\
galaxy-5 & violet & 387053 & 38.3 & 5.7 & 0.12 & 0.89 & 0.92 & 5.13 & 6.07 & 4.98 & 0.40 & 0.33 & 1 & 0.80\\
galaxy-6 & cyan & 204398 & 45.6 & 11.1 & 0.01 & 0.90 & 0.95 & 4.50 & 4.50 & 4.17 & 0.98 & 0.22 & 3 & 0.80\\
\enddata
\tablecomments{
(1) adopted galaxy name; (2) line color used in some graphs for the respective galaxy; (3) SubFind ID; (4) $r_{\rm
max}$, the radius inside which the galaxy is analyzed at $z=0$, see Sect.~\ref{sec:selection}; (5) $M(r_{\rm max})$,
the stellar mass inside $r_{\rm max}$ at $z=0$; (6) $f_{\rm g0}$, gas fraction at $z=0$; (7) $T$, triaxiality
parameter, where $T=0$ for the oblate shape and $T=1$ for the prolate one; (8) $L_{\rm min}$, the minimum value of
$(L_x/L_{\rm tot})^2$ inside $r_{\rm max}$, where $L_x$ is the major-axis component of the angular momentum, see
Sect.~\ref{sec:selection}; (9) $t_{\rm KT}$, the look-back time of the kinematic transition into prolate rotation; (10)
$t_{\rm ST}$, the look-back time of the shape transition towards prolate shape; (11) $t_{\rm M}$, the look-back time of
the last significant merger, see Sect.~\ref{sec:merger}; (12) $M_{\rm s}/M_{\rm p}$, the mass ratio of the merger; (13)
$f_{\rm gM}$, the postmerger gas fraction; (14) $N$, the number of significant mergers during last 10\,Gyr; (15) $P$,
probability to detect prolate rotation (for the default set of thresholds, see Sect.~\ref{sec:los}).
}
\end{deluxetable*}

Centers of the galaxies are computed iteratively using stellar particles inside a sphere with a gradually decreasing
radius until the sphere contains one third of the total stellar mass. We find principal axes of the galaxies using the
inertia tensor and compute the angular momentum of stellar particles in the coordinate system defined by the principal
axes, where $x$, $y$, and $z$ is the major, intermediate, and minor axis.
We perform this procedure inside a sphere of radius 0.25, 0.35, 0.5, 0.7, and 1 $r_{\rm max}$
so that the principal axes and angular momenta are determined using only stellar particles inside the respective radius.
We select galaxies with $(L_x/L_{\rm tot})^2>0.5$ for all five
radii, where $L_x$ is the $x$-component of the angular momentum and $L_{\rm tot}$ is its absolute value.
This way, we choose only well-established prolate rotators that have most of their rotation around the 3D major axis on
both, small and large galactocentric radii.
The criterion is satisfied by 59 out of 7697 galaxies and we select these
59 galaxies as our sample of Illustris prolate rotators. If we set the threshold for $(L_x/L_{tot})^2$, at all five
radii, to 0.4, 0.3, 0.2, and 0.1, we would obtain samples of 74, 110, 153, and 254 galaxies, respectively.

\begin{figure}
\includegraphics[width=\hsize]{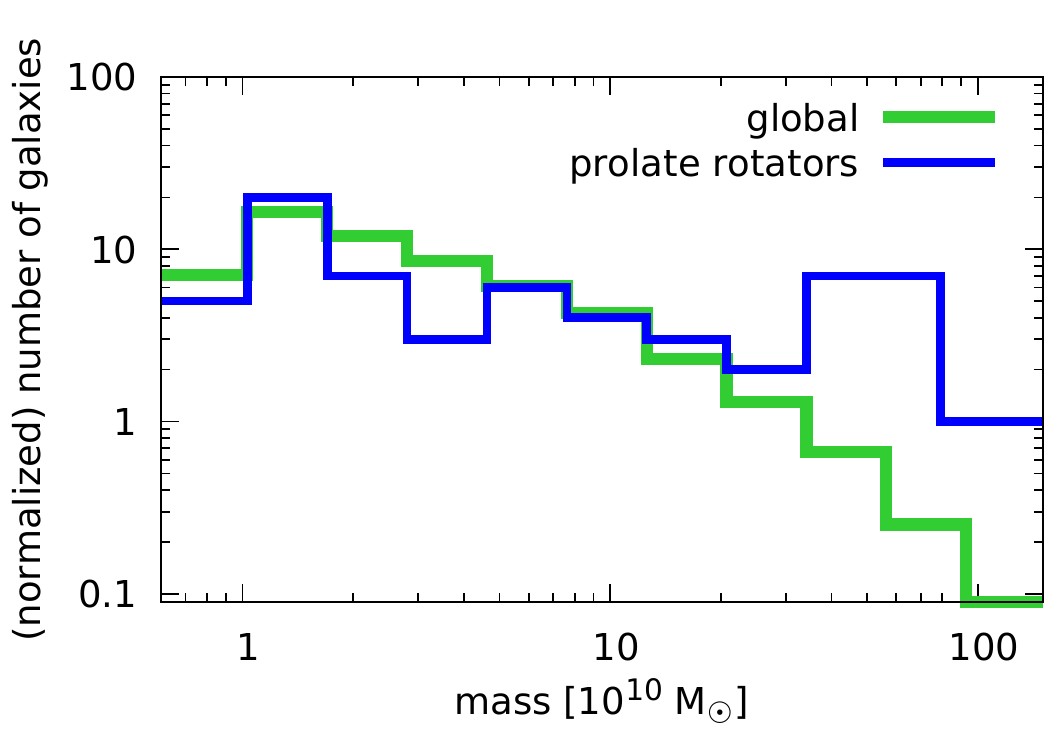}
\caption{
Comparison of the stellar mass distribution of the prolate rotators with the global sample. The distribution of the
global sample is normalized to the same total number as for the prolate rotators.
\label{fig:Mdist}}
\end{figure}

Fig.~\ref{fig:map-g1} shows an example of a well-ordered, regular prolate rotator from our sample of 59 galaxies,
which we name galaxy-1, viewed along all three principal axes. This galaxy, with $r_{\rm max}=32.5$\,kpc and $M(r_{\rm
max})=2.2\times10^{10}$\,M$_{\sun}$, belongs to the lower end of the mass range of the global sample.
Fig.~\ref{fig:map-g2-6} shows the surface density and kinematics in the $xy$ plane of additional 5 galaxies selected
from our sample of prolate rotators. The mean line-of-sight velocity and velocity dispersion along the major
and minor projected axis in $xy$ plane (computed in 11 bins as described in Sect.~\ref{sec:los}) for the six selected
galaxies is displayed in Fig.~\ref{fig:axvel}. Lower-mass galaxies are on the left-hand side, more massive galaxies on
the right. Both groups have no or little rotation along the major axis and clear rotational signal along the minor
axis.
The maximum value of this prolate rotation is similar for low- and high-mass galaxies but the dispersion
increases with the mass of a galaxy and it is much higher for massive galaxies.
According to the Faber-Jackson law, the relation between the luminosity and the central stellar velocity
dispersion of elliptical galaxies is a power-law with an index around 4 \citep{fj76}.
Basic properties of the galaxies are listed in Table~\ref{tab:g1-6}.

The six example galaxies are drawn from twelve galaxies with the highest values of $(L_x/L_{\rm tot})^2$. We performed
a visual inspection of the maps of the mean line-of-sight velocity of those twelve galaxies viewed along all three principal
axes. We selected galaxies that appear to have the most regular prolate rotation, with the requirement of having three
examples from lower-mass end of the global sample ((1--5)$\times10^4$ stellar particles) and three massive galaxies
(more than $5\times10^4$ stellar particles).

The first noticeable feature of our sample of prolate rotators is its mass distribution. Fig.~\ref{fig:Mdist} compares
the distribution of stellar mass inside $r_{\rm max}$ at redshift $z=0$ for the prolate rotators with the global sample
of 7697 galaxies. The global sample is normalized to the sample of prolate rotators. In comparison
with the global sample, the prolate rotators are clearly biased towards more massive galaxies.

\begin{deluxetable}{cccccccc}
\tablecaption{Evolution of the fraction of prolate rotators with redshift \label{tab:z}}
\tablecolumns{5}
\tablewidth{0pt}
\tablehead{
\colhead{(1)} & \colhead{(2)} & \colhead{(3)} & \colhead{(4)} & \colhead{(5)} \\
\colhead{$z$} & \colhead{$t$\,[Gyr]} & \colhead{$N_{\rm g}$} &
\colhead{$N_{\rm pr}$} & \colhead{$f_{\rm pr}$}
}
\startdata
0.0 & 0.00 & 7697 & 59 & 0.77\,\%\\
0.2 & 2.43 & 7061 & 57 & 0.81\,\%\\
0.4 & 4.34 & 6414 & 57 & 0.89\,\%\\
0.6 & 5.79 & 5788 & 58 & 1.00\,\%\\
0.8 & 6.89 & 5196 & 44 & 0.85\,\%\\
1.0 & 7.84 & 4671 & 52 & 1.11\,\%\\
\enddata
\tablecomments{
(1) redshift; (2) look-back time; (3) number of galaxies with more than $10^4$ stellar particles; (4) number of prolate
rotators; (5) fraction of prolate rotators.
}
\end{deluxetable}

\begin{figure*}[!ht]
\plotone{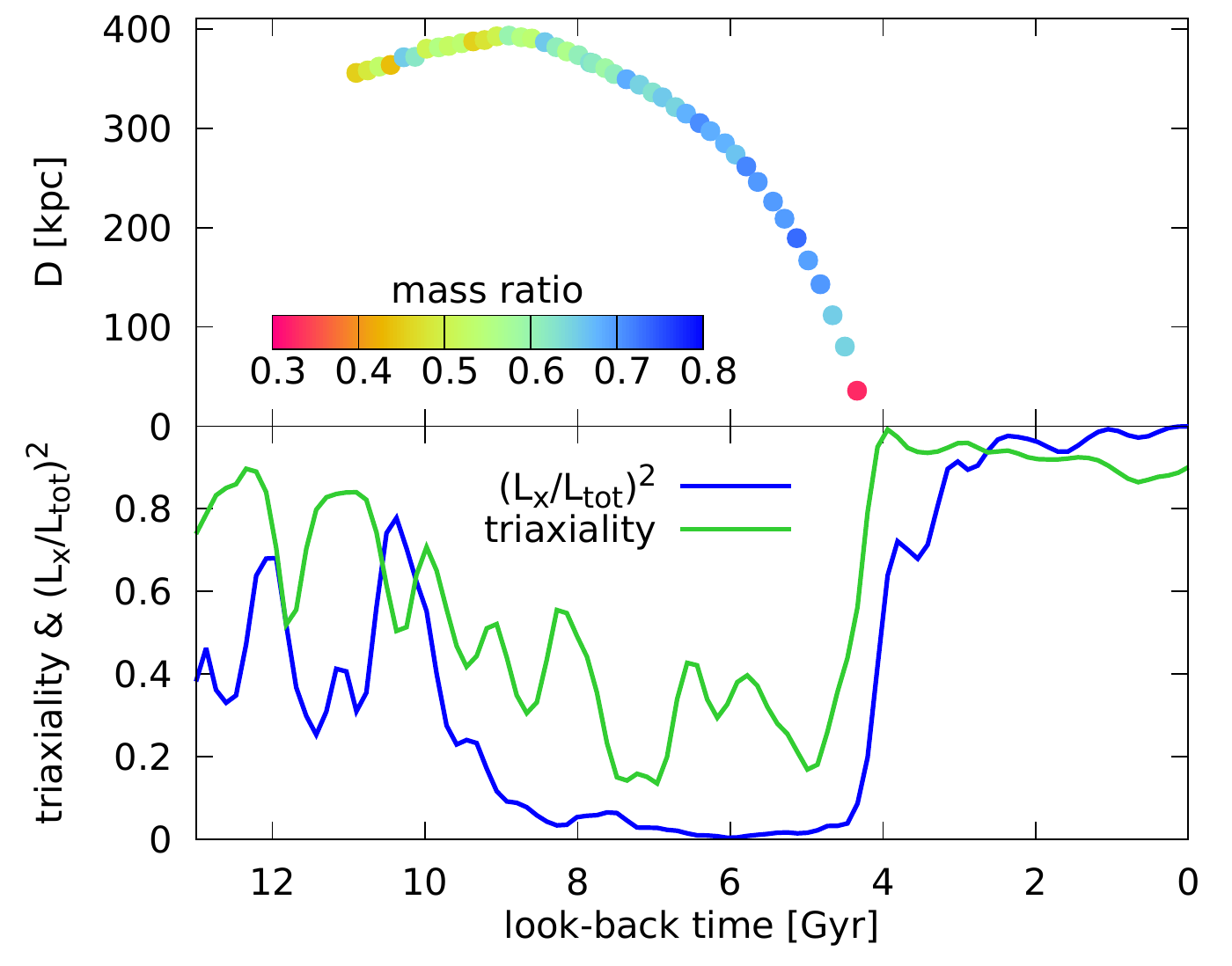}
\caption{
Evolution graph for galaxy-1 (see Fig.~\ref{fig:map-g1}), which went through just one significant merger during the last
10\,Gyr. Top: Relative distance of the primary and secondary galaxy. The color codes the stellar mass ratio at each
snapshot. Bottom: Evolution of $(L_x/L_{\rm tot})^2$ and triaxiality.
\label{fig:evo-g1}
}
\end{figure*}

We tested whether the mass distribution of prolate rotators is not affected by the fact that less massive
galaxies have inferior resolution of the shape and kinematics due to a lower number of particles. We repeated the
selection procedure (see above) on all 7697 galaxies, but this time with only $10^4$ random stellar particles for each
galaxy. With this reduced number of particles, 57 galaxies would be classified as prolate rotators and 54 of these
galaxies are identical to the galaxies selected using all stellar particles. We consider this to be a reasonable
agreement and keep our original sample of 59 prolate rotators for further analysis.

We repeated the procedure described above to select prolate rotators from several Illustris snapshots at higher
redshifts (up to $z=1$). The results are summarized in Table~\ref{tab:z}. The fraction of prolate rotators is slightly
increasing with redshift but not very convincingly with such a low numbers of galaxies. For the rest of the paper, we
restrict the analysis to the prolate rotators classified as such in the last snapshot ($z=0$).

\subsection{Evolution of prolate rotators} \label{sec:evo}

\begin{figure*}
\plotone{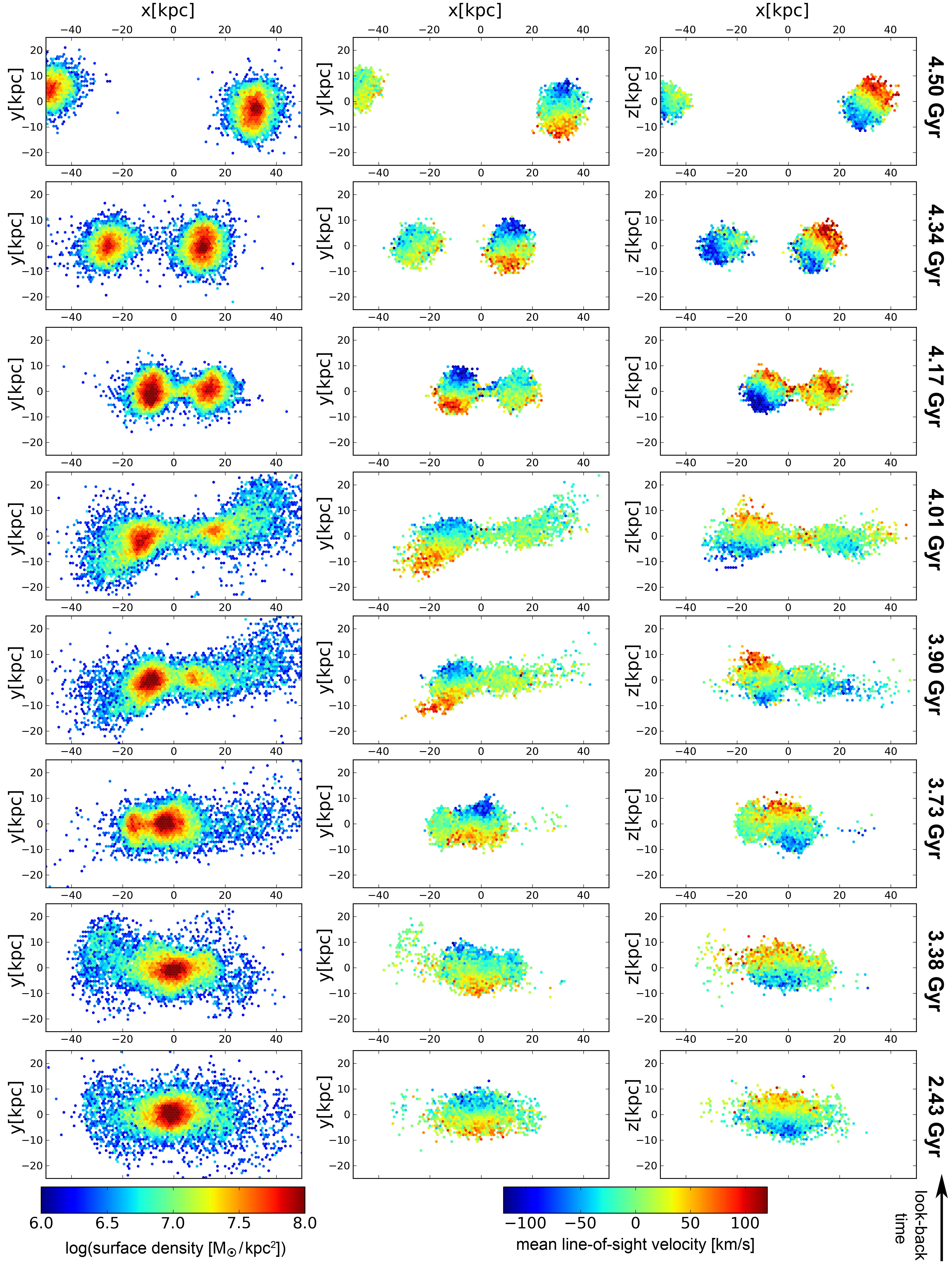}
\caption{
Different stages of the major merger suffered by galaxy-1 (see Fig.~\ref{fig:map-g1}). The $xy$ plane corresponds to
the collision plane. Left column: the surface density; middle and right column: the mean line-of-sight velocity in two
different projections. The look-back time for each snapshot is indicated on the right.
\label{fig:merger-g1}
}
\end{figure*}

We investigated the history of all our 59 prolate rotators in order to determine scenarios most likely leading
to the formation of galaxies with this property, in particular to see whether there is a connection to mergers and if
so, what are the properties of such mergers.

We analyze the properties of the main progenitor branch galaxies of the Illustris SubLink merger trees
\citep{rg15illmer} for the sample of 59 prolate rotators. The analyses are done on stellar particles associated with
the galaxy by SubFind algorithm as provided by the Illustris project. We compute $(L_x/L_{\rm tot})^2$ and the
triaxiality parameter $T$ inside a sphere of radius $r_{\rm max}$ for all outputs during the last 13\,Gyr. The
triaxiality parameter, $T$, is defined as $T=[1-(b/a)^2]/[1-(c/a)^2]$, where $b/a$ is the intermediate to major axis
ratio and $c/a$ is the minor to major axis ratio computed as the ratio of the respective eigenvalues of the tensor of
inertia. In this sense, our axis ratios do not refer to any particular isodensity contour but rather the overall stellar
mass distribution inside $r_{\rm max}$. When $T$ is lower than $1/3$ the galaxy is rather oblate, values greater than
$2/3$ indicate a prolate shape, and when $T$ falls between these values the galaxy has a triaxial shape.

From the branches of the next progenitors (secondaries) of the SubLink trees, we select only galaxies that: (1) have at
least 10\,\% of the baryonic mass of the primary in at least two subsequent snapshots during the last 20 snapshots
before the merger, and (2) merged with the primary during the last 10\,Gyr (e.g. between redshifts 1.8 and 0). The
look-back times are computed with the cosmological parameters consistent with the WMAP-9 measurements \citep{wmap9} as
these parameters are also adopted in the Illustris simulation. The average value of the time spacing between snapshots
is 156\,Myr during the last 10\,Gyr of the simulation.

Fig.~\ref{fig:evo-g1} shows the evolution of the shape and kinematics of galaxy-1 (Fig.~\ref{fig:map-g1}) in the lower
panel. Only one secondary satisfied our selection criteria and the evolution of its distance to the primary and their
stellar mass ratio is illustrated in the upper panel. Several stages of the merger are shown in
Fig.~\ref{fig:merger-g1}.

Following the evolution of all 59 galaxies back in time from redshift $z=0$, we define the time of kinematic (shape)
transition of a galaxy towards prolate rotation and shape as the time of the snapshot in which $(L_x/L_{\rm
tot})^2$ ($T$) drops under 0.4 (0.5) for the first time and stays under this value in the preceding snapshot as well.
We use the threshold value of 0.4 for $(L_x/L_{\rm tot})^2$ because it seems to correspond well to the transition from
disky rotation. Two galaxies have $T<0.5$ at $z=0$. We search for the transition time for those two galaxies back from
the last time where they had $T>0.5$. The time of the merger is defined as the look-back time of the first snapshot in
which the branch of the secondary does not exist in the merger tree anymore. Our galaxy-1 experienced the merger
4.17\,Gyr ago. The time of the merger is the same as the inferred time of kinematic transition and the shape transition
happened just 0.17\,Gyr earlier (see Table~\ref{tab:g1-6} and Fig.~\ref{fig:evo-g1}).

Premerger quantities are measured when the two progenitors are separated by a distance greater than $5r_{\rm max}$ for
the last time in their history. Such a distance roughly corresponds to the point of the first infall, when the galaxies
have not yet started to strip the baryonic mass from each other. For progenitors that never reached such a
separation, we take the measurements at the point of their greatest distance. As a measure of mass we take as usual the
stellar mass inside a sphere of radius $r_{\rm max}$. The mass ratio of the merger is the premerger mass of the less
massive progenitor divided by the mass of the other one, regardless of whether the more massive galaxy is formally
called a primary (i.e. comes from the main progenitor branch) or a secondary (i.e. comes from the branch of the next
progenitor).

\begin{figure*} [!htb]
\centering
\includegraphics[width=0.7\hsize]{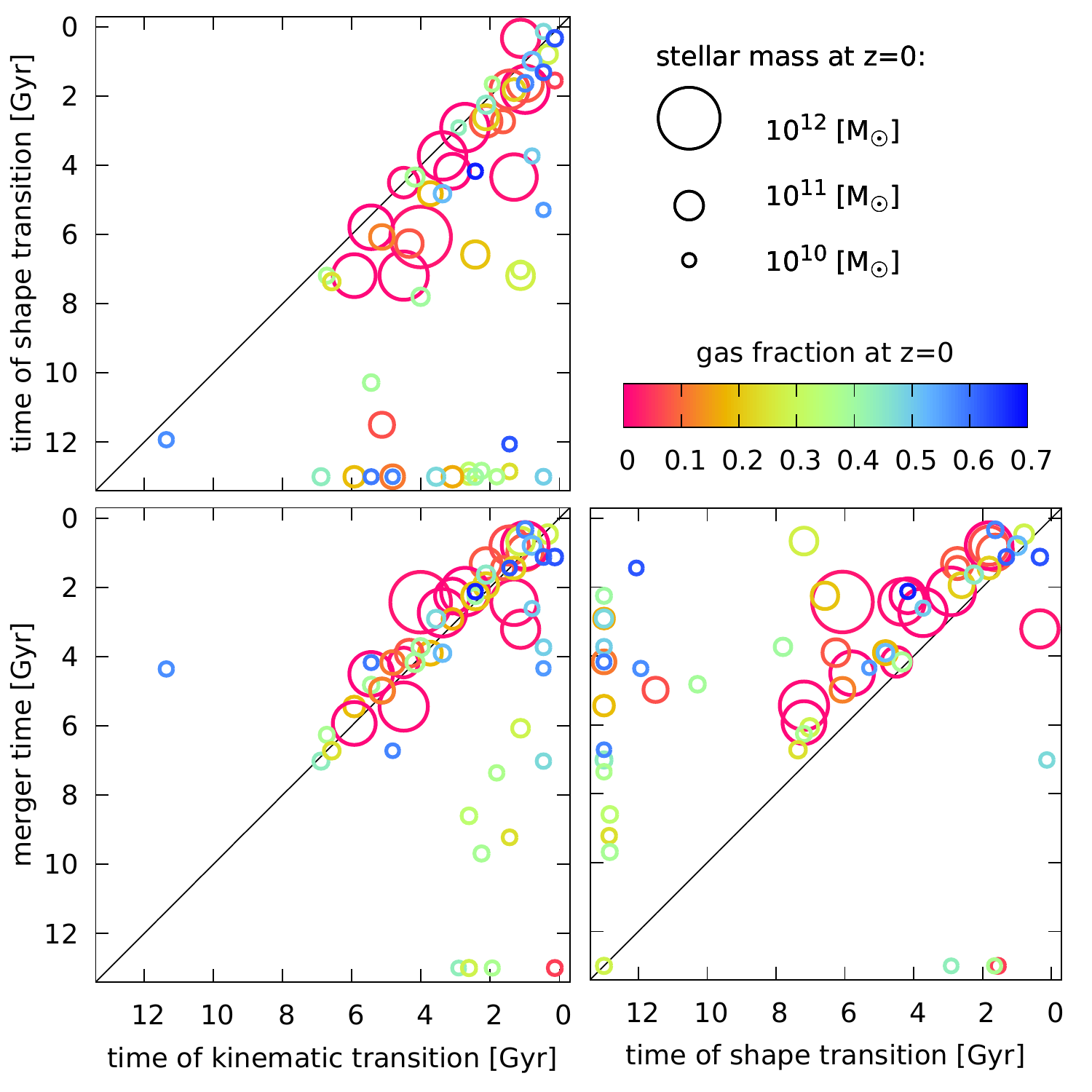}
\caption{
Correlations between the time of the shape and kinematic transition (towards prolate shape and rotation) and the time of
the last significant merger.
The time labels are expressed in terms of look-back time.
Circle sizes are proportional to the third root of the stellar mass of the galaxies at
redshift $z=0$. Colors reflect the gas fraction of the galaxies at $z=0$.
\label{fig:cor-KT-ST}
}
\end{figure*}

Fig.~\ref{fig:cor-KT-ST} shows correlations between the time of the shape and kinematic transition and the time of the
last significant merger for all 59 prolate rotators. As a significant merger we denote mergers with mass ratio 0.1 and
higher. None of the galaxies retains prolate rotation for the whole reference period. In fact, only one galaxy
maintains prolate rotation for longer than 7\,Gyr (specifically 11.4\,Gyr), while the majority of the sample rotate in
the prolate way for less than 3\,Gyr in a row. The circle size in the plots corresponds to the stellar mass and its
color to the gas fraction (i.e. the mass of gas particles divided by the mass of all baryonic particles). The massive
galaxies are practically devoid of gas and show a rather prominent correlation between the times of the shape and
kinematic transition. For the vast majority of the sample, the transition of the shape happens around the time of the
kinematic transition or earlier.

Four galaxies did not experience any merger with mass ratio of at least 0.1 during the
last 10\,Gyr and they are displayed at the merger time of 13\,Gyr. Apart from these four galaxies, additional 15 have
the time difference between the merger and the kinematic transition greater than 1\,Gyr. This leaves 40 out of 59
galaxies with the times of the merger and the kinematic transition well correlated. The other 19 cases are discussed
in detail in Sect.~\ref{sec:dis}.

The time of the shape transition is less tightly correlated with the kinematic transition
(top left panel of Fig.~\ref{fig:cor-KT-ST}) and it is also less correlated with the time of the merger (bottom
right panel of Fig.~\ref{fig:cor-KT-ST}). When the shape transition does not happen during the last significant merger,
it took place earlier (with just two or three exceptions).

There seem to be two populations of prolate rotators. The
first population was born with a rather prolate shape or transitioned to it very early (more than 10\,Gyr ago) and
managed to maintain the shape while the kinematics caught up later (mostly during a merger event). The other
population changed the shape less than 8\,Gyr ago usually during a merger but not necessarily during the same
merger as the kinematics. In some cases there is a shape transition during an earlier pericentric passage of the
secondary that eventually probably caused the kinematic transition.
In such cases, the shape transition can happen 1 or 2\,Gyr prior to the kinematic transition but both can still be related to the same merger event.
Apparently galaxies (at least in Illustris) are
often born with a little elongated shape and even when they are rather flat, it is then easier to make them change their
shape into prolate or triaxial than to change their kinematics. This means that often, even when the galaxy is
elongated, it still can have most of its rotation around the short axis.

In addition, some of our 59 galaxies went
through a period of prolate rotation in their past but gradually (in peaceful times) or abruptly (during a merger)
switched to the oblate rotation until they were disrupted again so they could become a prolate rotator in our sample.
Often, but not always, these changes from prolate to oblate rotation are followed also by the flattening of the shape.

The prolate or triaxial shape seems to be easier to evoke and maintain than the prolate kinematics. While the shape can
hold through some mergers, the prolate kinematics very rarely survives such a violent event. The prolate or triaxial
shape ($T>1/3$) seems to be a necessary condition for prolate rotation but it does not have to be accompanied by the
prolate kinematics and probably more often it is not. Thus we are going to concentrate on the kinematic transition that
seems to be more connected to the last significant merger.

\subsection{The last significant merger} \label{sec:merger}

In order to see which conditions most likely lead to the formation of prolate rotators, we examine the
last significant mergers (i.e. with the mass ratio of at least 0.1) in more detail. We have 55 of such events since 4
galaxies in our prolate sample did not experience any significant merger. For 15 galaxies the time of the last
significant merger does not agree with the time of the kinematic transition, leaving us with a subsample of 40 well-correlated mergers.

\begin{figure*}
\plotone{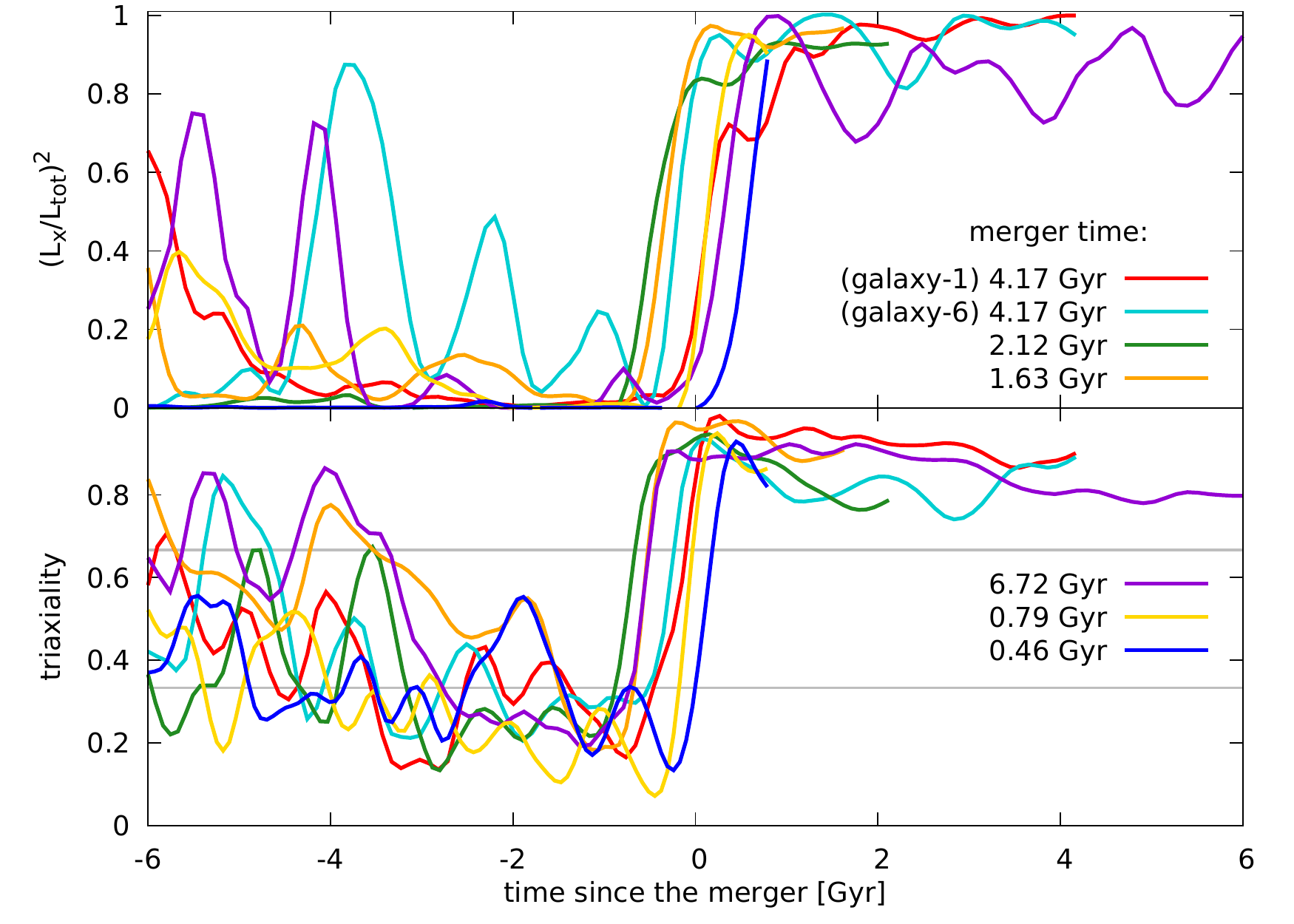}
\caption{
Evolution of $(L_x/L_{\rm tot})^2$ and triaxiality of the galaxies from the golden~7 sample. Timescales were shifted so
that zero matches the time of the last significant merger. The true look-back time of the merger for each galaxy is
listed in the legend.
\label{fig:golden7}
}
\end{figure*}

For some galaxies, even when there is a good time correlation between the merger and the kinematic transition, there are
hints that the last merger is not (solely) responsible for the transition. In approximately 9 cases, there are actually
two (even three in one case) secondaries merging during the same period and it would be next to impossible to
disentangle the role of each progenitor in this incident. The progenitors also probably highly alter their initial
direction and velocities during such a merger. For about the same number of cases, there are subsequent mergers more
separated in time and space but it seems that the primary was pre-processed by the previous disturbance of its
shape and kinematics so it could eventually become the prolate rotator during the last merger. Moreover, in two or
three cases the primary had an established prolate shape and rotation before the last merger. The merger just caused a
temporary drop in the values of $(L_x/L_{\rm tot})^2$ (and $T$) so that the time of the transition coincides with the
time of the last merger.

\begin{figure}
\plotone{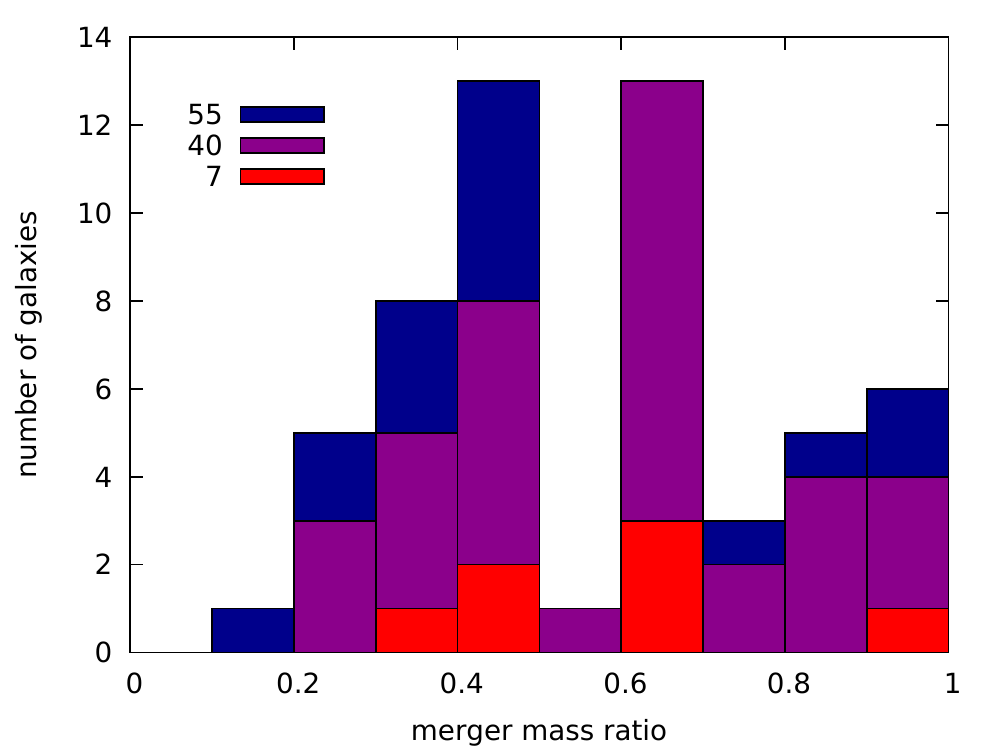}\\
\plotone{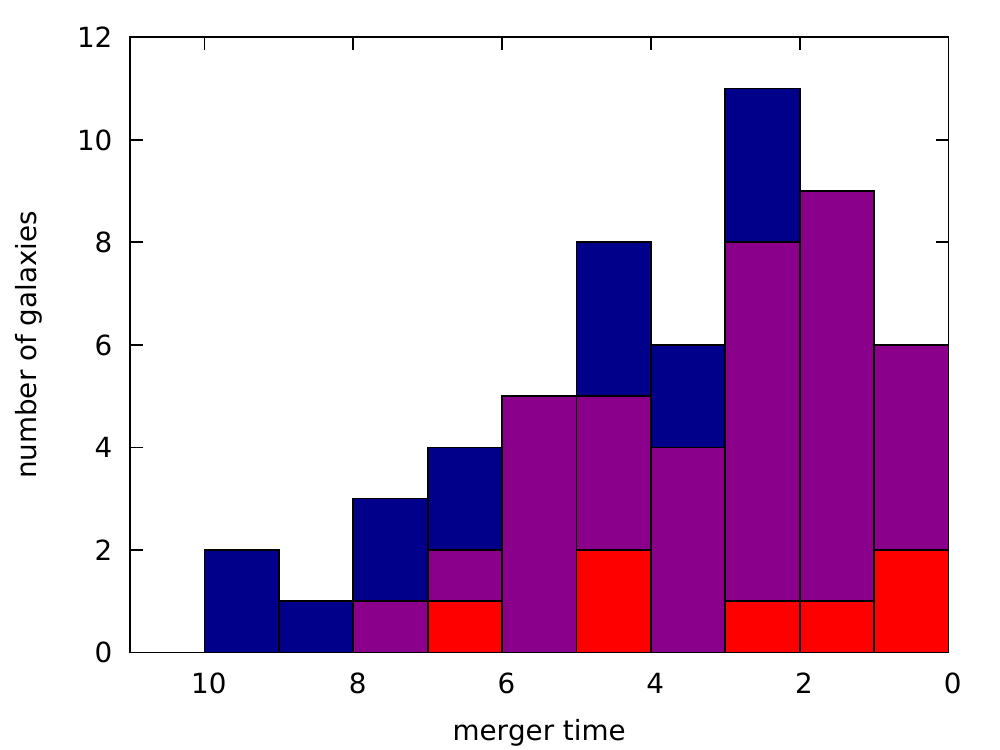}\\
\plotone{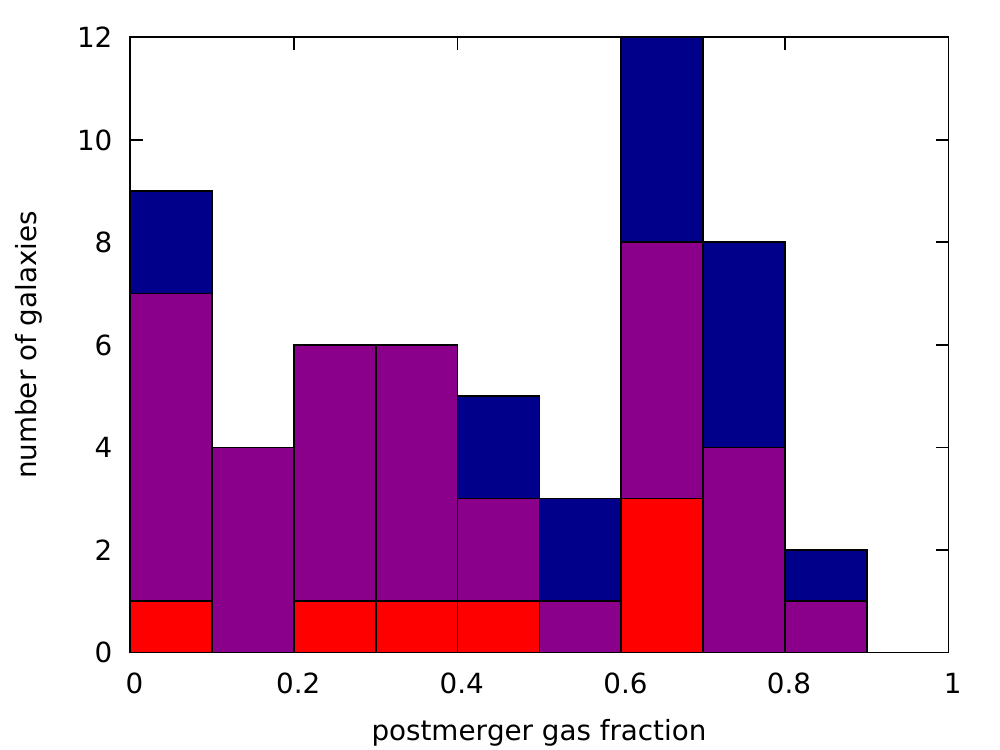}
\caption{
Histograms of the properties of all 55 last significant mergers for the prolate rotators with the subsamples of 40 well-correlated mergers and the golden~7 overplotted.
\label{fig:histograms1}
}
\end{figure}

\begin{figure}
\plotone{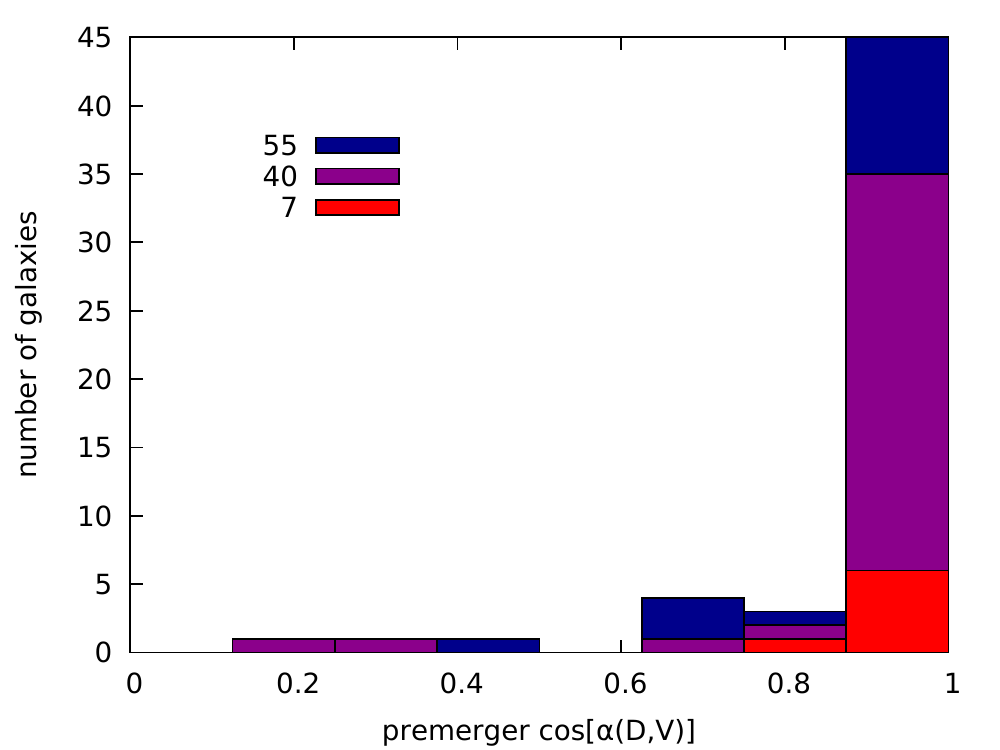}\\
\plotone{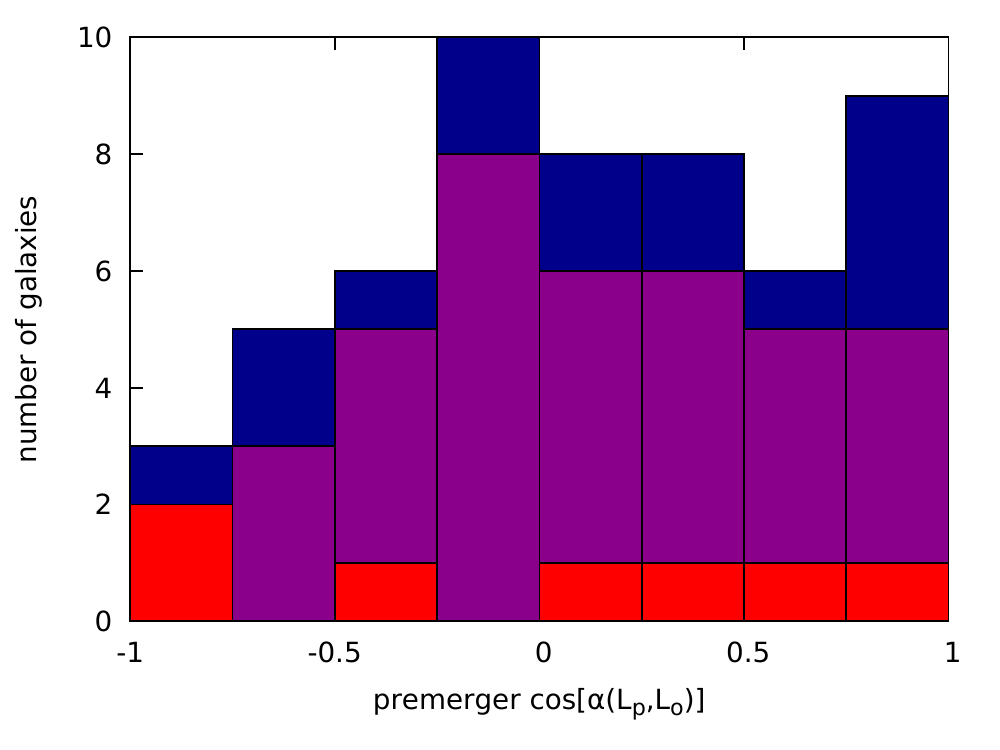}\\
\plotone{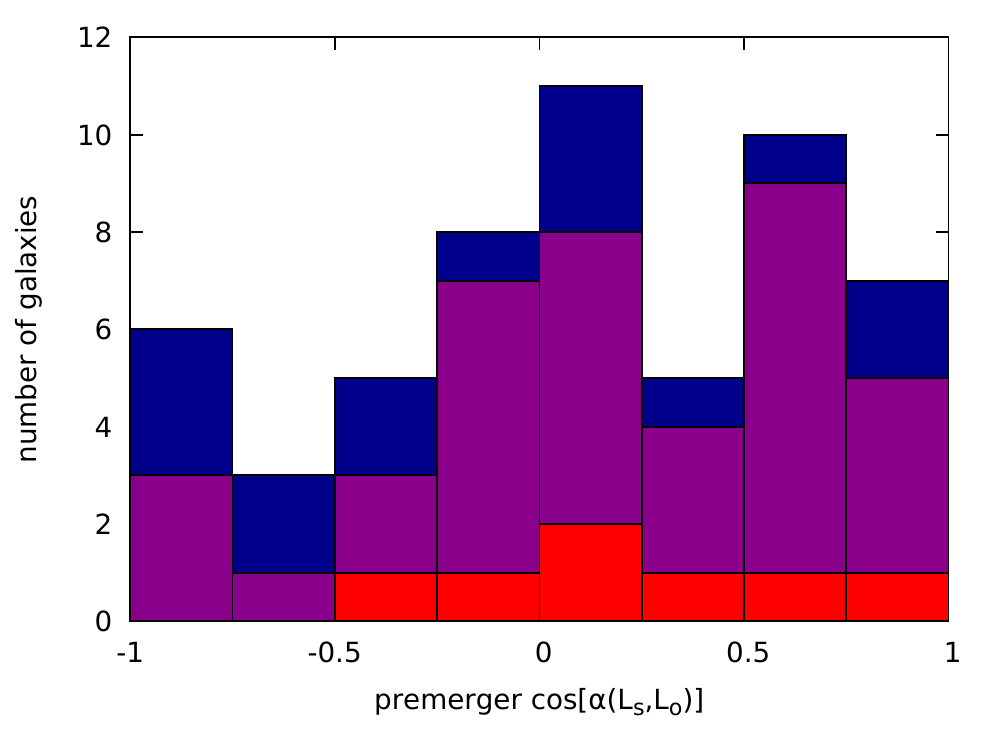}
\caption{
Same as Fig.~\ref{fig:histograms1}, but for other properties of the mergers.
\label{fig:histograms-alpha}
}
\end{figure}

For these reasons, we also selected a small subsample of 7 isolated mergers associated with an abrupt transition from
a rather oblate shape and practically zero $(L_x/L_{\rm tot})^2$ to a conspicuous prolate shape and rotation. We
refer to them as `golden~7'. These seven mergers seem to most convincingly engender the prolate rotators without
excessive disturbance from any other nearby significantly massive galaxy. Two of our six initial example galaxies
(Table~\ref{tab:g1-6}), galaxy-1 and galaxy-6, are also in the golden~7 sample. The other four of these six have more
complicated formation/merger histories but also end up as well-ordered prolate rotators (see
Fig.~\ref{fig:map-g2-6}). Fig.~\ref{fig:golden7} illustrates the evolution of $(L_x/L_{\rm tot})^2$ and triaxiality of
golden~7 galaxies. The timescales of the plot were shifted so that zero matches the time of the last significant
merger. Since all 7 galaxies had no significant merger for at least 3.5\,Gyr before the last one, there is an apparent
general trend for the galaxies to form a progressively flatter and flatter disk during this period.

Figs.~\ref{fig:histograms1} and~\ref{fig:histograms-alpha} present histograms of the properties of all last significant
mergers with overplotted histograms of the two subsamples: 40 well-correlated mergers and the golden~7. Except for the
merger time and gas fraction, the properties are calculated from the stellar particles at the pre-merger moment when
galaxies are still well separated as described in Sect.~\ref{sec:evo}.
The postmerger gas fraction is taken from the first snapshot in which the secondary progenitor does not exist anymore.
In the histogram of merger mass ratio, the bin
of 0--0.1 is not occupied by construction (since we exclude such minor mergers), but there is still apparent lack of
mergers at the lower mass ratio end while generally minor mergers should be more frequent. The well-correlated sample
is slightly shifted towards more recent mergers in comparison with the sample of all 55 last significant mergers. This
is naturally reflected in the gas fraction being on average higher for non-correlated mergers. Otherwise, the gas
fraction during the merger does not seem to play a key role in the creation of prolate rotators. It seems to follow
expected trends: for lower-mass galaxies and early mergers the gas fraction is high independently of whether the
merger time correlates with the kinematic transition. Massive galaxies tend to have low gas fraction quite early in
their evolution. All six most massive galaxies in our sample (stellar mass $>4\times10^{11}$\,M$_{\sun}$) have the
postmerger gas fraction of only 1--2\,\% with the last significant merger taking place 0.8--5.4\,Gyr ago.

In Fig.~\ref{fig:histograms-alpha} we examine three angles: $\alpha(D,V)$, the angle between the line connecting
the centers of the merger progenitors and their relative velocity;  $\alpha(L_{\rm p},L_{\rm o})$, the angle between the spin of the
primary (i.e. the angular momentum of stars in the galaxy) and the orbital angular momentum of the merging progenitors;
$\alpha(L_{\rm s},L_{\rm o})$, the same as previous but for the secondary progenitor. The histograms are made for bins of equal
width in the cosine of the angles to account for the solid angle of the same size.

\begin{figure}
\plotone{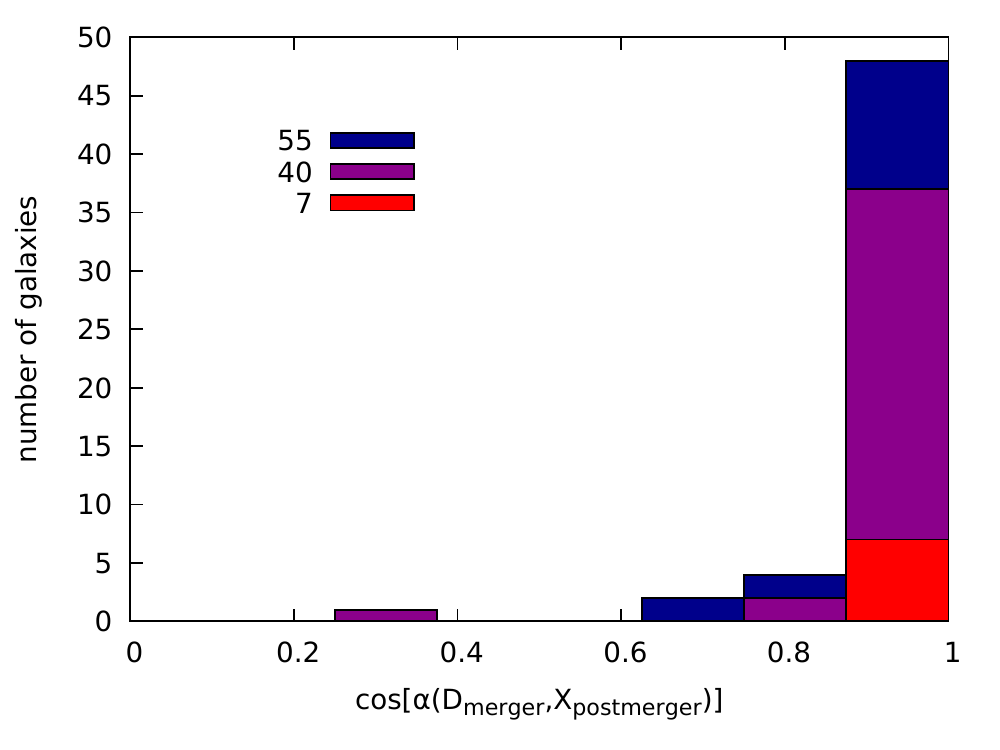}\\
\plotone{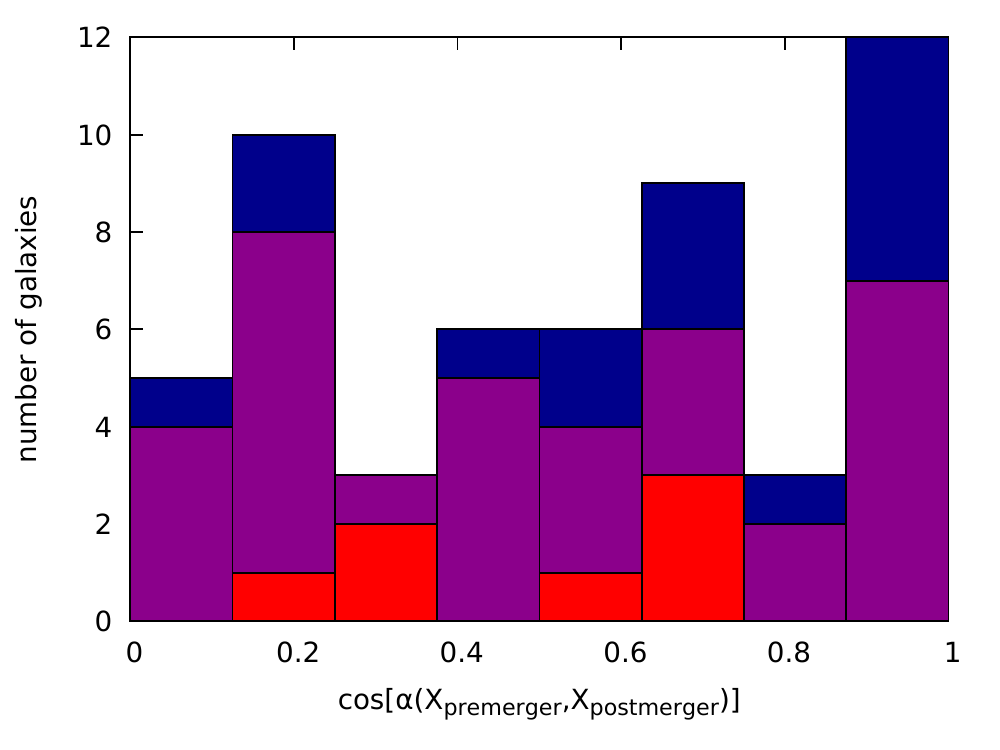}\\
\plotone{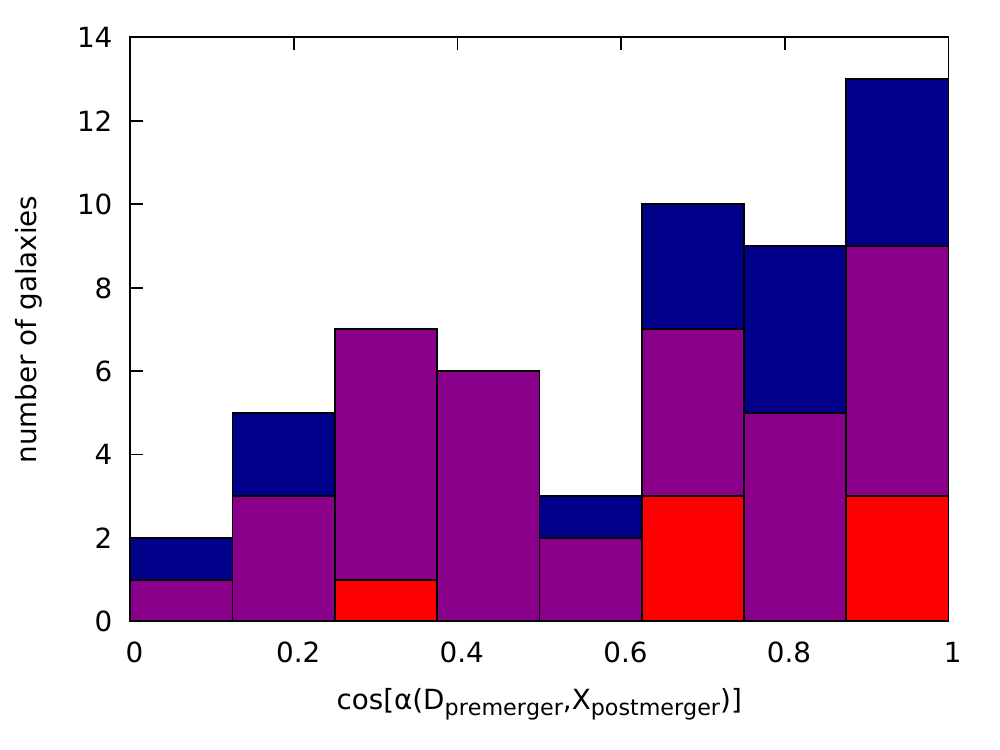}
\caption{
Same as Fig.~\ref{fig:histograms1}, but for other properties of the mergers.
\label{fig:DXL}
}
\end{figure}

Generally in cosmological simulations, mergers with high eccentricity are favored, but our sample seems to be even more
biased towards radial mergers (see Sect.~\ref{sec:others} and the top left panel of Fig.~\ref{fig:hisother}). There is
no clear trend for the orientation of galactic spins with respect to the orbital angular momentum of the two merging
galaxies. The anti-parallel arrangement seems to be slightly disfavored but it could be just a fluctuation or
an imprint of larger-scale structures. For the golden~7, the angles are distributed practically uniformly.

There is a strong correlation between the direction of the line connecting the centers of the progenitors at the last
snapshot before the merger, $D_{\rm merger}$, and the direction of the major axis of the galaxy two snapshots after the
merger, $X_{\rm postmerger}$, as illustrated in the top panel of Fig.~\ref{fig:DXL}. At the same time, as shown
by the middle panel of Fig.~\ref{fig:DXL}, most of the galaxies significantly changed the orientation of their major
axis during the merger. This means that the major axis of the merger remnant is set by the direction from
which the progenitors approach each other at the end of the merger. It is actually something that can be expected
especially for a set of mostly major mergers with highly eccentric orbits. Apart from a few cases, the major
axis is well established just after the merger and only slowly changes its direction (with respect to the coordinate
system of the simulation). In spite of the fact that the orbits of the merger progenitors are strongly radially biased,
there is no strong correlation between the incoming direction of the secondary before the first approach, $D_{\rm
premerger}$, and the future major axis of the merger remnants, $X_{\rm postmerger}$, as seen in the bottom panel of
Fig.~\ref{fig:DXL}. We also compared the orientation of spins of the progenitors before the merger with the orientation
of the major axis just after the merger. One would expect that prolate rotation emerges when the spin of the progenitor
is rather parallel to the future major axis, but our data do not seem to confirm these expectations. Especially the
golden~7 cover a wide variety of spin angles.

In other words, the formation of prolate rotators via a merger event seems to be related to the final stages of the
merger that depends on the details of the merger evolution and are not easily inferable from the initial conditions of
the merger.

\subsection{Comparison with a twin sample} \label{sec:others}

\begin{figure*}
\plottwo{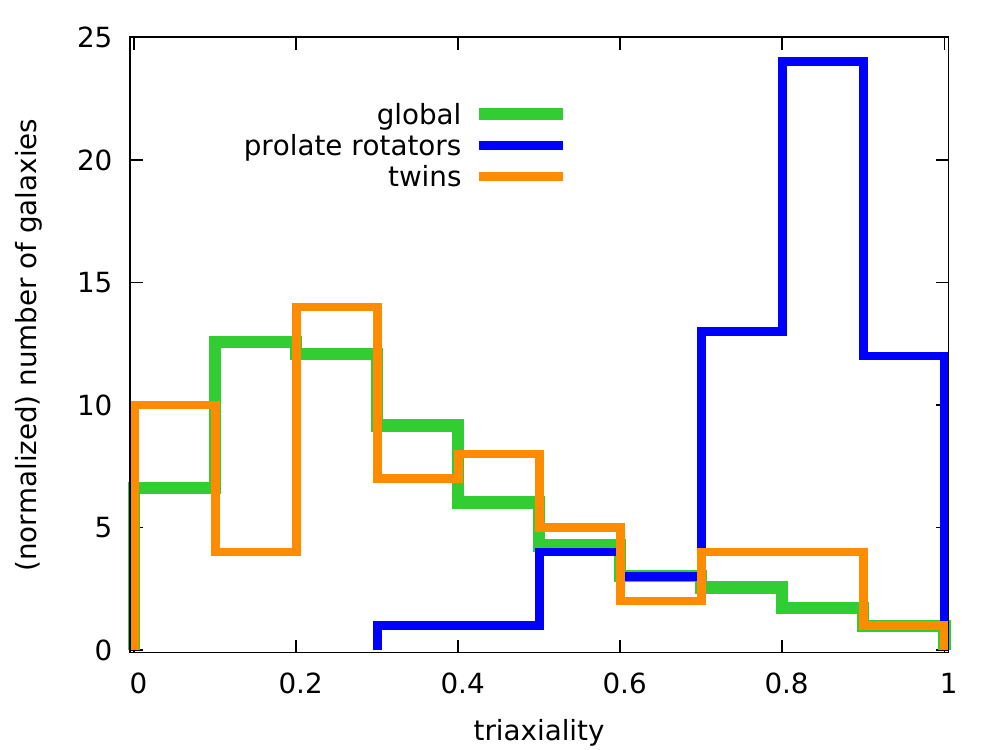}{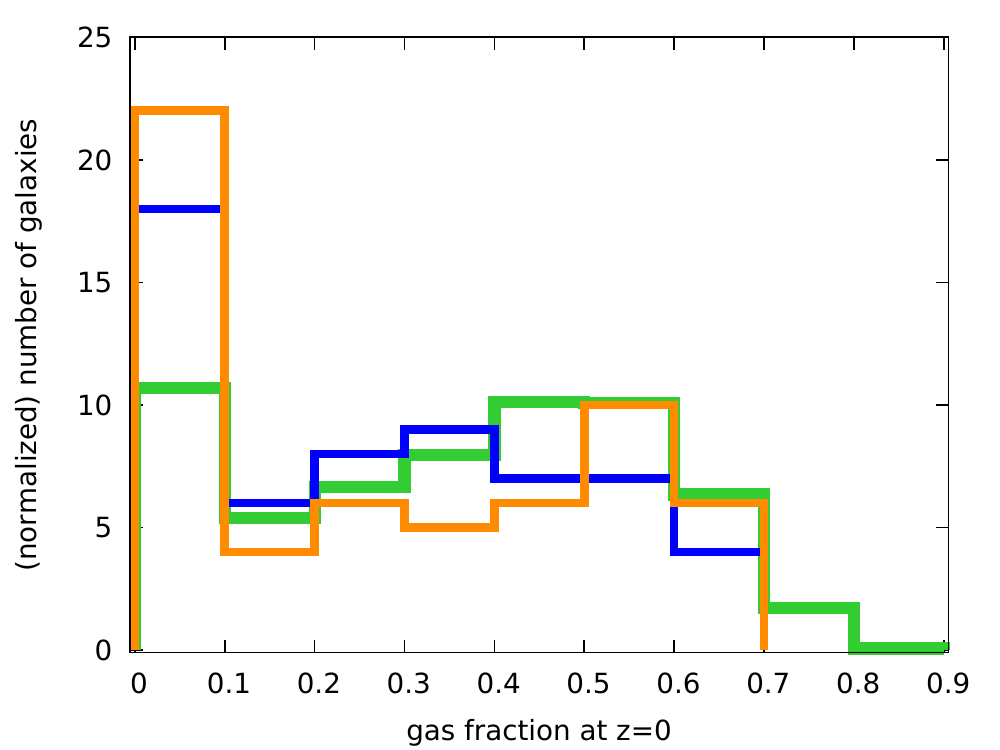}
\caption{
Distribution of the triaxiality parameter and the gas fraction at $z=0$ for the global sample, prolate rotators, and
the twin sample. The global sample is normalized to have the same total number of galaxies as the other two samples.
\label{fig:hist0other}
}
\end{figure*}

In order to investigate whether the mergers leading to prolate rotation are special in some sense, we need to
compare them with those happening to the global sample of galaxies. Since the mass distribution of the sample of prolate
rotators differs significantly from the mass distribution of all 7697 galaxies (Fig.~\ref{fig:Mdist}), it is not
sufficient to compare the properties of these two samples. Most differences would be probably caused just by the
general differences in the evolution of galaxies of different masses. In addition, since we analyze full merger trees
and the particle data of primary galaxies and many secondaries for many outputs, it is not feasible to repeat the whole
procedure for a large sample of galaxies. This led us to construct a twin sample of 59 Illustris galaxies. For each
galaxy from the sample of prolate rotators, we pick a galaxy from the remaining 7638, that is similar in mass, size,
and minor-to-major axis ratio.

During the processing of the stellar particle data from the last snapshot ($z=0$) for all 7697 galaxies, we also
performed simple fitting of 3D density with both Hernquist and Jaffe profiles \citep{her90,jaff83}. Only some fraction
of galaxies has reasonable values of the fitted scale radius (about 3/4 for Hernquist profile and 1/2 for Jaffe
profile). The values of Hernquist and Jaffe scale radius, when expressed in units of the effective radius, are
consistent for the same galaxy (within 20\,\% difference) only for about 26\,\% of the global sample. In general, the
unsightly value of the fitted scale radius can indicate, at least in some cases, that something odd is happening to
the galaxy. The sample of prolate rotators has quite reasonable scale radii for all galaxies for Hernquist fits and
all but three have reasonable values for Jaffe profiles (although they are not always within 20\,\% difference). For
these reasons, we require for the twin galaxies to have a believable scale radius at least for the Hernquist profile.

In a bin of $\pm2$\,\% stellar mass difference centered on the mass of a given prolate rotator, we choose the best-matching galaxy within a 10\,\% difference in minor-to-major axis ratio and 20\,\% difference in $r_{\rm max}$ value.
In a few cases we were not able to find any suitable twin galaxy and we had to expand these boundaries.

\begin{figure*}
\plottwo{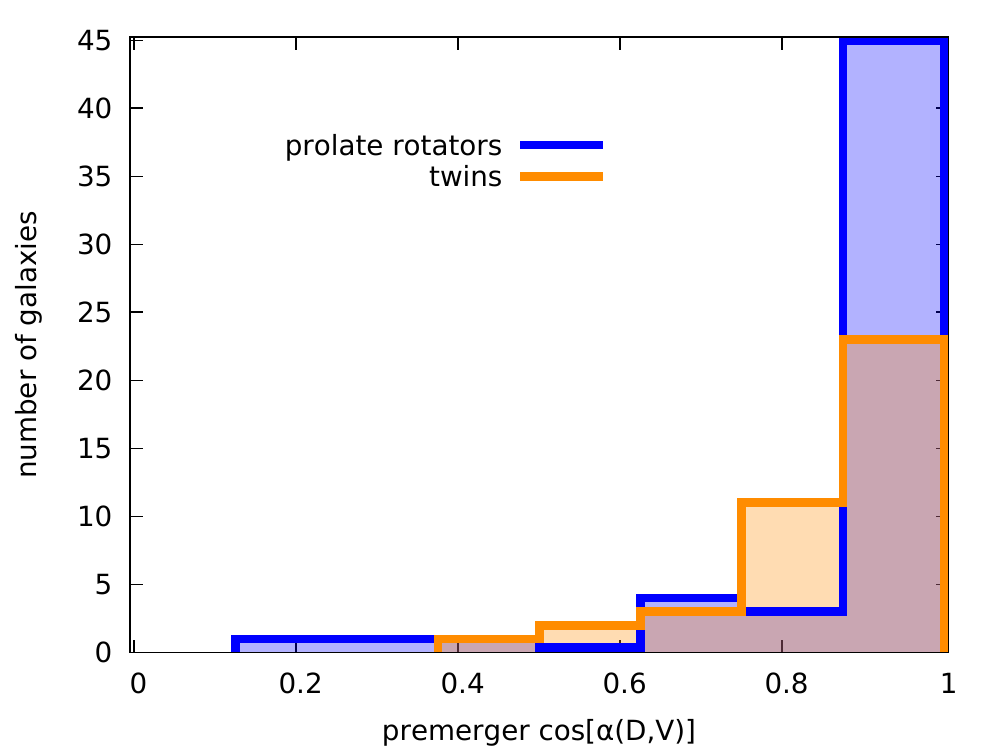}{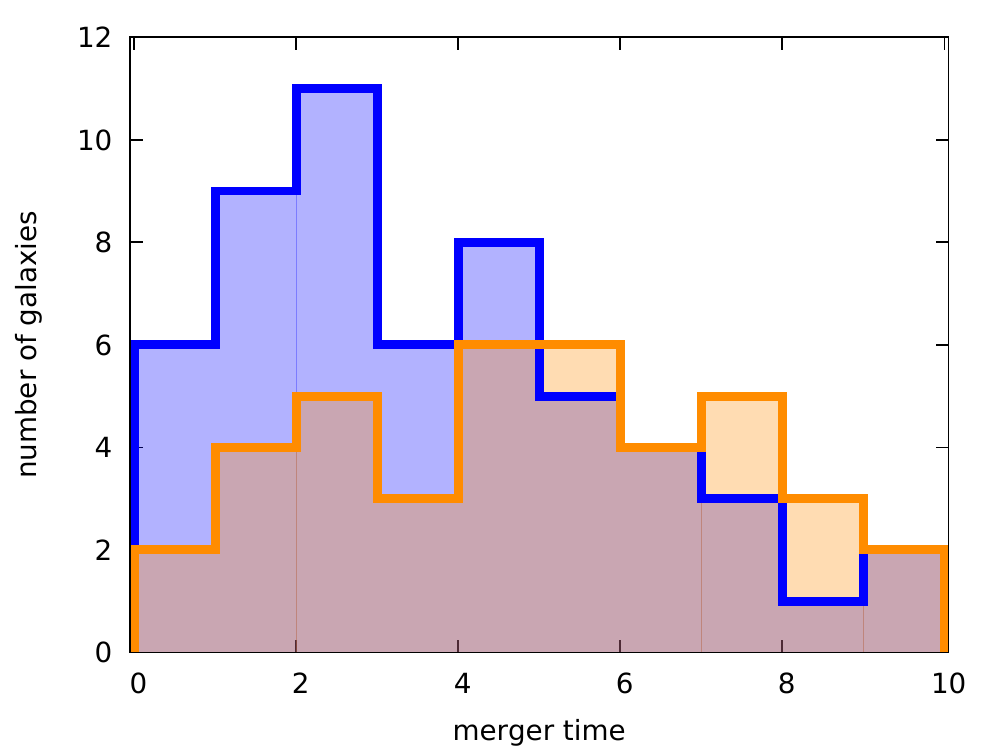}\\
\plottwo{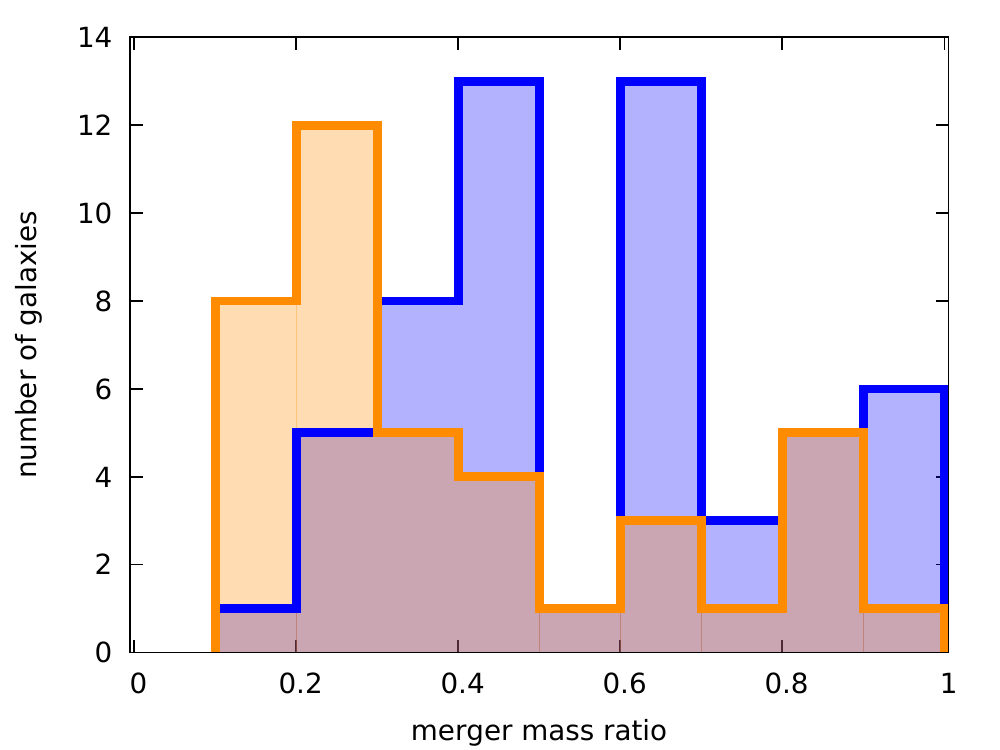}{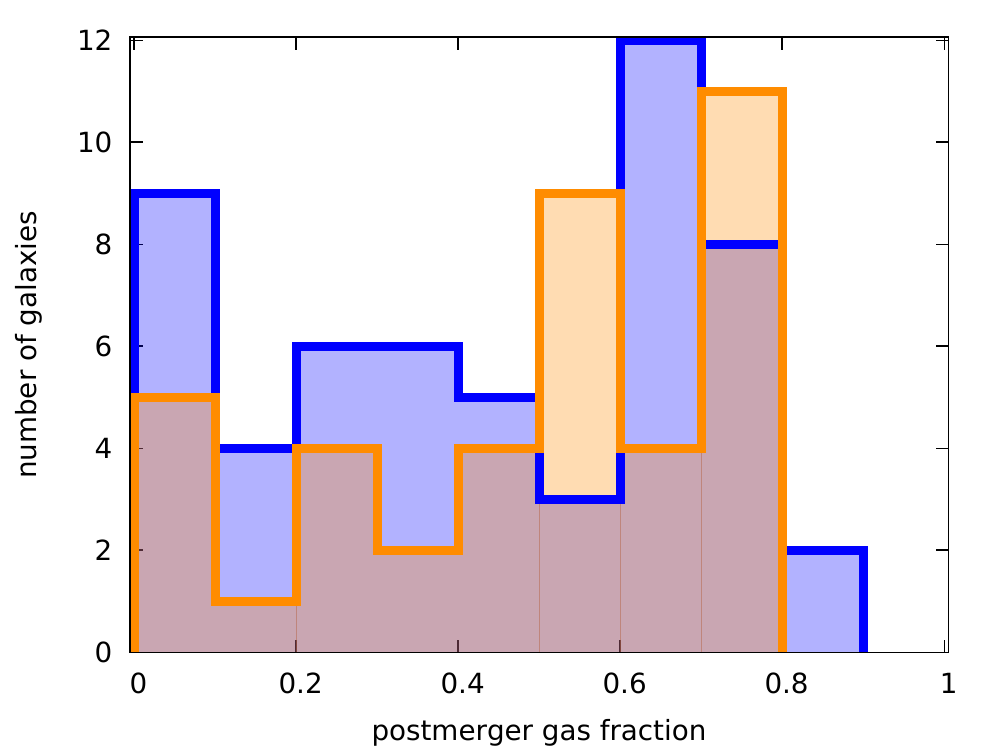}
\caption{
Histograms of the properties of all 55 last significant mergers for the prolate rotators compared to the 40 last
significant mergers of the twin sample.
\label{fig:hisother}
}
\end{figure*}

By construction, the twin sample and the sample of prolate rotators have similar mass distributions with a higher
percentage of massive galaxies in comparison to the global sample (Fig.~\ref{fig:Mdist}). The distribution of the
triaxiality parameter, $T$, and the gas fraction in the last snapshot ($z=0$) is depicted in Fig.~\ref{fig:hist0other}.
Prolate rotators have also a prolate shape with a few exceptions of triaxiality around 0.5. The twin sample covers
the whole range of shapes with slightly more prolate galaxies when compared with the global sample. Both, the prolate
rotators and their twins, have higher ratio of gas-poor galaxies since the more massive galaxies are more likely to be
gas-poor.

For the twin sample, there are less mergers satisfying criteria described in Sect.~\ref{sec:evo}. Prolate rotators
experienced 139 such mergers while the twins just 100. This supports the idea that some prolate rotators were created
by adding impacts of two (or more) subsequent mergers. Fig.~\ref{fig:hisother} compares the quantities for the last
significant merger of the two samples. The twin sample has only 40 such mergers compared to 55 of prolate rotators. As
expected from the general distribution of merger parameters in cosmological simulations, the twin sample is also biased
towards more radial mergers but the fraction of galaxies in the most radial bin is significantly higher for prolate
rotators. The prolate rotators show also more recent mergers and less mergers with lower mass ratio. Accordingly, the
postmerger gas fraction is on average higher for the twin sample: 44\,\% of prolate rotators and  60\,\% of twins have
the gas fraction grater than 0.5 at the end of the merger.

We do not show histograms of the orientation of spins of progenitors with respect to the orbital angular momenta. The
twin sample also does not show any clear trend and has a similar distribution to the prolate rotators. Both samples seem
to mildly disfavor antiparallel orientation, at least for the spin of the primary galaxy. Similarly to prolate
rotators (Fig.~\ref{fig:DXL}), the twin sample shows a strong correlation between the direction of the postmerger major
axis and the direction to the secondary at the end of the merger, but with lower fraction of the smallest angles
(87\,\% for prolate rotators and 70\,\% for twins).

\subsection{Disentangling contributions to prolate rotation} \label{sec:Lx}

Here we investigate the origin of the stellar particles that contribute to the prolate rotation at $z=0$. For the 40
galaxies with well-correlated time of the kinematic transition and the time of the last significant merger, we divide
the stellar particles taking part in this merger into four groups: (1)/(2) particles belonging to the
\textit{primary}/\textit{secondary} before the merger when the galaxies were still well separated, (3) \textit{new}
particles formed after the end of the merger, and (4) the \textit{rest}, which consists of particles mostly born
during the merger but this category may also include stars created before the merger outside the primary and the
secondary and only later accreted.

We compute the relative contribution of these groups to the $x$-component of the angular momentum as $\pm(L_i/L_x)^2$.
The quantity is negative for counter-rotating particles, $x$ is coincident with the 3D major axis of the galaxy, and
$L_i$ is the $x$-component of the angular momentum of the respective group. We find that 26 galaxies have
$(L_i/L_x)^2>0.5$ for one group: 13 primary, 3 secondary, 9 new, and 1 rest. For the galaxies where new stars dominate,
the prolate rotation still could have emerged from older stars at the time of the merger, but new stars were formed on
circular orbits consistent with the prolate rotation and gradually took over.

One would expect that when the premerger angular momentum of the primary progenitor is aligned with the postmerger
major axis, the prolate rotation would come mostly from the particles of the primary. Indeed, 7 galaxies with those
quantities most aligned also have $(L_1/L_x)^2>0.5$. On the other hand, the galaxy with the most perpendicular
orientation (89.1$^{\circ}$) has, counterintuitively, $(L_1/L_x)^2=0.95$. This galaxy, however, has lost most of its
rotation during the merger and shows only weak overall rotation at $z=0$. Three galaxies with dominant primary
particles even have $(L_1/L_x)^2>1$ due to counter-rotating particles from the secondary ($-0.14<(L_2/L_x)^2<-0.46$)
and, in one case, also from the new stars ($(L_1/L_x)^2=-0.11$). Among all 40 galaxies, there are 12 with one
counter-rotating group with $(L_i/L_x)^2<-0.05$ and one with two such groups. Interestingly, 5 of the golden~7 galaxies
belong to this collection.

\begin{figure*}
\resizebox{\hsize}{!}{\includegraphics{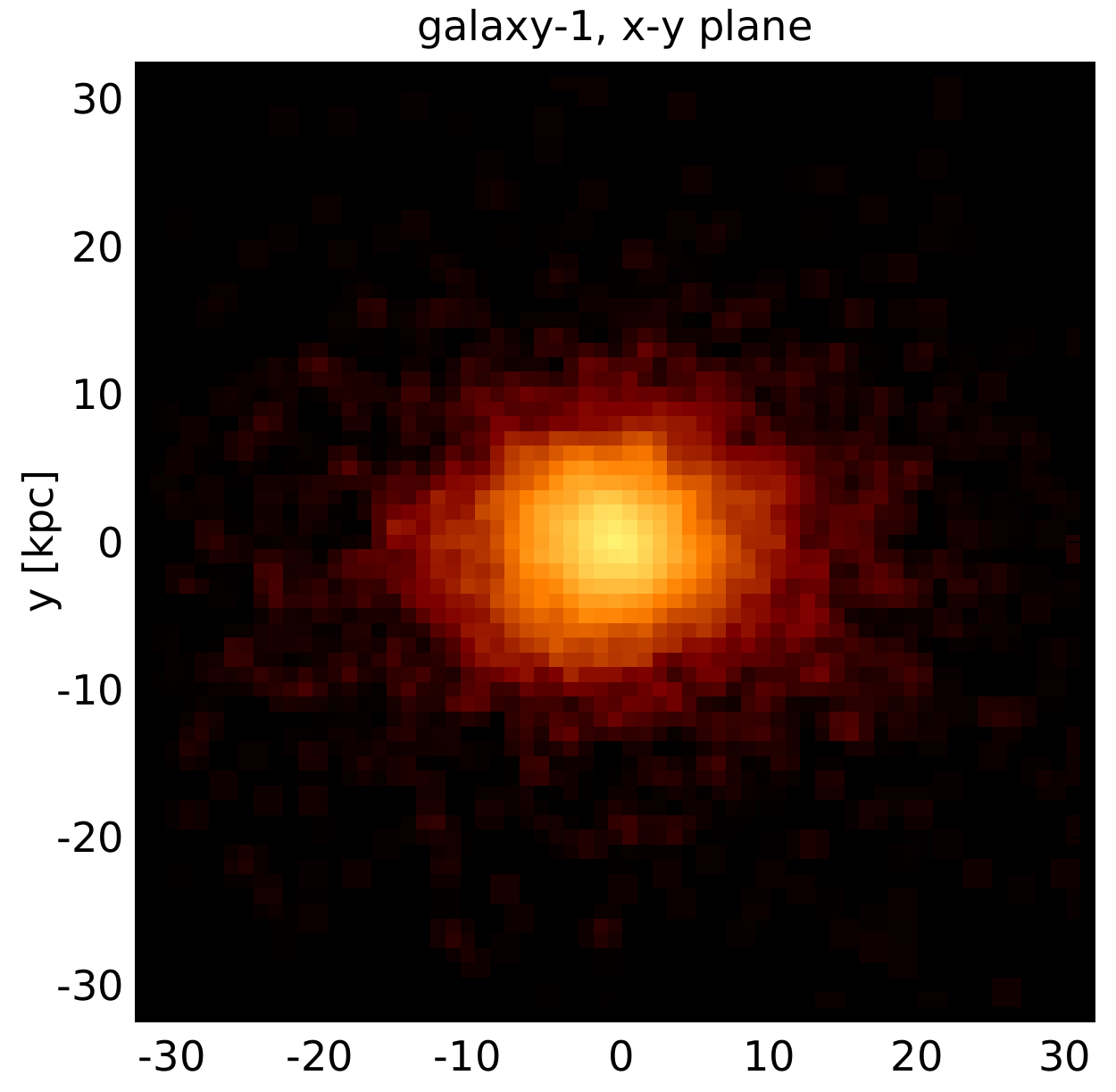}\includegraphics{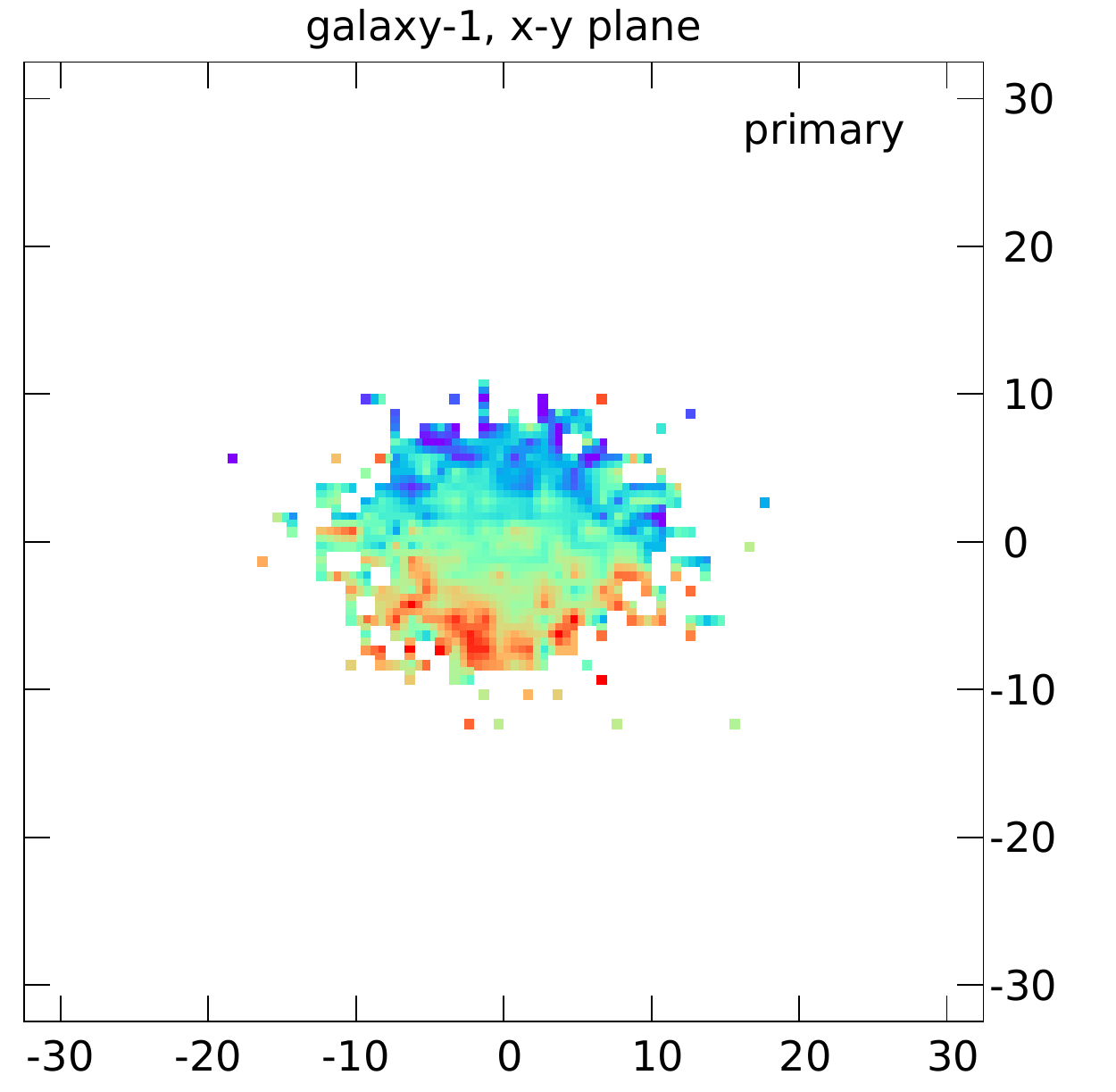}\includegraphics{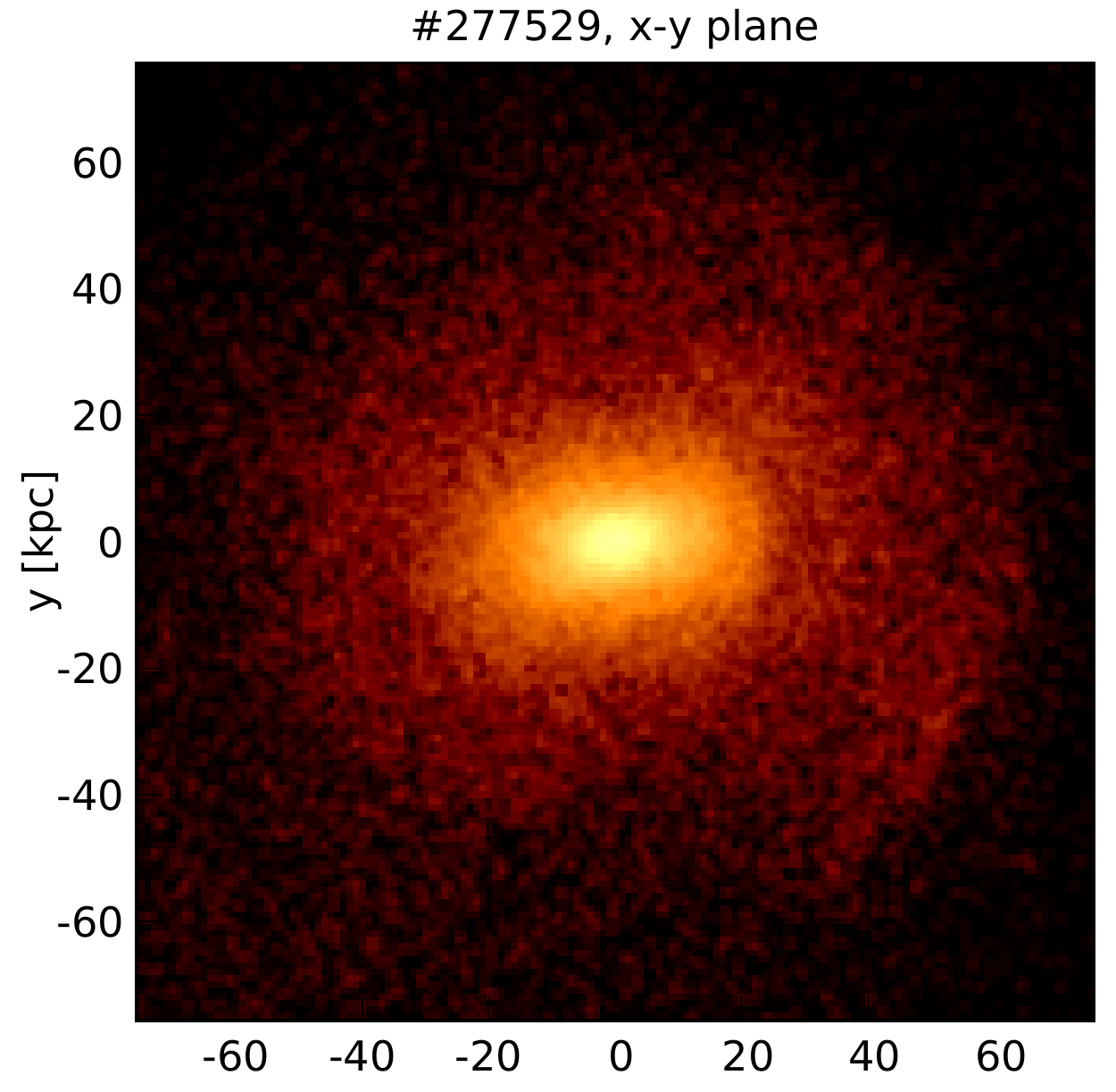}\includegraphics{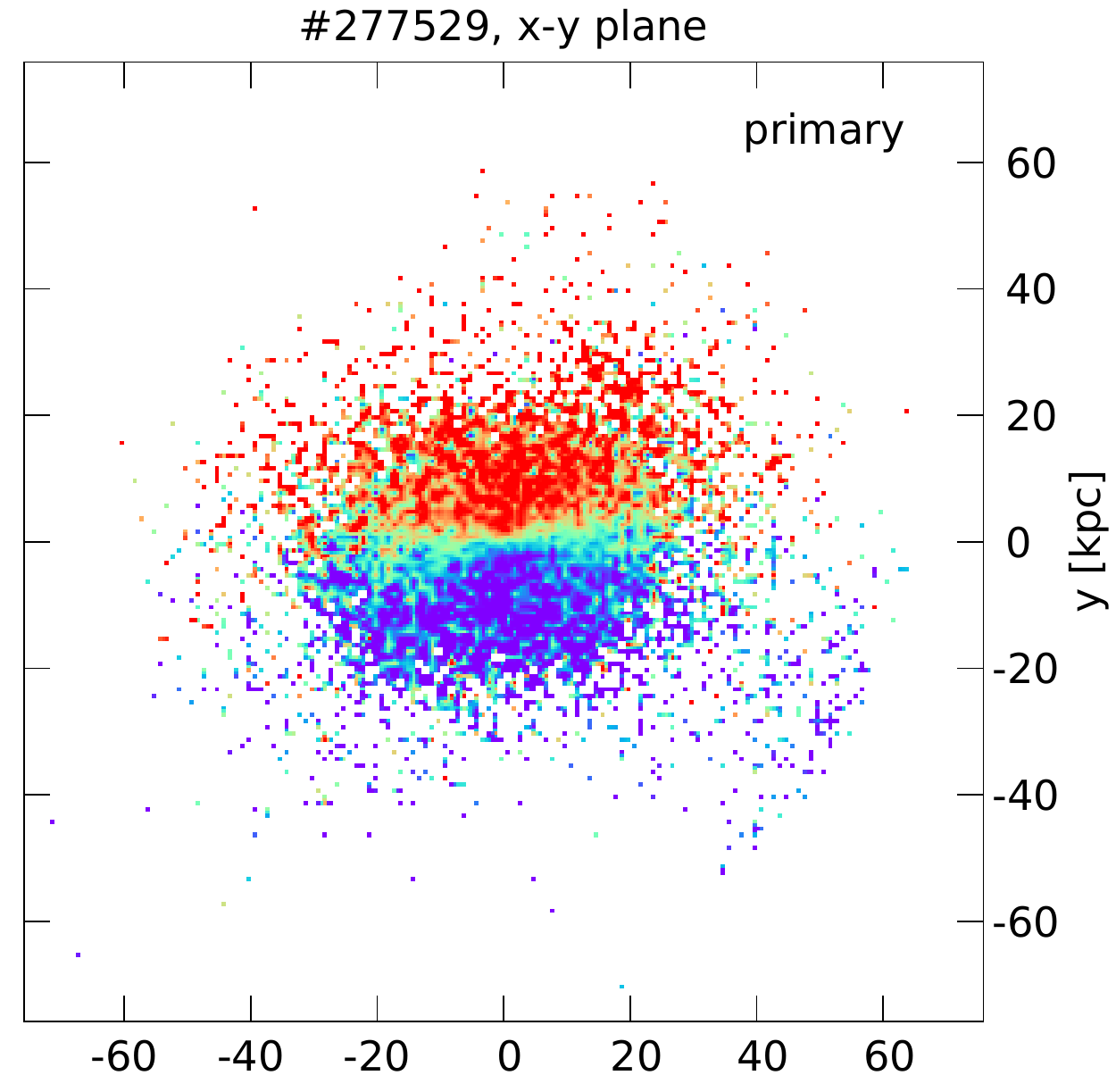}}
\resizebox{\hsize}{!}{\includegraphics{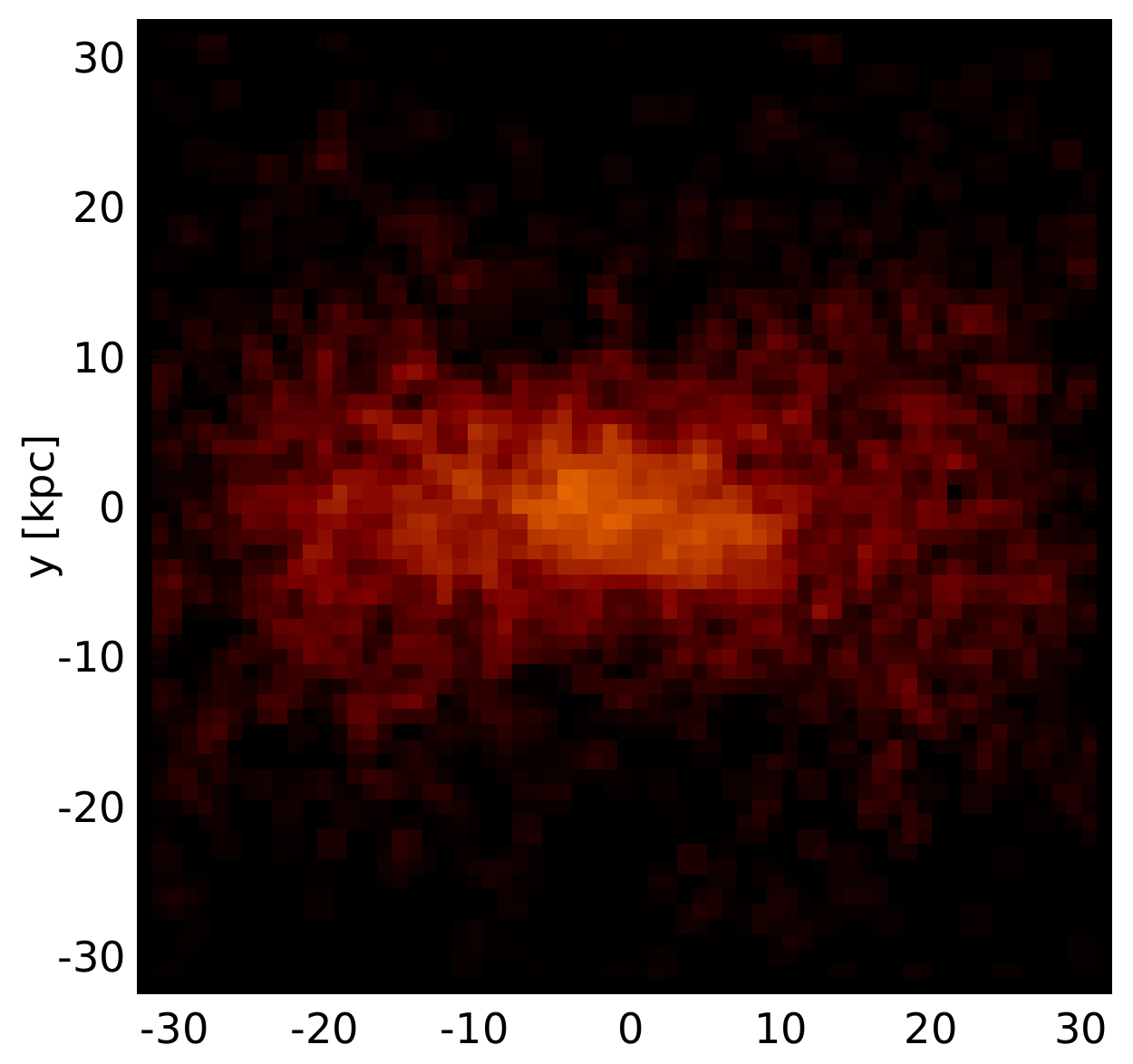}\includegraphics{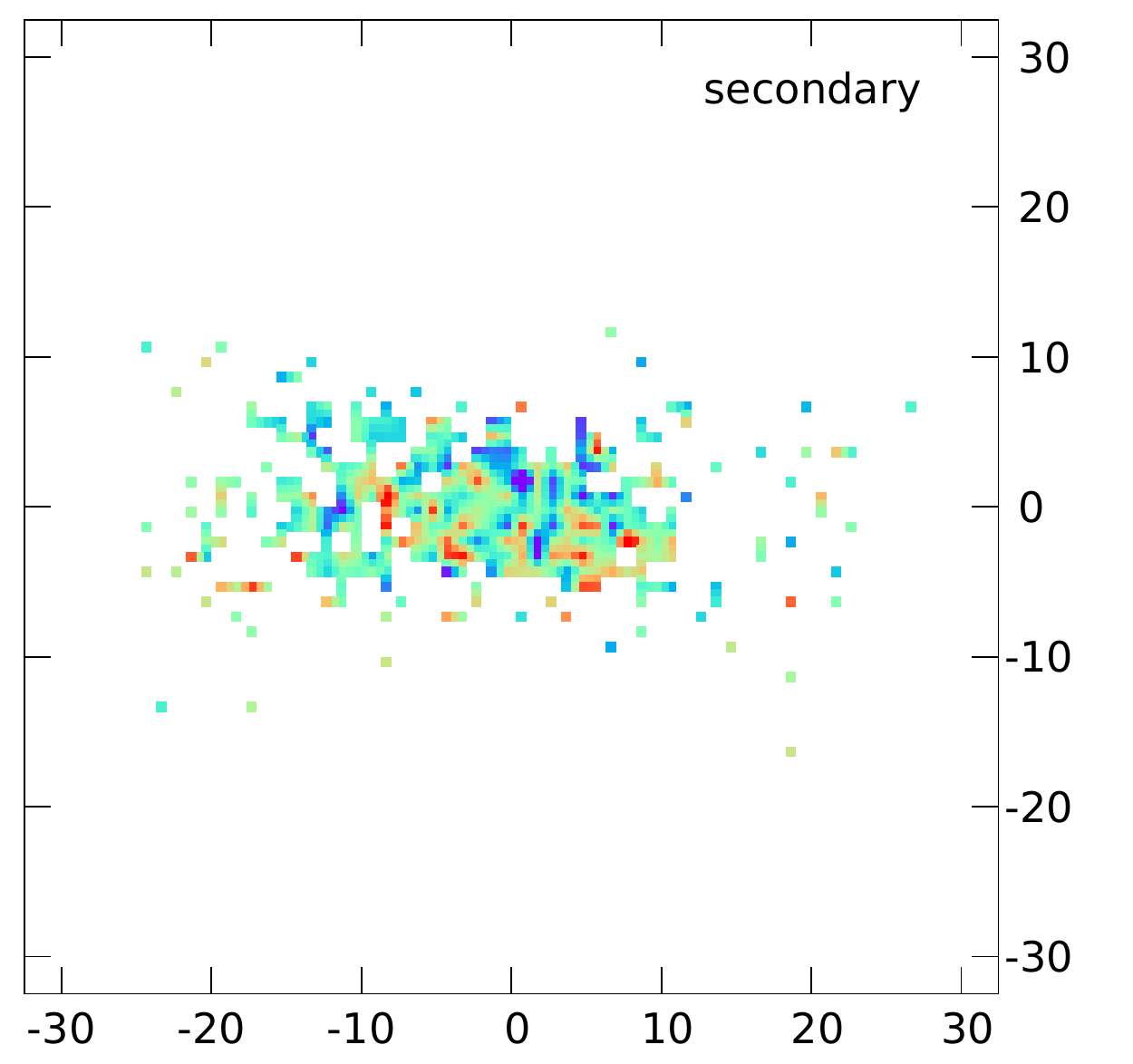}\includegraphics{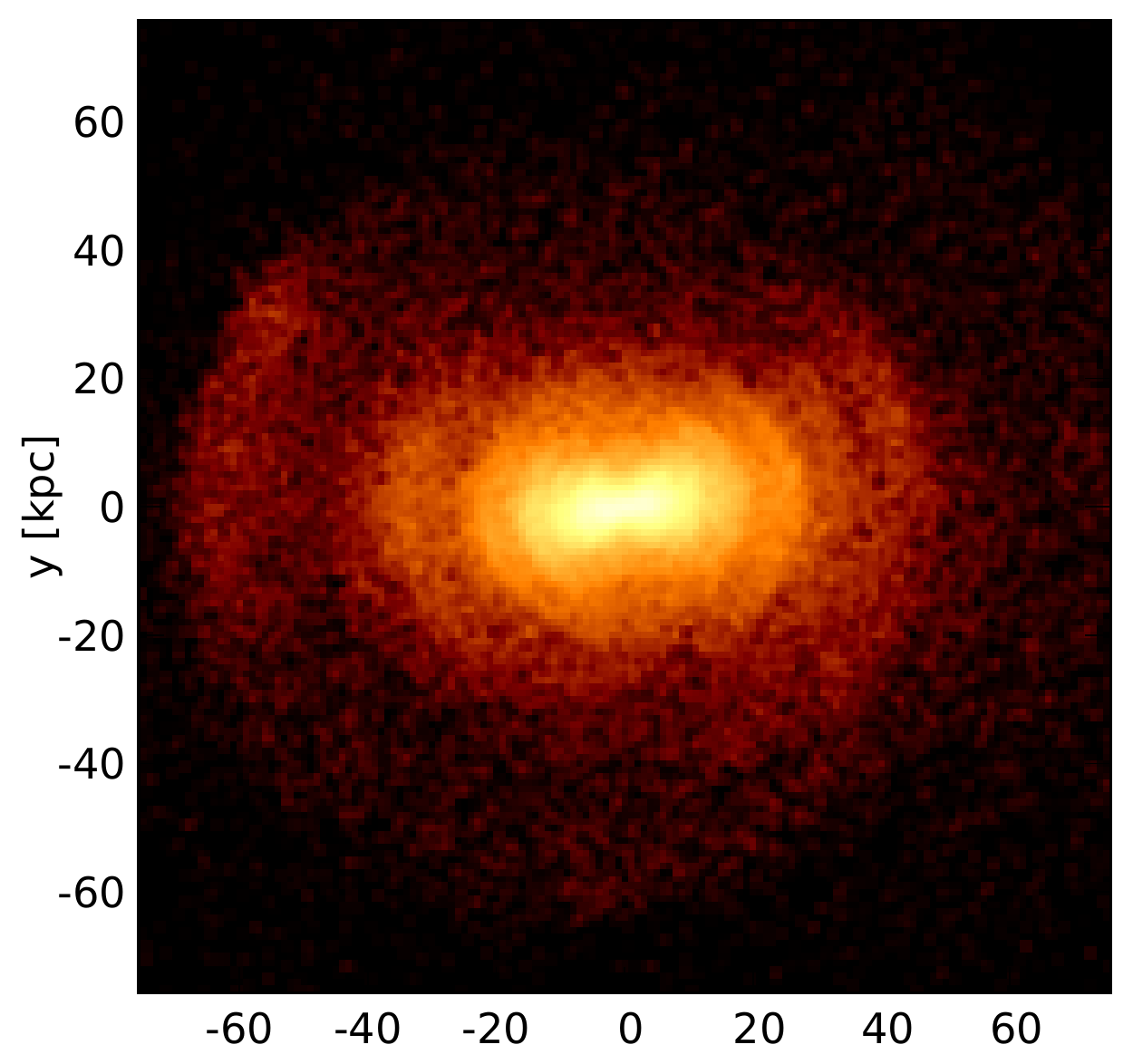}\includegraphics{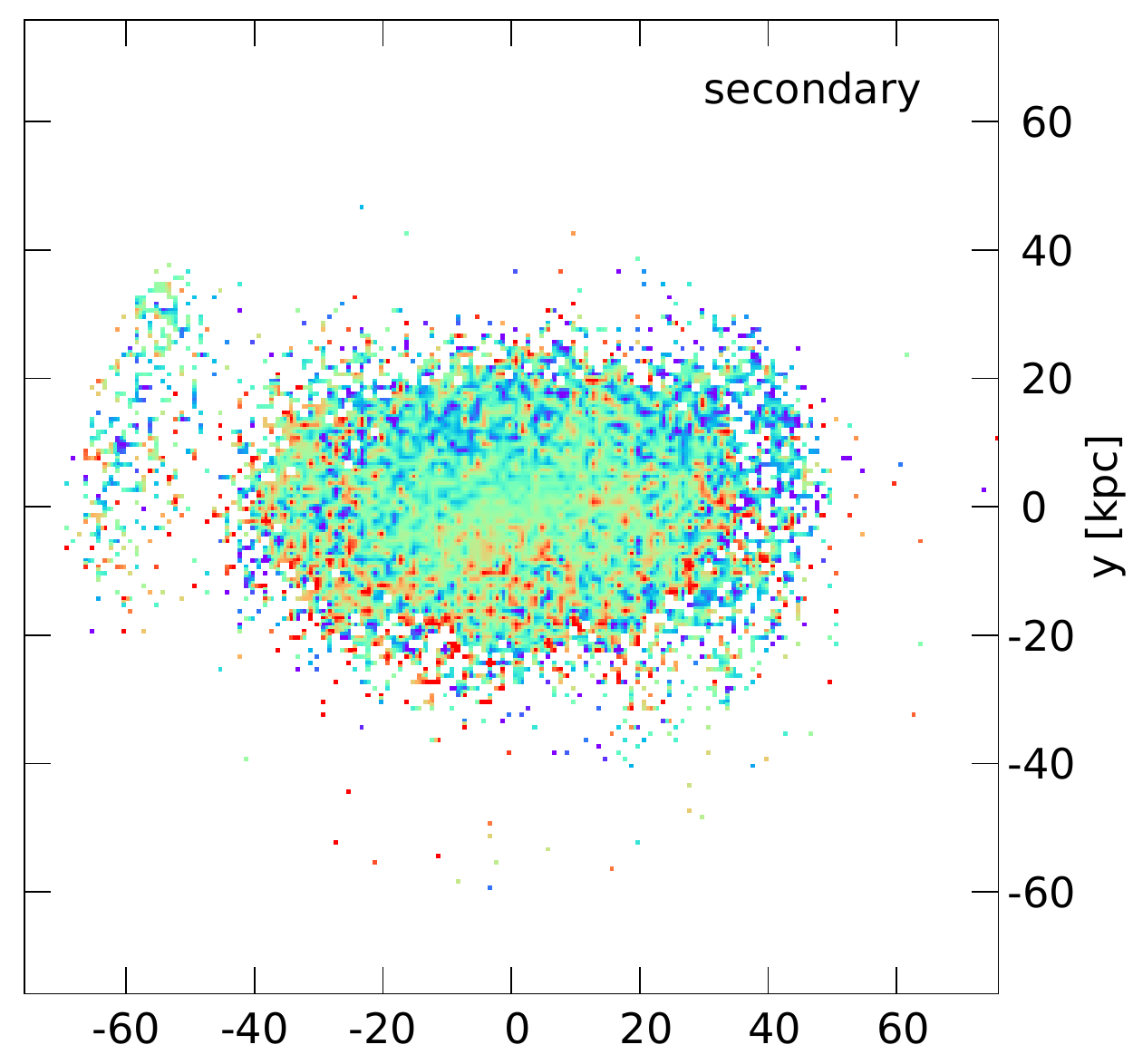}}
\resizebox{\hsize}{!}{\includegraphics{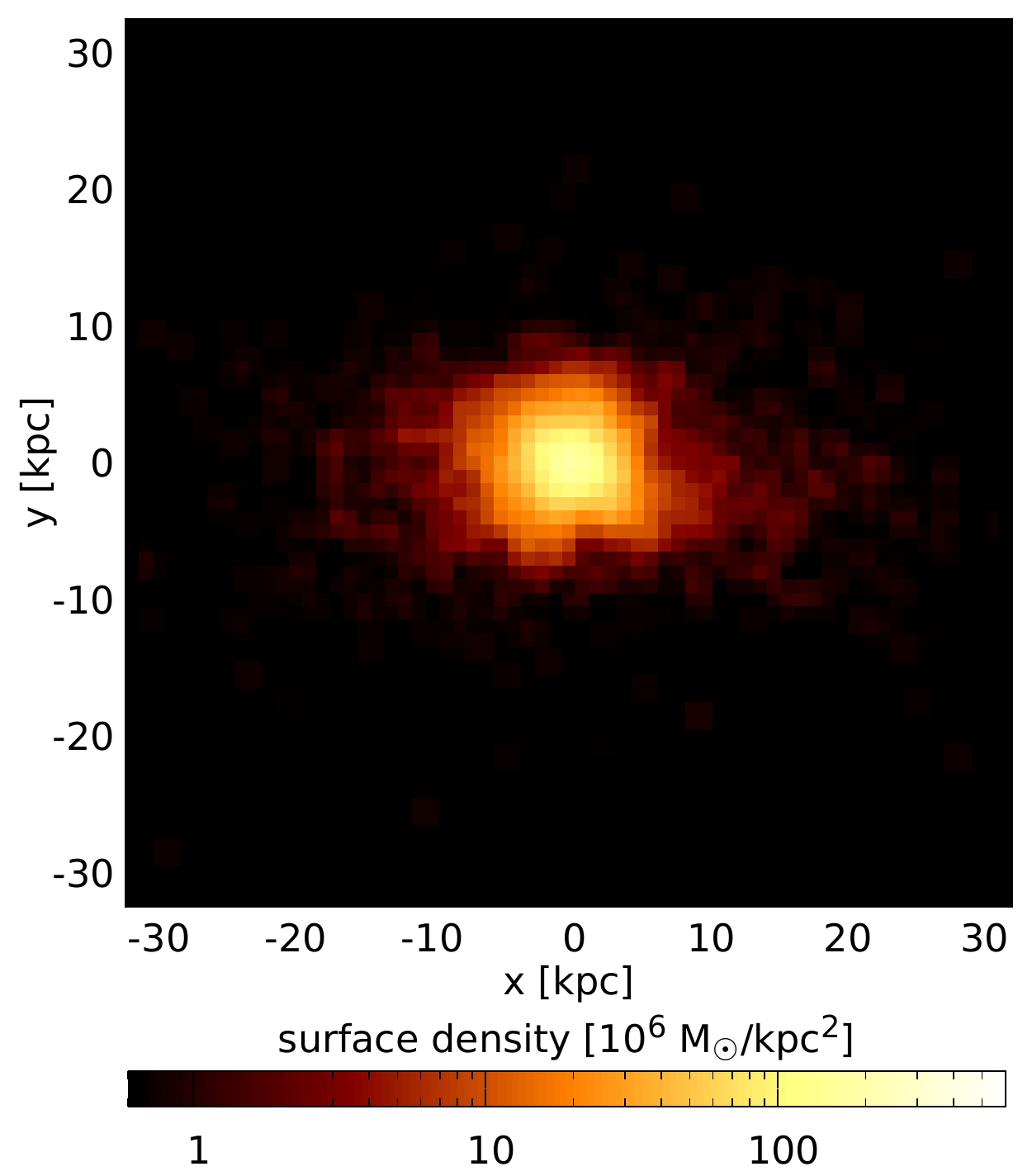}\includegraphics{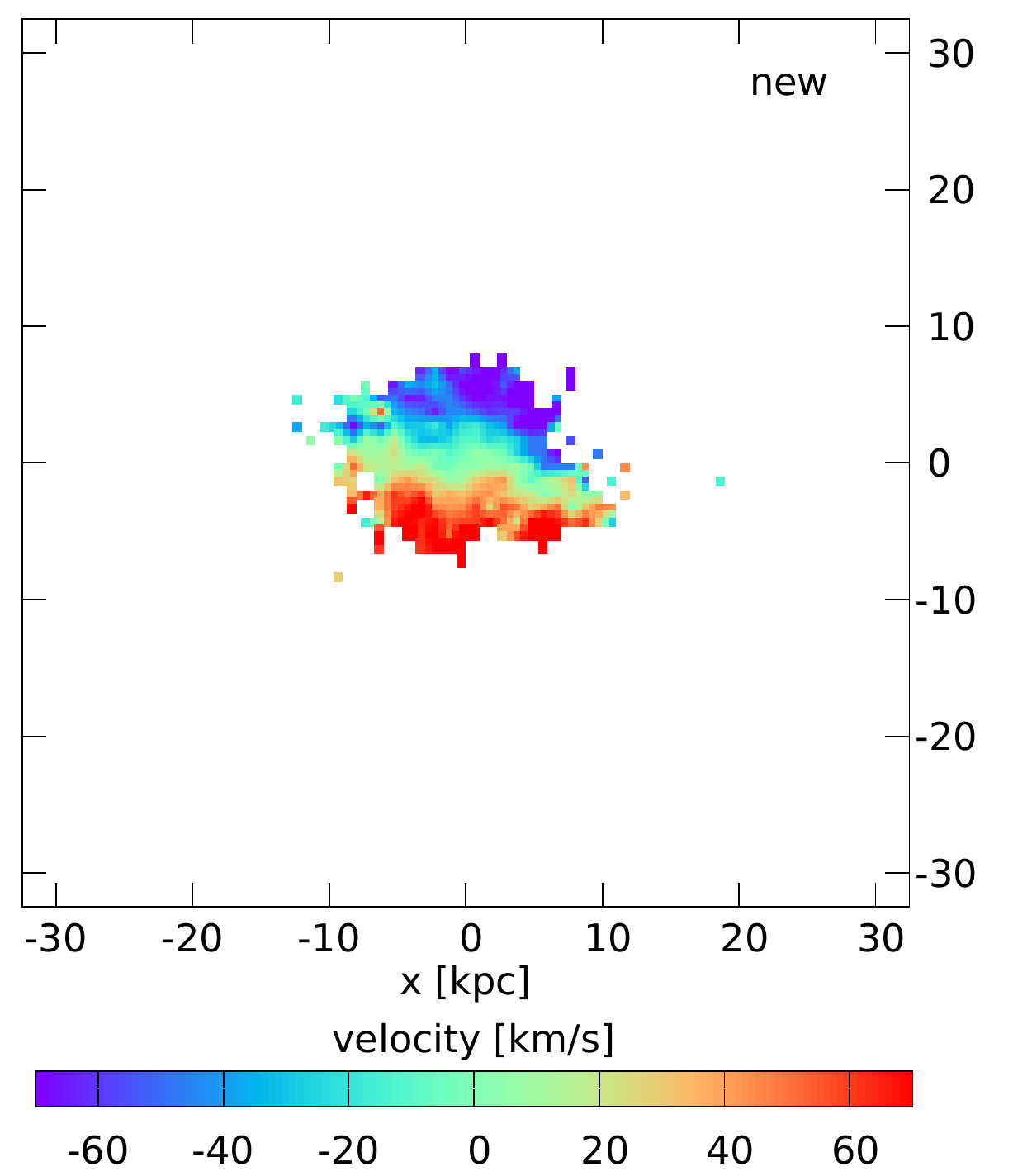}\includegraphics{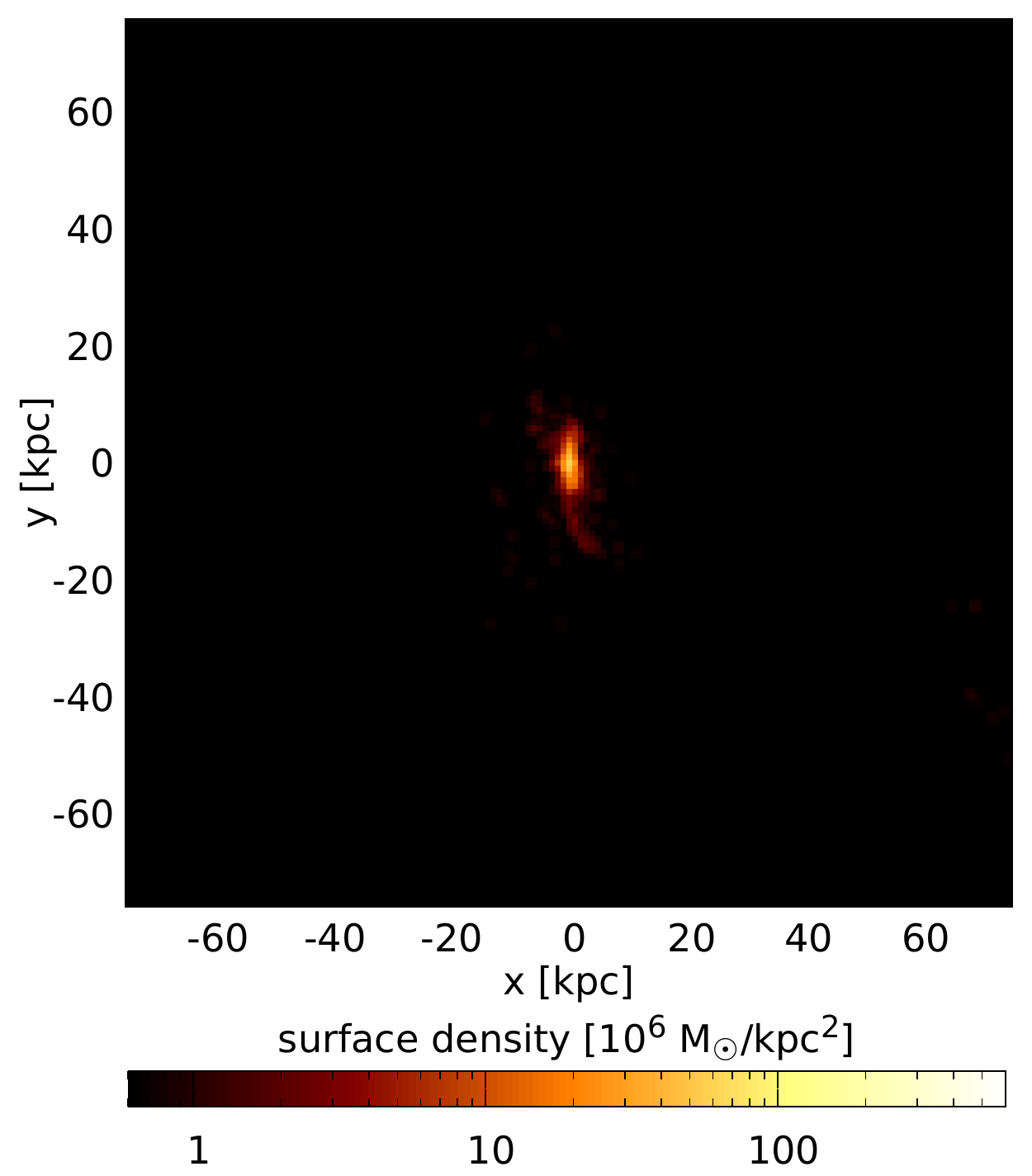}\includegraphics{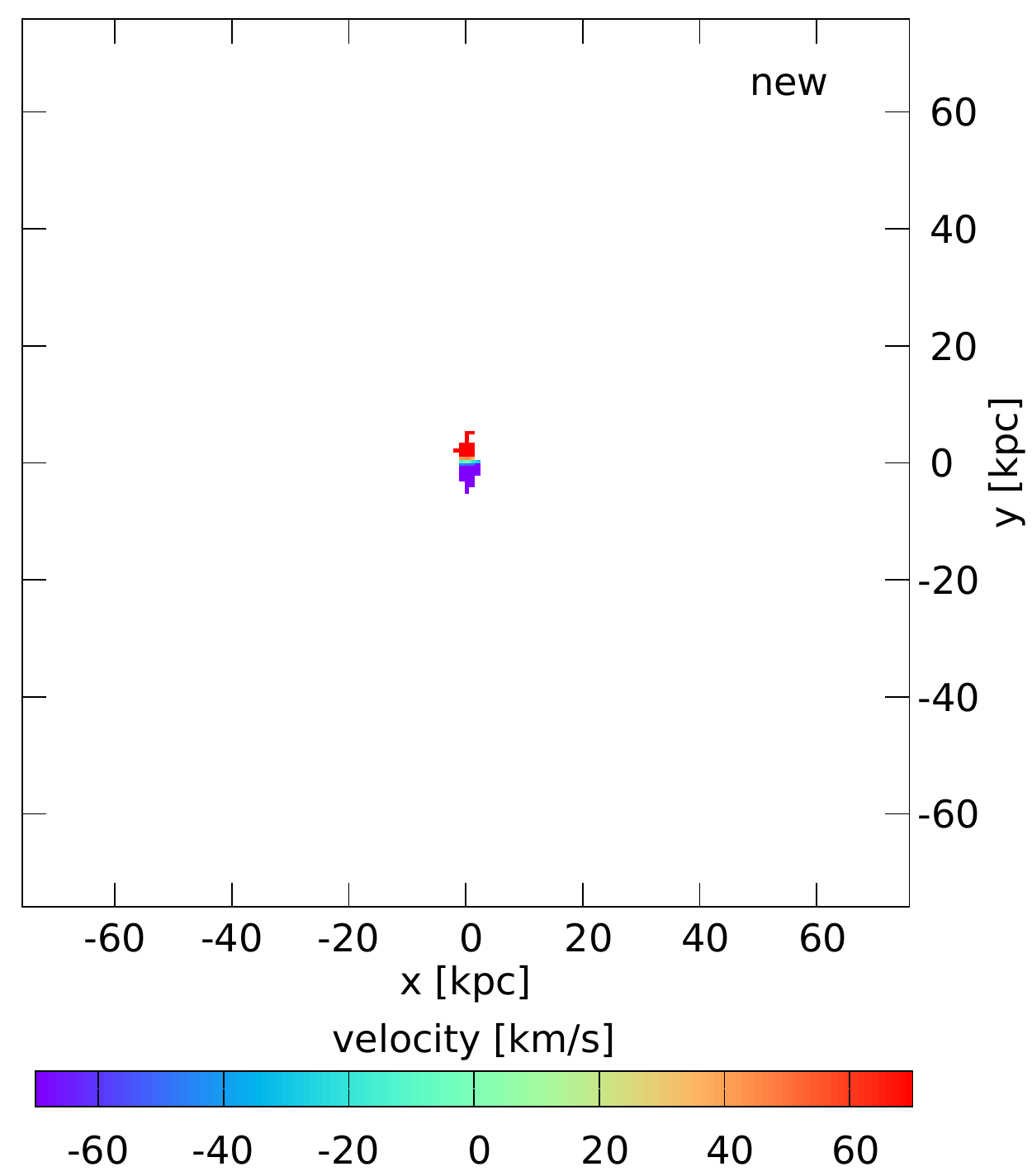}}
\caption{
Maps of the stellar surface density and the mean line-of-sight velocity for galaxy-1 and \#277529 viewed along the
3D minor axis. The stellar particles were divided according to their origin.
The range on the axes corresponds to [-$r_{\rm max}$, $r_{\rm max}$] for the respective galaxy.
\label{fig:map-psn}
}
\end{figure*}

Fig.~\ref{fig:map-psn} shows the separated contributions of different groups of particles for galaxy-1 and for one
of our 59 prolate rotators with a weak counter-rotating component (\#277529). For galaxy-1 (see also
Fig.~\ref{fig:map-g1} and Table~\ref{tab:g1-6}), new stars contribute the most to the prolate rotation at $z=0$. Stars
from the primary progenitor also show significant prolate rotation, while stars from the secondary rotate only weakly
but their distribution significantly supports the resulting elongated shape. The galaxy \#277529, shown in the
right-hand panels of Fig.~\ref{fig:map-psn}, is a massive galaxy with $r_{\rm max}=75.9$\,kpc and stellar mass
$M(r_{\rm max})=24.0\times10^{10}$\,M$_{\sun}$. It experienced a recent major merger 0.79\,Gyr ago with the mass ratio
of 0.84. The kinematic transition happened already during the penultimate pericenter passage of the secondary 1.44\,Gyr
ago. The prolate rotation is almost entirely provided by stars from the primary, while the secondary stars rotate
weakly in the opposite direction. New stars display strong rotation, co-rotating with the primary stars, but they are
restricted to the small central disk perpendicular to the major axis of the galaxy. A similar feature, in the form of
an elongated distribution of young stars in the central region aligned with the projected minor axis, was reported
for the Phoenix dwarf galaxy in addition to the clear presence of prolate rotation \citep{kach17}.

\subsection{Prolate rotators and shell galaxies} \label{sec:shells}

As demonstrated by Fig.~\ref{fig:map-psn}, galaxy \#277529 possesses noticeable shells.
Shell galaxies have been believed to result from close-to-radial minor mergers for decades
\citep{q84, dc86, hq88, e12sg}, but recently the view concerning their origin is shifting from minor mergers to
intermediate-mass \citep{duc15} or even major ones \citep{illsg17}.
Among observed galaxies, stellar shells occur in roughly 10\,\% of ETGs  \citep[e.g.,][]{mc83,at13}.
Interestingly, similarly to prolate rotators, shell galaxies also
seem to be more frequent among the high-mass ETGs \citep{tal09}.
Also in Illustris shell galaxies show this trend \citep{illsg17}.

Because both shells and prolate rotation are (1) more common in massive galaxies and (2) more
likely to emerge from close-to-radial mergers, we can expect prolate
rotators to be often shell galaxies and vice versa.
Indeed, at least 10 of 25 known prolate rotators, or candidates for
such, listed in \cite{tsa17} are shell galaxies.
Most of them are known shell galaxies \citep{arp66,mc83,js88,for94,for95,duc11,wil13}, a few more can be identified by the visual inspection of SDSS images.

Among our 59 Illustris prolate rotators, at $z=0$, there are about 14
quite clear cases of shell galaxies and other 16 possible candidates.
About one third of observed shell galaxies have the shells
well aligned with the major axis of the galaxy \citep{pri90}.
Almost all shell galaxies from our sample of Illustris prolate
rotators have shells aligned with the major axis, since both, the
major axis and the axis of the shell system, are set by the direction
of the collision at the final stages of the merger.

The visibility of the shells in Fig.~\ref{fig:map-psn} is enhanced
because the mass of the galaxy is separated into contributions from the
primary and secondary progenitor, but they are easily distinguishable
even when all stellar particles are included in one image.
Let us note that for galaxy \#277529, there are noticeable shells in stars from
both, the primary and the secondary progenitor. However, a vast majority of the Illustris
prolate rotators with shells have the shells created only by stars of
the secondary or have the shells from primary particles much less
visible.

\subsection{Observing prolate rotators} \label{sec:los}

Although our study was not designed for a direct comparison with the observed fraction and properties of prolate
rotators, we try to sketch the connection to observations by studying detectability of prolate rotation for our
sample of prolate rotators in different projection planes. When selecting our sample of prolate rotators, we took
advantage of the full 3D information available from the simulation. In observations only projected quantities are
accessible and prolate rotation will not be detected for all lines of sight. To quantify the detectability we need to
define some projected quantities.

For each projection, we include all stellar particles inside a sphere of radius $3r_{\rm max}$ which have the
projected radius lower than $r_{\rm max}$. Using 2D inertia tensor, we find the principal axes of the projected
galaxy and using the tensor eigenvalues, we calculate the ellipticity as $e=1-b/a$, where $a$ and $b$ are proportional
to the length of the major and minor 2D axis, respectively.
Thereafter we divide the particles into 11 bins equally
spaced along the minor axis ranging from $-r_{\rm max}$ to $+r_{\rm max}$.
The width of these bins is 2.5\,kpc for the smallest of our 59 prolate rotators and 24.0\,kpc for the biggest one.
We retrieve the maximum from these 11 values of the mean velocity, $V_X$, and the velocity
dispersion, $\sigma_X$, along the minor axis (i.e. the rotation \textit{around} the major axis). We repeat the
procedure with the major axis to obtain the maximum value of the mean velocity along the major axis, $V_Y$.

When the galaxy is projected onto the plane of its 3D major and intermediate axes, the maximum $V_X$ ranges between 9
and 85\,km s$^{-1}$ for our 59 prolate rotators, with the average at 31\,km s$^{-1}$. The maximum dispersion
$\sigma_X$ goes from 52 to 263\,km s$^{-1}$ with the average value of 100\,km s$^{-1}$.

\begin{deluxetable}{ccc|c}[!htb]
\tablecaption{Detectability of prolate rotation \label{tab:flos}}
\tablecolumns{4}
\tablehead{
\colhead{} & \colhead{thresholds} & \colhead{} & \colhead{fraction}
}
\startdata
$e>0.05$, & $V_X/\sigma_X>0.1$, & $V_X/V_Y>1.0$ & 0.72 \\
$e>0.05$, & $V_X/\sigma_X>0.1$, & $V_X/V_Y>1.2$ & 0.63 \\
$e>0.05$, & $V_X/\sigma_X>0.1$, & $V_X/V_Y>1.5$ & 0.49 \\
$e>0.05$, & $V_X/\sigma_X>0.2$, & $V_X/V_Y>1.0$ & 0.61 \\
$e>0.05$, & $V_X/\sigma_X>0.2$, & $V_X/V_Y>1.2$ & 0.54 \\
$e>0.05$, & $V_X/\sigma_X>0.2$, & $V_X/V_Y>1.5$ & 0.44 \\
$e>0.1$, & $V_X/\sigma_X>0.1$, & $V_X/V_Y>1.0$ & 0.66 \\
$e>$ \textbf{0.1}, & $V_X/\sigma_X>$ \textbf{0.1}, & $V_X/V_Y>$ \textbf{1.2} & \textbf{0.58} \\
$e>0.1$, & $V_X/\sigma_X>0.1$, & $V_X/V_Y>1.5$ & 0.46 \\
$e>0.1$, & $V_X/\sigma_X>0.2$, & $V_X/V_Y>1.0$ & 0.56 \\
$e>0.1$, & $V_X/\sigma_X>0.2$, & $V_X/V_Y>1.2$ & 0.50 \\
$e>0.1$, & $V_X/\sigma_X>0.2$, & $V_X/V_Y>1.5$ & 0.41 \\
\enddata
\tablecomments{
Fraction of different lines of sight satisfying given thresholds for the ellipticity, $e$, the ratio of the maximum
velocity along the minor axis to the maximum velocity dispersion along the minor axis, $V_X/\sigma_X$, and the ratio of
the maximum velocity along the minor axis to the maximum velocity along the major axis, $V_X/V_Y$. We give the average
value over 59 prolate rotators for each combination of thresholds. The fraction for the default set of thresholds is
highlighted in bold.
}
\end{deluxetable}

For each of our 59 prolate rotators we generate 6414 different viewing angles equally spaced in the solid angle
covering the whole hemisphere. We tag the respective angle of view as a detection if the projected galaxy satisfies
three criteria:
(1) it has sufficient ellipticity, $e$;
(2) the rotation around the major axis, $V_X$, is stronger than around the minor one, $V_Y$; and
(3) the rotation around the major axis, $V_X$, is reasonably strong with respect to the dispersion, $\sigma_X$.
Similar calculations were performed analytically by \cite{bin85}, who
derived probabilities for models of galaxies with a given triaxial
shape to be observed with a given projected ellipticity and a given
degree of minor-axis rotation.

\begin{figure}
\plotone{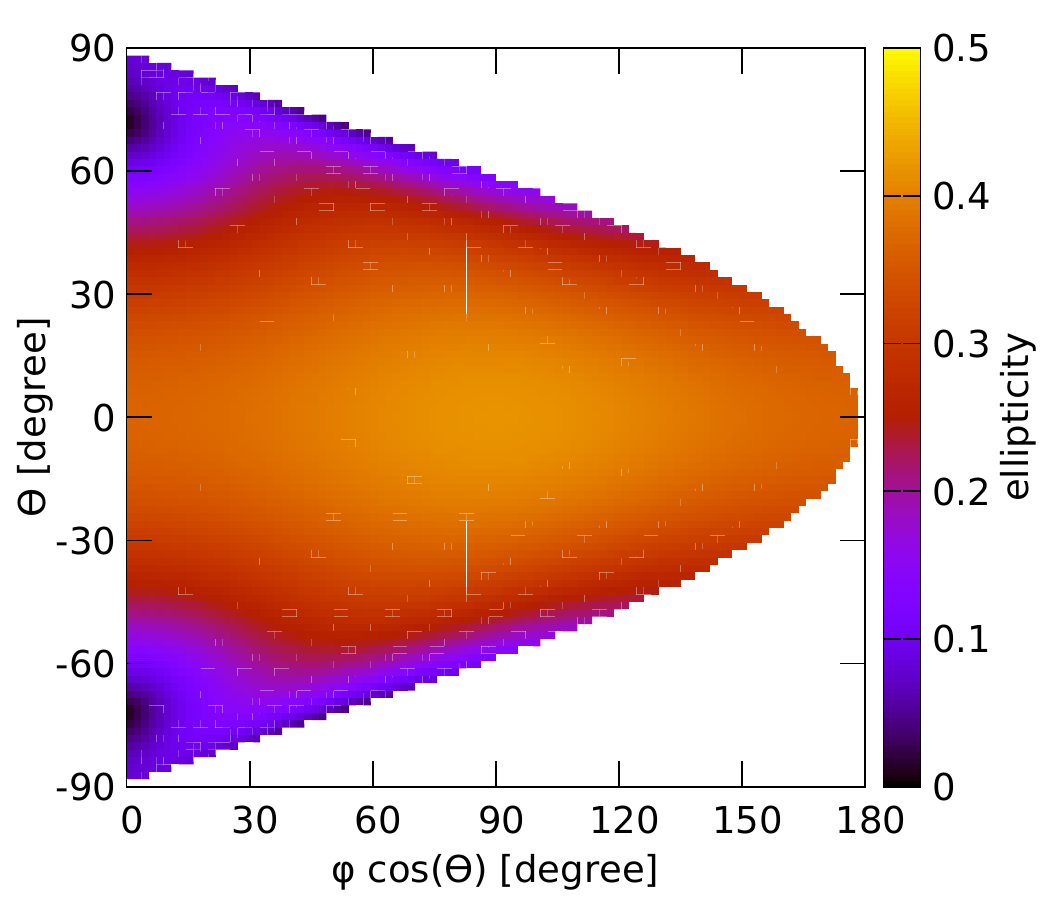}\\
\plotone{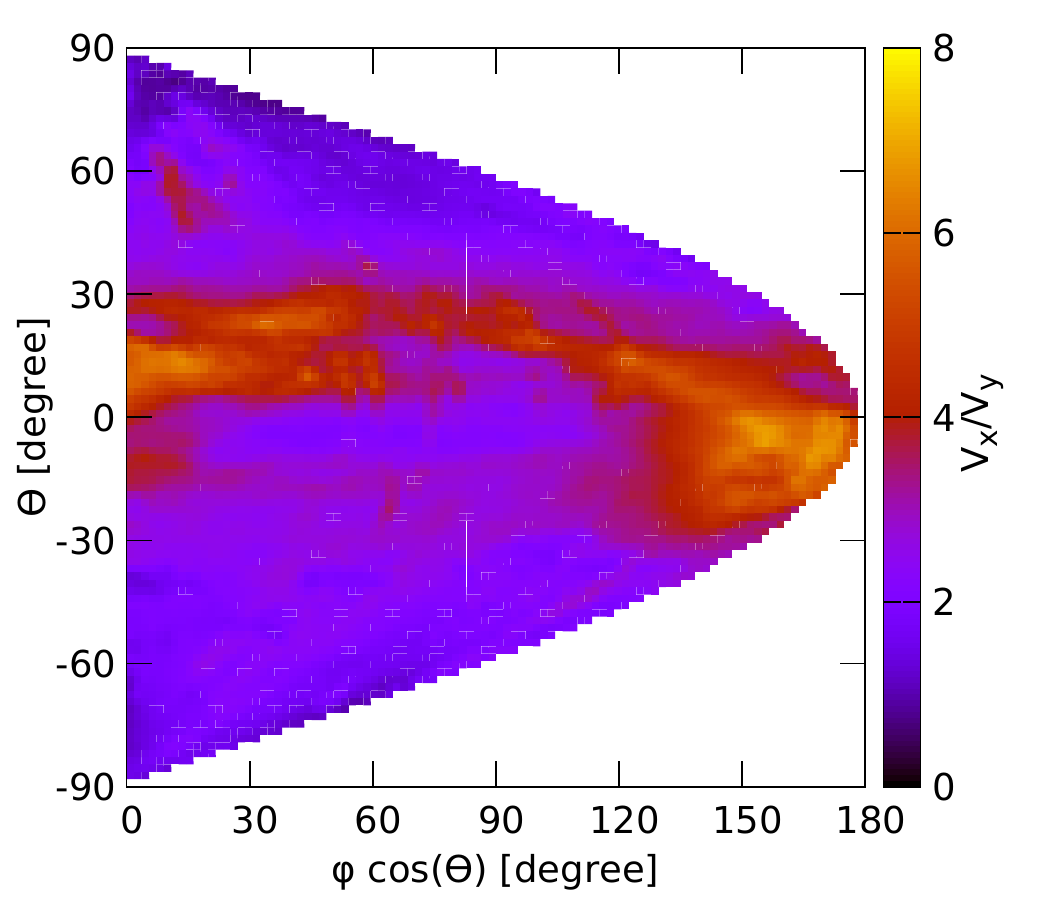}\\
\plotone{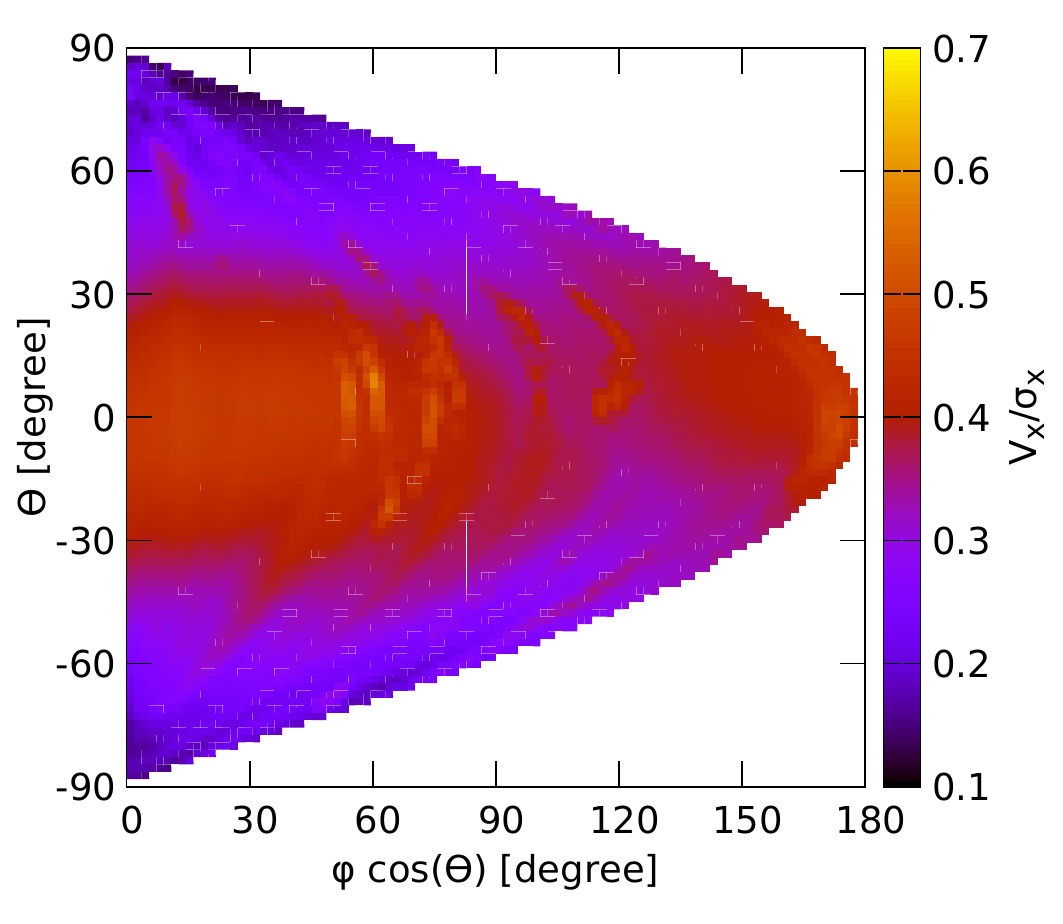}\\
\caption{
Maps of the values of the ellipticity, the ratio of the rotation around the major axis to the rotation around the minor
one, $V_X/V_Y$, and the ratio of the rotation around the major axis to the dispersion, $V_X/\sigma_X$, for lines of
sight covering the whole hemisphere for galaxy-1.  The angles ($\Theta$,$\phi$)=(0,0), (0,90) and ($\pm$90,0)
correspond to the view along the 3D minor, intermediate, and major axis, respectively.
\label{fig:los-maps}
}
\end{figure}

Fig.~\ref{fig:los-maps} shows the maps of values of $e$, $V_X/V_Y$, and $V_X/\sigma_X$ for lines of sight covering the
whole hemisphere for galaxy-1. The angles ($\Theta$,$\phi$)=(0,0), (0,90) and ($\pm$90,0) correspond to
the view along the 3D minor, intermediate, and major axis, respectively. Table~\ref{tab:flos} shows the average fraction
of positive detections for all 59 prolate rotators for different sets of thresholds. The success rate for galaxy-1
ranges from 0.98 to 0.89 for the combinations of thresholds listed in the table. Fractions of positive detections for
the default set of cuts ($e>0.1$, $V_X/V_Y>1.2$, and $V_X/\sigma_X>0.1$) for all six selected prolate rotators are
listed in the last column of Table~\ref{tab:g1-6}. For these thresholds, the detection fraction values for all 59
galaxies range from 0 (for a galaxy with too low ellipticity) to 0.93 (for galaxy-1), while the average is 0.58.

\section{Discussion} \label{sec:dis}

\subsection{Fraction of prolate rotators}

We constructed our sample selection in order to obtain a set of Illustris galaxies with a well-established rotation
around the 3D major axis, therefore our fraction of prolate rotators is not easily comparable to the observed one. Our
sample accounts for less than 1\,\% of examined Illustris galaxies. \cite{tsa17} found a volume-corrected fraction of
about 9\,\% of prolate rotators  in the CALIFA kinematic sub-sample of 81 galaxies \citep{califa300}, and they derived
about 12\,\% fraction for ATLAS$^{3{\rm D}}$ sample of ETGs \citep{a3d2}.

Our fraction of prolate rotators would be
only reduced when observed due to the fact that prolate rotation cannot be detected from all viewing angles
(Sect.~\ref{sec:los}). However, we found that additional 193 galaxies with the minimum $(L_x/L_{\rm tot})^2$ inside
$r_{\rm max}$ (see Sect.~\ref{sec:selection}) between 0.5 and 0.1 would have the average fraction of detection equal to
0.34 (for our default thresholds $e>0.1$, $V_X/V_Y>1.2$, and $V_X/\sigma_X>0.1$).

There are even several cases, where the galaxy has the minimum $(L_y/L_{\rm tot})^2$ inside $r_{\rm max}$ higher than
0.5 and such a galaxy would be seen as a prolate rotator from many different lines of sight. The $y$-axis in this case
corresponds to the intermediate axis of the galaxy.
Since this type of rotation is not supposed to be stable in a triaxial system, we do not expect it to survive
for long. Indeed, the three galaxies
we found that look like quite regular prolate rotators when viewed along the major axis,
all maintain the rotation around the intermediate axis for less than 0.5\,Gyr.

Moreover, some galaxies, that show oblate rotation in the innermost or outermost parts but prolate
rotation for the rest of the body, can be classified as prolate rotators as well
(e.g., NGC\,6338, \citealp{tsa17}; or NGC\,4365, \citealp{dav01}).
Our selection criteria are rather strict and account for galaxies that could be called `regular prolate rotators'.
Thus the fraction derived from our sample is substantially lower than the observed one.
Also galaxies with low ellipticities can be included when the position angle of the photometric axis is well determined
\citep[e.g. NGC\,5216 and NGC\,4874,][]{tsa17}.

Furthermore, we examined all galaxies with at least $10^4$
stellar particles regardless of their morphology, while the percentage of prolate rotators in CALIFA and ATLAS$^{3{\rm
D}}$ is derived only for early-type galaxies. \cite{illfrsr} used the method of \cite{sal12} to measure
the fraction of kinetic energy invested in ordered rotation for all 4591 Illustris galaxies with more than
$2\times10^4$ stellar particles and classified 3207 galaxies as elliptical, corresponding to a quite high fraction
of 70\,\%. They also stress difficulties in assigning the morphological type for lower-mass galaxies in Illustris
as described in Section~2.3 of \cite{illfrsr}. To mimic the selection procedure of ETGs in the CALIFA or ATLAS$^{3{\rm
D}}$ sample with Illustris galaxies would be quite demanding and we did not attempt it here.

\subsection{Sample mass distribution}

The mass distribution of the sample of Illustris prolate rotators seems to follow the same trend as the observed ones do.
As pointed out in \cite{tsa17}, when considering only ETGs with the stellar mass higher than $2\times10^{11}$\,M$_{\sun}$, the
fraction of prolate rotators increases to 27\,\% for the CALIFA sample and to 23\,\% for ATLAS$^{3{\rm D}}$. The
first results of the M3G project (MUSE Most Massive galaxies) suggest that more than a half of very-high-mass galaxies
rotate in a rather prolate manner (8 prolate, 5 oblate, and 1 non-rotators; Emsellem, conference
presentation\footnote{\url{http://www.astroscu.unam.mx/galaxies2016/presentaciones/Miercoles/Cozumel2016_Emsellem.pdf}}).
Fig.~\ref{fig:Mdist} shows that in Illustris massive galaxies are more likely to have prolate rotation than the less
massive ones.

\subsection{Origin of prolate rotation}

For at least 40 out of 59 Illustris galaxies with well-established rotation around the 3D major axis, the emergence of
prolate rotation is well correlated with the time of the last significant merger (Fig.~\ref{fig:cor-KT-ST}). Four
galaxies did not experience any significant merger (i.e. with the mass ratio above 0.1). One of them is about to be
swallowed by a galaxy approximately 10 times more massive and the prolate rotation is probably caused by the tidal
deformation in the vicinity of the massive galaxy acting on a galaxy that already had a prolate shape.
The other three have rather prolate shapes and rather high but
fluctuating values of $(L_x/L_{\rm tot})^2$ for many Gyr; one seems to be experiencing an increase of prolate
rotation due to an ongoing merger with a slightly less massive galaxy, for the other two we were unable to identify the
reason for the increase.

Concerning the 15 galaxies with weak correlation between the time of the merger and the time kinematic transition, two
experienced several simultaneous or subsequent mergers during a short period of time and the prolate rotation emerged
already during the penultimate merger and was strengthened or at least not destroyed by the last one; one galaxy has
had a prolate shape and rotation for more than 10\,Gyr but a merger 4.3\,Gyr ago seems to be crucial for the galaxy to
maintain the prolate rotation until the end of the simulation; 1--4 galaxies have prolate rotation emerged during a
merger event but something, most likely nearby galaxies, caused a temporary drop in $(L_x/L_{\rm tot})^2$ shifting the
time of kinematic transition to more recent times; 4--7 galaxies seem to be strongly influenced by a close flyby, by an
ongoing merger or by a `near-merger' at the time when the prolate rotation was created or strengthened (where by
`near-merger' we mean a situation when a smaller galaxy was almost swallowed by our prolate rotator but its small core
survived and managed to escape, so the galaxy does not appear in the merger tree of the examined galaxy); for several
galaxies the origin of prolate rotation remains unknown.

All these galaxies, including the four galaxies without a
significant merger but excluding the two with prolate rotation emerging during the preceding merger, have a common
feature: the galaxies already had a rather prolate shape, usually for several Gyr (with possible shape
fluctuations) before prolate rotation arose. The prolate shape was built during an early formation epoch or during
a significant merger. Some of these galaxies already experienced periods of prolate rotation before the final kinematic
transition.

\subsection{Merger parameters}

Even though mergers appear to be the main cause for the emergence of prolate rotation, there seem to be no strongly
favored initial conditions of such a merger. In comparison with the twin sample (Sect.~\ref{sec:others}), prolate
rotators result from more radial mergers. They also experience slightly more mergers with higher mass ratio and more
recent mergers resulting in a slightly higher fraction of remnants with lower postmerger gas fraction
(Fig.~\ref{fig:hisother}). Despite these trends, the mergers, which are probably responsible for prolate
rotation, cover a wide range of initial conditions in terms of the mass ratio (0.2--1), merger time (0-8\,Gyr ago),
postmerger gas fraction (0-0.9), radiality of the orbit and the relative orientations of spins of the progenitors and
the orbital angular momenta (Sect.~\ref{sec:merger}, Figs.~\ref{fig:histograms1} and \ref{fig:histograms-alpha}).

\begin{figure}
\plotone{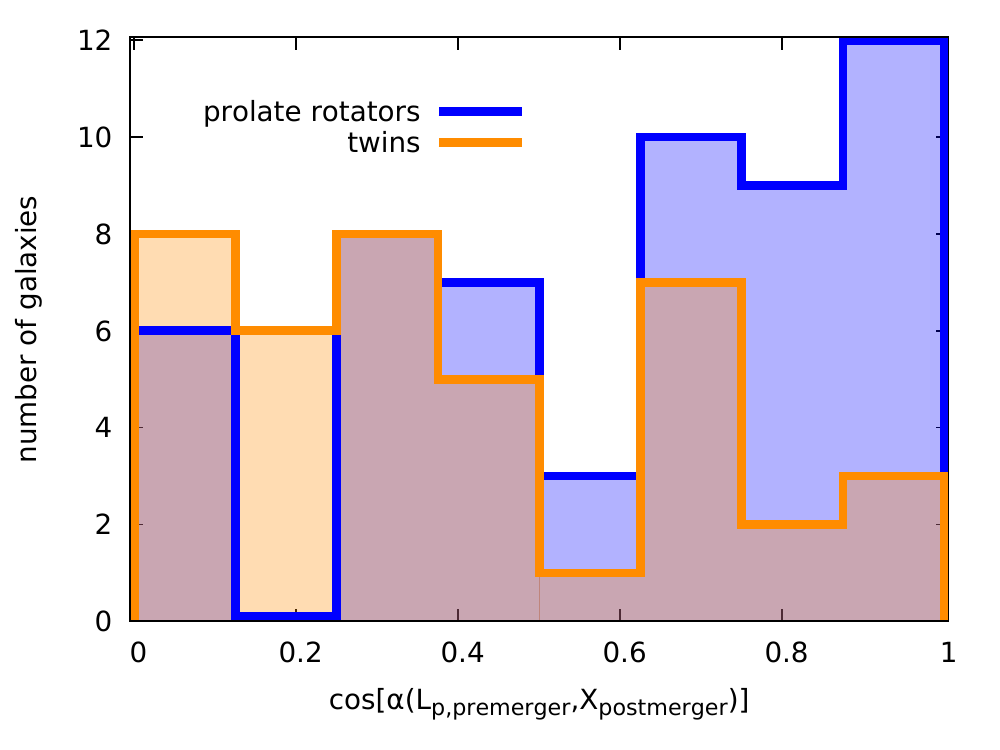}
\caption{
Histograms of the orientation of the spin of the primary progenitor before the merger with respect to the postmerger
major axis for all 55 last significant mergers for the prolate rotators compared to the 40 last significant mergers of
the twin sample.
\label{fig:LX}
}
\end{figure}

The major axis of the emerging prolate rotators is set by the direction from which the progenitors approach each other at
the end of the merger (Fig.~\ref{fig:DXL}). However, even the orientation of the spin of the primary progenitor before
the merger with respect to the postmerger major axis, $\alpha(L_{p,\rm premerger},X_{\rm postmerger})$, covers a full
range of possible values. It probably just slightly prefers more parallel orientations but more interestingly, for the
twin sample the parallel orientations seem to be disfavored, as illustrated in Fig.~\ref{fig:LX}. This suggests that
prolate rotators can emerge from a variety of possible $\alpha(L_{p,\rm premerger},X_{\rm postmerger})$, but if the
orientation is parallel, the creation of the prolate rotation is favored. This would be possible even if the parallel
orientation is generally (in Illustris) disfavored, at least for the given set of masses and flattening of the
galaxies. Our results thus indicate that the `polar merger' scenario
recently advocated by \cite{tsa17} may be only one of many possible channels for the formation of prolate rotators.

We emphasize that Illustris data show that, contrary to previous conjectures \citep[e.g.,][]{tsa17,li17},
prolate rotators can be created in wet, not only dry, mergers: 44\,\% of our prolate rotators have gas fractions higher than
0.5 at the end of the last significant merger (bottom panel of Fig.~\ref{fig:histograms1}). In some cases, new stars
formed during the merger can be seen as an edge-on disk in $xy$ plane (plane of the 3D major and intermediate axes)
aligned or mildly tilted with respect to the minor projected axis (see the bottom right panel of
Fig.~\ref{fig:map-psn}). In many cases, the new stars even follow the prolate shape of the galaxy and sometimes
they account for most of the $x$-component of the angular momentum at $z=0$. Only for two galaxies the rotation of new
stars is in the opposite direction to the majority of rotation around the 3D major axis.

\section{ Conclusions } \label{sec:con}

We used the publicly available data of the large-scale cosmological hydrodynamical simulation Illustris and analyzed
all 7697 galaxies with more than $10^4$ stellar particles in the final output (redshift $z=0$) of the
Illustris-1 run.

We identified a sample of 59 galaxies with a well-established rotation around the 3D major axis. All but one galaxy
maintain prolate rotation for less than 7\,Gyr and the majority of the sample for less than 3\,Gyr.

The mass distribution of the sample shows a clear trend for prolate rotators to be more abundant among more massive
galaxies (Fig.~\ref{fig:Mdist}), which is consistent with observations \citep{tsa17,a3d2}.

The time of emergence of prolate rotation is strongly correlated with the time of the last significant merger
(i.e. a merger with the mass ratio of at least 0.1; Fig.~\ref{fig:cor-KT-ST}). Some prolate rotators seem to be related
to an ongoing merger, a non-merger interaction, or they have unclear origin.
In the cases when prolate rotation occurs without an associated merger event, a pre-existing rather prolate shape seems to be a necessary condition.

The last significant mergers have slightly more radial orbits, higher mass ratios, and more recent times than the
mergers in a comparison sample of twin galaxies (Fig.~\ref{fig:hisother}). However, the mergers of prolate rotators
cover a wide range of initial conditions in terms of the mass ratio, merger time, radiality of the orbit, and the
relative orientations of spins of the progenitors and the orbital angular momenta (Figs.~\ref{fig:histograms1} and
\ref{fig:histograms-alpha}).

The formation of prolate rotators via a merger event seems to be related to the final stages of the merger, in which the
major axis of the arising prolate rotators is being established (Fig.~\ref{fig:DXL}). The actual creation of the
prolate rotation seems to depend on the details of the merger evolution that are not easily inferable from the initial
conditions of the merger.

About half of our sample of Illustris prolate rotators were created during gas-rich mergers. New stars usually support the prolate rotation.
They preferentially settle in the plane of the 3D minor and intermediate axes in a form of a disk or, more often, in a form of a thicker
ellipsoid.

There is a high incidence of shell galaxies, with the axis of stellar shells well-aligned with the major projected axis, among both,
observed (about 40\,\%) and Illustris (24--50\,\%) prolate rotators (Sect.~\ref{sec:shells}), which also supports their merger origin.
\\



\acknowledgments
We thank the anonymous referee for the thorough review that helped us improve this manuscript. 
We are grateful to Athanasia Tsatsi for useful comments. 
This research was supported in part by the Polish National Science Centre under grant 2013/10/A/ST9/00023.


\bibliographystyle{aasjournal}
\bibliography{prolate}

\begin{thebibliography}{}
\expandafter\ifx\csname natexlab\endcsname\relax\def\natexlab#1{#1}\fi
\providecommand{\url}[1]{\href{#1}{#1}}

\bibitem[{{Arp}(1966)}]{arp66}
{Arp}, H. 1966, \apjs, 14, 1

\bibitem[{{Atkinson} {et~al.}(2013){Atkinson}, {Abraham}, \& {Ferguson}}]{at13}
{Atkinson}, A.~M., {Abraham}, R.~G., \& {Ferguson}, A.~M.~N. 2013, \apj, 765,
  28

\bibitem[{{Bertola} {et~al.}(1999){Bertola}, {Corsini}, {Beltr{\'a}n},
  {Pizzella}, {Sarzi}, {Cappellari}, \& {Funes}}]{ber99}
{Bertola}, F., {Corsini}, E.~M., {Beltr{\'a}n}, J.~C.~V., {et~al.} 1999, \apjl,
  519, L127

\bibitem[{{Binney}(1985)}]{bin85}
{Binney}, J. 1985, \mnras, 212, 767

\bibitem[{{Davies} {et~al.}(2001){Davies}, {Kuntschner}, {Emsellem}, {Bacon},
  {Bureau}, {Carollo}, {Copin}, {Miller}, {Monnet}, {Peletier}, {Verolme}, \&
  {de Zeeuw}}]{dav01}
{Davies}, R.~L., {Kuntschner}, H., {Emsellem}, E., {et~al.} 2001, \apjl, 548,
  L33

\bibitem[{{de Zeeuw}(1985)}]{dz85}
{de Zeeuw}, T. 1985, \mnras, 216, 273

\bibitem[{{Deason} {et~al.}(2014){Deason}, {Wetzel}, \&
  {Garrison-Kimmel}}]{dea14}
{Deason}, A., {Wetzel}, A., \& {Garrison-Kimmel}, S. 2014, \apj, 794, 115

\bibitem[{{del Pino} {et~al.}(2017){del Pino}, {{\L}okas}, {Hidalgo}, \&
  {Fouquet}}]{dp17}
{del Pino}, A., {{\L}okas}, E.~L., {Hidalgo}, S.~L., \& {Fouquet}, S. 2017,
  \mnras, 469, 4999

\bibitem[{{Duc} {et~al.}(2011){Duc}, {Cuillandre}, {Serra}, {Michel-Dansac},
  {Ferriere}, {Alatalo}, {Blitz}, {Bois}, {Bournaud}, {Bureau}, {Cappellari},
  {Davies}, {Davis}, {de Zeeuw}, {Emsellem}, {Khochfar}, {Krajnovi{\'c}},
  {Kuntschner}, {Lablanche}, {McDermid}, {Morganti}, {Naab}, {Oosterloo},
  {Sarzi}, {Scott}, {Weijmans}, \& {Young}}]{duc11}
{Duc}, P.-A., {Cuillandre}, J.-C., {Serra}, P., {et~al.} 2011, \mnras, 417, 863

\bibitem[{{Duc} {et~al.}(2015){Duc}, {Cuillandre}, {Karabal}, {Cappellari},
  {Alatalo}, {Blitz}, {Bournaud}, {Bureau}, {Crocker}, {Davies}, {Davis}, {de
  Zeeuw}, {Emsellem}, {Khochfar}, {Krajnovi{\'c}}, {Kuntschner}, {McDermid},
  {Michel-Dansac}, {Morganti}, {Naab}, {Oosterloo}, {Paudel}, {Sarzi}, {Scott},
  {Serra}, {Weijmans}, \& {Young}}]{duc15}
{Duc}, P.-A., {Cuillandre}, J.-C., {Karabal}, E., {et~al.} 2015, \mnras, 446,
  120

\bibitem[{{Dupraz} \& {Combes}(1986)}]{dc86}
{Dupraz}, C., \& {Combes}, F. 1986, \aap, 166, 53

\bibitem[{{Ebrov{\'a}} {et~al.}(2012){Ebrov{\'a}}, {J{\'{\i}}lkov{\'a}},
  {Jungwiert}, {K{\v r}{\'{\i}}{\v z}ek}, {B{\'{\i}}lek}, {Barto{\v
  s}kov{\'a}}, {Skalick{\'a}}, \& {Stoklasov{\'a}}}]{e12sg}
{Ebrov{\'a}}, I., {J{\'{\i}}lkov{\'a}}, L., {Jungwiert}, B., {et~al.} 2012,
  \aap, 545, A33

\bibitem[{{Ebrov{\'a}} \& {{\L}okas}(2015)}]{e15andii}
{Ebrov{\'a}}, I., \& {{\L}okas}, E.~L. 2015, \apj, 813, 10

\bibitem[{{Faber} \& {Jackson}(1976)}]{fj76}
{Faber}, S.~M., \& {Jackson}, R.~E. 1976, \apj, 204, 668

\bibitem[{{Falc{\'o}n-Barroso} {et~al.}(2003){Falc{\'o}n-Barroso}, {Balcells},
  {Peletier}, \& {Vazdekis}}]{fa03}
{Falc{\'o}n-Barroso}, J., {Balcells}, M., {Peletier}, R.~F., \& {Vazdekis}, A.
  2003, \aap, 405, 455

\bibitem[{{Falc{\'o}n-Barroso} {et~al.}(2017){Falc{\'o}n-Barroso}, {Lyubenova},
  {van de Ven}, {Mendez-Abreu}, {Aguerri}, {Garc{\'{\i}}a-Lorenzo},
  {Bekerait{\'e}}, {S{\'a}nchez}, {Husemann}, {Garc{\'{\i}}a-Benito}, {Mast},
  {Walcher}, {Zibetti}, {Barrera-Ballesteros}, {Galbany},
  {S{\'a}nchez-Bl{\'a}zquez}, {Singh}, {van den Bosch}, {Wild}, {Zhu},
  {Bland-Hawthorn}, {Cid Fernandes}, {de Lorenzo-C{\'a}ceres}, {Gallazzi},
  {Gonz{\'a}lez Delgado}, {Marino}, {M{\'a}rquez}, {P{\'e}rez}, {P{\'e}rez},
  {Roth}, {Rosales-Ortega}, {Ruiz-Lara}, {Wisotzki}, {Ziegler}, \& {Califa
  Collaboration}}]{califa300}
{Falc{\'o}n-Barroso}, J., {Lyubenova}, M., {van de Ven}, G., {et~al.} 2017,
  \aap, 597, A48

\bibitem[{{Forbes} {et~al.}(1995){Forbes}, {Franx}, \& {Illingworth}}]{for95}
{Forbes}, D.~A., {Franx}, M., \& {Illingworth}, G.~D. 1995, \aj, 109, 1988

\bibitem[{{Forbes} {et~al.}(1994){Forbes}, {Thomson}, {Groom}, \&
  {Williger}}]{for94}
{Forbes}, D.~A., {Thomson}, R.~C., {Groom}, W., \& {Williger}, G.~M. 1994, \aj,
  107, 1713

\bibitem[{{Fouquet} {et~al.}(2017){Fouquet}, {{\L}okas}, {del Pino}, \&
  {Ebrov{\'a}}}]{f17andii}
{Fouquet}, S., {{\L}okas}, E.~L., {del Pino}, A., \& {Ebrov{\'a}}, I. 2017,
  \mnras, 464, 2717

\bibitem[{{Hernquist}(1990)}]{her90}
{Hernquist}, L. 1990, \apj, 356, 359

\bibitem[{{Hernquist} \& {Quinn}(1988)}]{hq88}
{Hernquist}, L., \& {Quinn}, P.~J. 1988, \apj, 331, 682

\bibitem[{{Hinshaw} {et~al.}(2013){Hinshaw}, {Larson}, {Komatsu}, {Spergel},
  {Bennett}, {Dunkley}, {Nolta}, {Halpern}, {Hill}, {Odegard}, {Page}, {Smith},
  {Weiland}, {Gold}, {Jarosik}, {Kogut}, {Limon}, {Meyer}, {Tucker}, {Wollack},
  \& {Wright}}]{wmap9}
{Hinshaw}, G., {Larson}, D., {Komatsu}, E., {et~al.} 2013, \apjs, 208, 19

\bibitem[{{Ho} {et~al.}(2012){Ho}, {Geha}, {Munoz}, {Guhathakurta}, {Kalirai},
  {Gilbert}, {Tollerud}, {Bullock}, {Beaton}, \& {Majewski}}]{ho12}
{Ho}, N., {Geha}, M., {Munoz}, R.~R., {et~al.} 2012, \apj, 758, 124

\bibitem[{{Jaffe}(1983)}]{jaff83}
{Jaffe}, W. 1983, \mnras, 202, 995

\bibitem[{{Jarvis} {et~al.}(1988){Jarvis}, {Dubath}, {Martinet}, \&
  {Bacon}}]{ja88}
{Jarvis}, B.~J., {Dubath}, P., {Martinet}, L., \& {Bacon}, R. 1988, \aaps, 74,
  513

\bibitem[{{Jedrzejewski} \& {Schechter}(1988)}]{js88}
{Jedrzejewski}, R., \& {Schechter}, P.~L. 1988, \apjl, 330, L87

\bibitem[{{Kacharov} {et~al.}(2017){Kacharov}, {Battaglia}, {Rejkuba}, {Cole},
  {Carrera}, {Fraternali}, {Wilkinson}, {Gallart}, {Irwin}, \&
  {Tolstoy}}]{kach17}
{Kacharov}, N., {Battaglia}, G., {Rejkuba}, M., {et~al.} 2017, \mnras, 466,
  2006

\bibitem[{{Klimentowski} {et~al.}(2010){Klimentowski}, {{\L}okas}, {Knebe},
  {Gottl{\"o}ber}, {Martinez-Vaquero}, {Yepes}, \& {Hoffman}}]{kli10}
{Klimentowski}, J., {{\L}okas}, E.~L., {Knebe}, A., {et~al.} 2010, \mnras, 402,
  1899

\bibitem[{{Krajnovi{\'c}} {et~al.}(2011){Krajnovi{\'c}}, {Emsellem},
  {Cappellari}, {Alatalo}, {Blitz}, {Bois}, {Bournaud}, {Bureau}, {Davies},
  {Davis}, {de Zeeuw}, {Khochfar}, {Kuntschner}, {Lablanche}, {McDermid},
  {Morganti}, {Naab}, {Oosterloo}, {Sarzi}, {Scott}, {Serra}, {Weijmans}, \&
  {Young}}]{a3d2}
{Krajnovi{\'c}}, D., {Emsellem}, E., {Cappellari}, M., {et~al.} 2011, \mnras,
  414, 2923

\bibitem[{{Li} {et~al.}(2018){Li}, {Mao}, {Emsellem}, {Xu}, {Springel}, \&
  {Krajnovi{\'c}}}]{li17}
{Li}, H., {Mao}, S., {Emsellem}, E., {et~al.} 2018, \mnras, 473, 1489

\bibitem[{{{\L}okas} {et~al.}(2014){{\L}okas}, {Ebrov{\'a}}, {Pino}, \&
  {Semczuk}}]{lo14andii}
{{\L}okas}, E.~L., {Ebrov{\'a}}, I., {Pino}, A.~d., \& {Semczuk}, M. 2014,
  \mnras, 445, L6

\bibitem[{{{\L}okas} {et~al.}(2015){{\L}okas}, {Semczuk}, {Gajda}, \&
  {D'Onghia}}]{lok15ts}
{{\L}okas}, E.~L., {Semczuk}, M., {Gajda}, G., \& {D'Onghia}, E. 2015, \apj,
  810, 100

\bibitem[{{Malin} \& {Carter}(1983)}]{mc83}
{Malin}, D.~F., \& {Carter}, D. 1983, \apj, 274, 534

\bibitem[{{Mayer} {et~al.}(2001){Mayer}, {Governato}, {Colpi}, {Moore},
  {Quinn}, {Wadsley}, {Stadel}, \& {Lake}}]{ma01}
{Mayer}, L., {Governato}, F., {Colpi}, M., {et~al.} 2001, \apj, 559, 754

\bibitem[{{Naab} {et~al.}(2014){Naab}, {Oser}, {Emsellem}, {Cappellari},
  {Krajnovi{\'c}}, {McDermid}, {Alatalo}, {Bayet}, {Blitz}, {Bois}, {Bournaud},
  {Bureau}, {Crocker}, {Davies}, {Davis}, {de Zeeuw}, {Duc}, {Hirschmann},
  {Johansson}, {Khochfar}, {Kuntschner}, {Morganti}, {Oosterloo}, {Sarzi},
  {Scott}, {Serra}, {Ven}, {Weijmans}, \& {Young}}]{naa14}
{Naab}, T., {Oser}, L., {Emsellem}, E., {et~al.} 2014, \mnras, 444, 3357

\bibitem[{{Nelson} {et~al.}(2015){Nelson}, {Pillepich}, {Genel},
  {Vogelsberger}, {Springel}, {Torrey}, {Rodriguez-Gomez}, {Sijacki}, {Snyder},
  {Griffen}, {Marinacci}, {Blecha}, {Sales}, {Xu}, \&
  {Hernquist}}]{nel15illpub}
{Nelson}, D., {Pillepich}, A., {Genel}, S., {et~al.} 2015, Astronomy and
  Computing, 13, 12

\bibitem[{{Penoyre} {et~al.}(2017){Penoyre}, {Moster}, {Sijacki}, \&
  {Genel}}]{illfrsr}
{Penoyre}, Z., {Moster}, B.~P., {Sijacki}, D., \& {Genel}, S. 2017, \mnras,
  468, 3883

\bibitem[{{Pop} {et~al.}(2017){Pop}, {Pillepich}, {Amorisco}, \&
  {Hernquist}}]{illsg17}
{Pop}, A.-R., {Pillepich}, A., {Amorisco}, N.~C., \& {Hernquist}, L. 2017,
  ArXiv e-prints, arXiv:1706.06102

\bibitem[{{Prieur}(1990)}]{pri90}
{Prieur}, J.-L. 1990, in Dynamics and Interactions of Galaxies, ed.
  R.~{Wielen}, 72--83

\bibitem[{{Quinn}(1984)}]{q84}
{Quinn}, P.~J. 1984, \apj, 279, 596

\bibitem[{{Rodriguez-Gomez} {et~al.}(2015){Rodriguez-Gomez}, {Genel},
  {Vogelsberger}, {Sijacki}, {Pillepich}, {Sales}, {Torrey}, {Snyder},
  {Nelson}, {Springel}, {Ma}, \& {Hernquist}}]{rg15illmer}
{Rodriguez-Gomez}, V., {Genel}, S., {Vogelsberger}, M., {et~al.} 2015, \mnras,
  449, 49

\bibitem[{{Sales} {et~al.}(2012){Sales}, {Navarro}, {Theuns}, {Schaye},
  {White}, {Frenk}, {Crain}, \& {Dalla Vecchia}}]{sal12}
{Sales}, L.~V., {Navarro}, J.~F., {Theuns}, T., {et~al.} 2012, \mnras, 423,
  1544

\bibitem[{{Snyder} {et~al.}(2015){Snyder}, {Torrey}, {Lotz}, {Genel},
  {McBride}, {Vogelsberger}, {Pillepich}, {Nelson}, {Sales}, {Sijacki},
  {Hernquist}, \& {Springel}}]{syn15illmorph}
{Snyder}, G.~F., {Torrey}, P., {Lotz}, J.~M., {et~al.} 2015, \mnras, 454, 1886

\bibitem[{{Springel}(2010)}]{sp10}
{Springel}, V. 2010, \mnras, 401, 791

\bibitem[{{Tal} {et~al.}(2009){Tal}, {van Dokkum}, {Nelan}, \&
  {Bezanson}}]{tal09}
{Tal}, T., {van Dokkum}, P.~G., {Nelan}, J., \& {Bezanson}, R. 2009, \aj, 138,
  1417

\bibitem[{{Tsatsi} {et~al.}(2017){Tsatsi}, {Lyubenova}, {van de Ven}, {Chang},
  {Aguerri}, {Falc{\'o}n-Barroso}, \& {Macci{\`o}}}]{tsa17}
{Tsatsi}, A., {Lyubenova}, M., {van de Ven}, G., {et~al.} 2017, \aap, 606, A62

\bibitem[{{Vogelsberger} {et~al.}(2014{\natexlab{a}}){Vogelsberger}, {Genel},
  {Springel}, {Torrey}, {Sijacki}, {Xu}, {Snyder}, {Bird}, {Nelson}, \&
  {Hernquist}}]{vog14illpreintro}
{Vogelsberger}, M., {Genel}, S., {Springel}, V., {et~al.} 2014{\natexlab{a}},
  \nat, 509, 177

\bibitem[{{Vogelsberger} {et~al.}(2014{\natexlab{b}}){Vogelsberger}, {Genel},
  {Springel}, {Torrey}, {Sijacki}, {Xu}, {Snyder}, {Nelson}, \&
  {Hernquist}}]{vog14illintro}
---. 2014{\natexlab{b}}, \mnras, 444, 1518

\bibitem[{{Willett} {et~al.}(2013){Willett}, {Lintott}, {Bamford}, {Masters},
  {Simmons}, {Casteels}, {Edmondson}, {Fortson}, {Kaviraj}, {Keel}, {Melvin},
  {Nichol}, {Raddick}, {Schawinski}, {Simpson}, {Skibba}, {Smith}, \&
  {Thomas}}]{wil13}
{Willett}, K.~W., {Lintott}, C.~J., {Bamford}, S.~P., {et~al.} 2013, \mnras,
  435, 2835

\end{thebibliography}

\end{document}